\documentclass[preprint]{imsart}

\usepackage{amstext}
\arxiv{arXiv:0000.0000}

\usepackage{amsmath}
\usepackage{amssymb}
\usepackage{amsbsy} 
\usepackage{amsthm}
\usepackage{mathrsfs}
\usepackage{eufrak}
\usepackage{mathrsfs}
\usepackage{color}
\usepackage{verbatim}
\usepackage{graphicx}
\usepackage{bm}
\usepackage{enumerate}
\usepackage{natbib}
\RequirePackage[colorlinks,citecolor=blue,urlcolor=blue]{hyperref}
\usepackage{subfig}
\usepackage[final]{pdfpages}

\usepackage{algorithm}  %@subhajit
\usepackage{algpseudocode} %@subhajit
\usepackage{algorithmicx}     %@subhajit

\newcommand{\bx}{\bm{x}}
\newcommand{\bX}{\bm{X}}
\newcommand{\by}{\bm{y}}
\newcommand{\bZ}{\bm{Z}}

\newcommand{\btheta}{\bm{\theta}}
\newcommand{\Bpi}{\boldsymbol{\pi}}
\newcommand{\thetabf}{\boldsymbol{\theta}}

\newcommand{\HaarMu}{\mu}

\newcommand{\ConstOne}{K}

\newcommand{\bM}{\bm{M}}
\newcommand{\bD}{\bm{D}}
\newcommand{\bV}{\bm{V}}
\newcommand{\loglikmix}{\mathcal{L}_{\bM,\bD,\bV}}
\newcommand{\hypdc}{{}_0F_1\left(\frac{n}{2},\frac{D_c^2}{4}\right)}

\usepackage{xstring}
\usepackage[normalem]{ulem}
\definecolor{ultramarine}{RGB}{38,29,163}

\newcommand{\qedwhite}{\hfill \ensuremath{\Box}}
\newcommand{\SpaceD}{\mathcal{S}_p}
\newcommand{\SpaceM}{\widetilde{\mathcal{V}}_{n,p}}
\newcommand{\SpaceV}{\mathcal{V}_{p,p}}

\newcommand{\StiefelS}{\mathcal{V}_{n,p}}
\newcommand{\SpacePi}{\mathbb{S}_{\pi}}
\newcommand{\ML}{{\cal{ML}}}
\newcommand{\ProdSpace}{\boldsymbol{\Theta}}
\newcommand{\ThetaAndPi}{\Xi}
\newcommand{\ClassML}{\mathcal{C}_{\ML}}

\newcommand{\bbeta}{\bm{\beta}}
\newcommand{\bEta}{\bm{\eta}}
\newcommand{\bd}{\bm{d}}
\newcommand{\BoEta}{\boldsymbol{\eta}}
\newtheorem{theorem}{Theorem}
\newtheorem{lemma}{Lemma}
\newtheorem{defn}{Definition}
\newcommand{\pdv}[2][]{\frac{\partial#1}{\partial#2}}
\newcommand{\pdvtwo}[2][]{\frac{\partial^2#1}{{\partial#2}^2}}

\newcommand{\priorXzero}{\Psi}
\newcommand{\iMat}{\mathbf{I}_{p}}

\newcommand{\bpi} {\boldsymbol{\pi}}
\newcommand{\bphi} {\boldsymbol{\phi}}

\newcommand{\normtwo}[1]{{\left\lVert#1\right\rVert}_2}

\newcommand{\fthetapi}[1]{f_{\btheta_{#1},\bpi_{#1}}}
\newcommand{\hyp}{{}_0F_1\left(\frac{n}{2},\frac{D^2}{4}\right)}
\newcommand{\hypinline}{{}_0F_1\left({n}/{2},{D^2}/{4}\right)}

\newcommand{\partialhyp}[1]{\frac{\partial}{\partial\,{d_{#1}}}\,\left[\hyp\right]}

\newcommand{\fracProbZ}[1]{\frac{\langle Z_{ic} \rangle \, #1}{\sum_{i=1}^{N} \langle Z_{ic}\rangle  } }
\newcommand{\EmVar}[1]{\widetilde{#1}^{(c)}}

\newcommand{\IMDY}{{\it{IMDY}}}
\newcommand{\JMDY}{{\it{JMDY}}}

\newcommand{\DYlang}{\frac{\exp(\nu\,\bEta^T\bd)}{{\left[{}_0F_1\left(\frac{n}{2},\frac{D^2}{4}\right)\right]}^{\nu}}}

\newcommand{\logDYlang}{\nu\,\bEta^T\bd - \nu\,\log\left({}_0F_1\left(\frac{n}{2},\frac{D^2}{4}\right)\right)}

\newcommand{\lhyp}{\log\left({}_0F_1\left(\frac{n}{2},\frac{D^2}{4}\right)\right)}

\newcommand{\diam}[1]{{{#1}^{\ast}}}

\definecolor{auburn}{rgb}{0.53, 0.1, 0.5}
\newcommand{\MLDensity}{f_{\ML}}
\newcommand{\posterior}

\newcommand{\variableX}{\bd}

\newcommand{\IndVzero}[1]{\mathbb{I}({X\in \mathcal{V}^{#1}_0})}
\newcommand{\Rnp}{\mathbb{R}^{n \times p}}
\newcommand{\Rpp}{\mathbb{R}^{p \times p}}
\newcommand{\vecnorm}[1]{\lVert #1\rVert}

\newcommand{\etapsiD}{\eta_{\priorXzero}}
\newcommand{\BoEtapsiD}{\BoEta_{\priorXzero}}

\newcommand{\Rplus}{\mathbb{R}_{+}}
\newcommand{\prodMeasure}{\Upsilon}

\newcommand{\m}{{\bf m_{\BoEta}}} 
\newcommand{\SetWithMode}{\mathcal{S}}
\newcommand{\SetWithModePrime}{\mathcal{S}}
\newcommand{\TargetComp}{\mathcal{S}^{\star}}

\newcommand{\ConstCondDen}{K_{\nu, \BoEta}} 

\newcommand{\hyparam}[2]{
    \IfEqCase{#1}{
        {M}{\xi^{#2}_c}
        {V}{\gamma^{#2}_c}%
        
    }
  }
\newcommand{\threepartdef}[6]
{
	\left\{
		\begin{array}{lll}
			#1 & \mbox{if } #2 \\
			#3 & \mbox{if } #4 \\
			#5 & \mbox{if } #6
		\end{array}
	\right.
}

\begin{document}
\begin{frontmatter}

\title{A Bayesian Mixture Model for Clustering on the Stiefel Manifold %\sss{shouldn't we focus on the first story?}
}
\runtitle{BMMCSM}
\thankstext{t1}{Equal contribution}
\thankstext{t2}{Corresponding author}
%%%%%%%%%%%%%%%%%%%%%
\begin{aug}
\author{\fnms{Subhajit} \snm{Sengupta}\thanksref{t1,m1}\ead[label=e1]{subhajit@uchicago.edu}},
\author{\fnms{Subhadip} \snm{Pal}\thanksref{t1,m2}\ead[label=e2]{subhadip.pal@louisville.edu}}
\author{\fnms{Riten} \snm{Mitra}\thanksref{m2}
\ead[label=e3]{r0mitr01@louisville.edu}
},
\author{\fnms{Ying} \snm{Guo}\thanksref{m3}\ead[label=e4]{yguo2@emory.edu}},
\author{\fnms{Arunava} \snm{Banerjee}\thanksref{t2,m4}\ead[label=e5]{ arunava@cise.ufl.edu}}
\and
\author{\fnms{Yuan} \snm{Ji}\thanksref{t2,m1,m5}\ead[label=e6]{jiyuan@uchicago.edu}}

\runauthor{Sengupta et al.}

\address{Program for Computational Genomics and Medicine, NorthShore University HealthSystem\thanksmark{m1}} 
\address{Department of Bioinformatics and Biostatistics
, University of Louisville\thanksmark{m2}} 
\address{Department of Biostatistics and Bioinformatics, Emory University\thanksmark{m3}}
\address{Department of Computer \& Information Science \& Engineering, University of Florida\thanksmark{m4}}
\address{Department of Public Health Sciences, The University of Chicago\thanksmark{m5}}

\address{E-mails of the corresponding authors\\
\printead*{e5}; %\phantom{E-mail:}
\printead*{e6}}

\end{aug}
%%%%%%%%%%%%%%%%%%%%

%%%%%%%%%%%%%%%%%%%%%%%%%%%%%%%%%%%%%%%%%%%%%%%%

\begin{abstract}
Analysis of a Bayesian mixture model for the Matrix Langevin distribution on the Stiefel manifold is presented. The model exploits a particular parametrization of the Matrix Langevin distribution, various aspects of which are elaborated on. A general, and novel, family of conjugate priors, and an efficient Markov chain Monte Carlo (MCMC) sampling scheme for the corresponding posteriors is then developed for the mixture model. Theoretical properties of the prior and posterior distributions, including posterior consistency, are explored in detail. Extensive simulation experiments are presented to validate the efficacy of the framework. Real-world examples, including a large scale neuroimaging dataset, are analyzed to demonstrate the computational
tractability of the approach.  
\end{abstract}

\medskip

%%%%%%%%%%%%%%%%%%%%%%%%%%%%%%%%%%%%%%%%%%%%%%%%%%%%

\begin{keyword}
\kwd{Matrix Langevin Mixture model}
\kwd{ Mixture model}
\kwd{ Orthonormal vectors}
\kwd{ Parametric model}
\kwd{ Stiefel manifold}
\end{keyword}

\end{frontmatter}

%%%%%%%%%%%%%%%%%%%%%%%%%%%%%%%%%%%%%%%%%%%%%%%%%
\section{Introduction}
\label{sec:intro}
%Problem description, motivation (DTI, Stiefel), Background. Previous work, Contribution, Paper organization.Openly discuss the difficulty (sampling / intractability ) and trivialness of the problem. usefulness of clustering on Stiefel (Earthlike object/DTMRI)
Analysis of directional data comprises a major sub-field of study in
Statistics. Directional data range from unit vectors in the simplest examples, to sets of ordered orthonormal frames in the general case. Since the associated sample space is not the Euclidean space, standard statistical methods developed for the Euclidean space for the analysis of univariate or multivariate data cannot be easily adapted for directional data. For example, it is often desirable to account for the geometric structure underlying the sample space in statistical inference.
%Furthermore, statistical inference can be significantly improved if the intrinsic geometric structure underlying the sample space is taken into consideration. 
Beyond those fashioned for simpler non-Euclidean spaces like the circle or the sphere, there is a pressing need for methodology development for general sample spaces such as the Stiefel or the Grassmann manifold to support modern applications, increasingly seen in the fields of computer vision~\citep{Turaga:2008,Turaga:2011,Anand:2016,Lui:2008,Zeng:2015}, medical
image analysis~\citep{Lui:2012}, astronomy~\citep{Mardia:2009,Lin:2017}, and, biology~\citep{Downs:1972,Mardia:1977}, to name but a few. In this article, we present a framework for Bayesian inference of a mixture model on the Stiefel manifold~\citep{James:1976,Chikuse:2012} that remains computationally tractable even at large data sizes. With ever-growing computational power, we argue that it is now feasible to apply Bayesian methods to real world large and directional data.
%This is the main focus of the paper.
%With digital computers becoming ever more powerful, the applicability of Bayesian methods to real world applications is on the rise. This work extends the Bayesian paradigm onto large data applications in directional statistics.

One of the most commonly used distributions on the Stiefel manifold is an exponential family distribution known as the Matrix Langevin ($\ML$) or the Von-Mises Fisher matrix distribution~\citep{Mardia:2009,Khatri:1977}, introduced first by~\cite{Downs:1972}. In early work~\cite{Mardia:1977} and~\cite{Jupp:1980} studied the properties of the maximum likelihood estimators for this distribution in the classical setting. In large measure, subsequent efforts at exploring the $\ML$
distribution~\citep{Chikuse:1991,Chikuse:1991:as, Chikuse:1998} were limited to asymptotic results on distributional or inferential problems. More recently,~\cite{Hoff:2009} has developed a rejection sampling based method to sample from a matrix Bingham-Von Mises-Fisher distribution on the Stiefel manifold. To date, Bayesian analysis on these general sample spaces have been very limited. A major obstacle for the development of efficient inference techniques for this family of distributions has been the intractability of the corresponding normalizing constant, a hypergeometric function of matrix
argument.% (in the case of $\ML$, it being ${}_0F_1(\cdot,\cdot)$).

The article that is most aligned to our overall objective is~\cite{Lin:2017}, where the authors have developed a rejection sampling based data augmentation strategy for Bayesian inference with the mixture of $\ML$  distribution.  However, it is well known that sampling techniques based on a data augmentation strategy often suffer from slow rates of convergence. With the additional detrimental impact of the rejection ratio, convergence can become painfully slow. Applicability
of their MCMC technique is therefore limited, particularly in terms of
scalability to large datasets. 

Our contribution begins with an exploration of the properties of the $\ML$ distribution, followed by the construction of a family of conjugate priors for %a mixture of 
$\ML$  distribution, which we then analyze in considerable detail. In the context of the natural exponential family, Diaconis and Ylvisaker~\citep{Diaconis:Ylvisaker:1979} laid the foundations for constructing conjugate prior distributions (the DY class) for natural exponential family
models. In our case, however, the DY construction can not be directly applied, and we therefore derive a modified construction. The resultant prior is flexible in the sense that one can incorporate information from data via appropriate hyperparameter selection, and furthermore, there is the provision to set the hyperparameters in the absence of any prior knowledge to a weakly non-informative prior. For the latter, the prior might become improper in which case we adopt a constrained mixture model. Using this novel prior we implement
a scalable posterior inference scheme by designing an efficient Gibbs sampler. We note in passing that in the expression for the posterior, the presence of ${}_0F_1(\cdot,\cdot)$ in the denominator make the inference procedure challenging. We also explore the weak and strong posterior consistency under the new class of priors. Finally, we extend the proposed framework for a single $\ML$ distribution to a finite mixture of $\ML$'s.

To identify the optimum number of clusters, often times deviance information criterion (DIC) has been used in the literature~\citep{Book:Gelman:Rubin:2003,Spiegelhalter:2002}. However, several studies have pointed to the weakness of the standard DIC measure in mixture models and have proposed alternatives. We perform extensive simulations to identify alternative schemes to computing DIC that would work best for a mixture of $\ML$ distributions. In order to demonstrate the scalability of our inference scheme, we then analyze a large-scale DTMRI dataset. Real datasets that have been analyzed in the literature come from astronomy (near-earth objects) or
vectorcardiography. In both cases the data is drawn from a matrix valued manifold where each element is a collection of two orthonormal vectors in $\mathbb{R}^3$. Realizing that most of the existing applications rely on an efficient computation of the matrix hypergeometric function on a $2\times 2$ matrix, we have also optimized our inference technique for this class of matrices.  We have tested our method on a moderate sized dataset of near earth objects (NEO) with the goal of clustering the data. Obtained results are very similar to that reported in the literature.

%\sss
%We compared our model with the mixture of Wisharts model ({\attn{ I don't think we will be able to compare}}) and explicitly showed that incorporating directional information do improve inference results. To our knowledge, this is the first application of clustering of DTI dataset which are viewed as objects on Stiefel manifold.
%\sse

In summary, we aim to achieve three objectives: 
(i) the construction of a new class of distributions for conjugate priors for ML distributions and the development of their theoretical properties, (ii) the design of an efficient MCMC sampling algorithm, and finally, (iii) successful application of the framework to a large-scale (DTI) dataset.

The remainder of the paper is organized as follows. In Section~\ref{sec:stiefel_distr}, we introduce the $\ML$ distribution defined on the Stiefel manifold ($\StiefelS$) and explore its theoretical properties as well as properties of the corresponding hypergeometric  constant. In Section~\ref{sec:bayesian_framework_single_ML}, we present the construction of the conjugate prior and the posterior for a single $\ML$ distribution, properties of which are then analyzed in considerable detail.  Generalization to a finite mixture model and inference are presented in Section~\ref{sec:bayesian_mixture_model}, as well as extended theoretical properties such as the weak and strong posterior consistency. Extensive simulation studies are presented and summarized in Section~\ref{sec:sim}. In Section~\ref{sec:real_data_app}, we provide experimental results from two real-world datasets. Conclusions and future work in presented in Section~\ref{sec:disc}.

%%%%%%%%%%%%%%%%%%%%%%%%%%%%%%%%%%%%%%%%%%%%%%%

%%%%%%%%%%%%%%%%%%%%%%%%%%%%%%%%%%%%%%%%%%%%%%%

\paragraph{Notational Convention}
%%%%%
\begin{itemize}
\item $\mathbb{R}^k$ = The $k$-dimensional real space.
\item $\mathcal{S}_p= \left\{\left(d_1, \ldots, d_p\right) \in \mathbb{R}_{+}^{p} : 0< d_p< \cdots < d_1 <\infty \right\}.$
\item ${\Rnp}$ = Space of all $n \times p$ real-valued matrices.
\item $\StiefelS$  = Stiefel Manifold.
\item $\SpaceM  = \{ X \in \StiefelS : X_{1,j} > 0 \;\;\forall\, j=1,2,\cdots,p \}$. 
%Element in $\StiefelS$ whose elements of the first row of are positive.
\item $\SpaceV$ = $O(p)$ = Space of Orthogonal matrices.
\item $\prodMeasure(\cdot)$ = Product measure defined on $\StiefelS \times \Rplus^p \times \SpaceV$.
\item $\iMat$ = $p \times p$ identity matrix.
\item $f(\cdot;\cdot)$ = Probability density function.
\item $g(\cdot;\cdot)$ = Unnormalized version of the probability density function.
\item $tr(A)$ = Trace of a square matrix A.
\item $etr(A)$ = Exponential of $tr(A)$.
\item $\mathbb{E}(X)$ = Expectation of the random variable $X$.
\item $\mathbb{I}(\cdot)$ = Indicator function.
\item We use $\bd$ and $D$ interchangeably. $D$ is the diagonal matrix
with diagonal $\bd$. We use matrix notation $D$ in the place of $\bd$
wherever needed, and vector $\bd$ otherwise.
\item $\normtwo{\cdot}$ = Matrix operator norm.
\end{itemize}

%%%%%%%%%%%%%%%%%%%%%%%%%%%%%%%%%%%%
\section{$\ML$ distribution on the Stiefel manifold ($\StiefelS$)}
\label{sec:stiefel_distr}

The Stiefel manifold, $\StiefelS$ is the space of all $p$ ordered orthonormal vectors (also known as $p$-frames) in $\mathbb{R}^n$ 
and is defined as 
\begin{equation*}
\StiefelS = \{ X \in \Rnp \,:\,X^TX = \iMat\}, 
\label{eq:manifold_stiefel}
\end{equation*}
where $\Rnp$ is the space of all $n \times p$ real-valued matrices and $\iMat$ is the $p \times p$ identity matrix~\citep{Mardia:2009, Absil:2009, Chikuse:2012, Edelman:1998, Downs:1972}. $\StiefelS$ is a compact Riemannian manifold of dimension $np - p(p + 1)/2$. For $p=1$, $\StiefelS$ is the $(n-1)$ hypersphere $\mathbb{S}^{n-1}$ and for $p=n$, $\StiefelS = O(p)$, the orthogonal group consisting all orthogonal $p \times p$ real-valued matrices, with the group operation being matrix multiplication. $\StiefelS$ may be embedded in the $np$-dimensional Euclidean space of $n \times p$ real-valued matrices with the inclusion map as a natural embedding, and is thus a submanifold of $\mathbb{R}^{np}$. Since $\StiefelS$ is an embedded submanifold of $\Rnp$, its topology is the subset topology induced by $\Rnp$~\citep{Absil:2009,Edelman:1998}.

The differential form $(H_1^T dH_1) = \bigwedge_{i=1}^p \bigwedge_{j=i+1}^n h_j^T dh_i$ where $H_1 \in \StiefelS$, is invariant under the transforms $H_1 \to QH_1$ and $H_1 \to H_1P$ where $Q \in \mathcal{V}_{n,n}$ and $P \in \mathcal{V}_{p,p}$, respectively. 
This defines an invariant measure on $\StiefelS$. The surface area or volume of $\StiefelS$ is $Vol(\StiefelS) := \int_{\StiefelS} (H_1^T dH_1 ) = {2^p {(\sqrt{\pi})}^{np}}/{\Gamma_{p}({n}/{2})}$ where $\Gamma_p(\cdot)$ is the multivariate Gamma function (page 70 in~\cite{Muirhead:2009}). The measure defined in this manner is called the invariant unnormalized or the Haar measure. This measure can be normalized to a probability measure by setting $\int_{\StiefelS}[dH] = 1$ where $[dH] = (H_1^T dH_1)/{Vol(\StiefelS)}$. Uniform distribution on $\StiefelS$ is denoted by $[dH]$ and is the unique probability measure which is invariant under rotations and reflections. For detail description of construction of the Haar measure on $\StiefelS$ and its properties please refer to~\cite{Muirhead:2009}.

$\ML$ distribution~\cite{Mardia:2009} is a widely used non-uniform distribution on $\StiefelS$~\citep{Khatri:1977,Mardia:2009,Chikuse:2012,Lin:2017}. This distribution is also known as Von Mises-Fisher Matrix Distribution~\citep{Khatri:1977}. 
%The one dimensional special case of this distribution is the von-Mises distribution on a hypersphere. 
%Based on the normalized Haar measure $[dH]$, an exponential family of probability distribution has been defined on $V_{n,p}$ in the following manner. 
The density function of the $\ML$ distribution  with respect to the normalized Haar measure $[dX]$ and parametrized by $F \in \Rnp$, defined in~\cite{Chikuse:2012}, is given by 
\begin{equation}
\MLDensity (X\,;\,F) = \frac{etr(F^T X)}{{}_0F_1\left( \frac{n}{2}, \frac{F^TF}{4}\right)}, 
\label{eq:MLDensity_F}
\end{equation}
where $etr(Z) = \exp(trace(Z)$ for any square matrix $Z$ and the normalizing constant, ${}_0F_1({n}/{2}, {F^TF}/{4})$, is a hypergeometric function with a matrix argument~\citep{Herz:1955,James:1964,Muirhead:1975,Gupta:1985,Gross:1987,Gross:1989,Butler:2003,Koev:2006,Chikuse:2012}. We consider a particular form of the unique singular value decomposition (SVD) (as defined in Equation 1.5.8 in~\cite{Chikuse:2012}) of the $n \times p$ parameter matrix $F = MDV^T$ where $M \in \SpaceM$, $V \in \SpaceV$ and the diagonal entries of $D$, $\bd = (d_1, d_2, \cdots, d_p) \in \SpaceD$ where $0 < d_p < \cdots < d_2 < d_1 < \infty$~\citep{Chikuse:2012}. See Notation for definitions of $\SpaceM,\SpaceV$ and $\SpaceD$. Here, $\SpaceM$ denotes the a subspace of $\StiefelS$ consisting of matrices in $\StiefelS$ whose elements of the first row of are positive. Note that, being a closed subspace of $\StiefelS$, $\SpaceM$ is also a compact space.

Plugging in the SVD form of $F$ , we rewrite the $\ML$ density function as 
\begin{equation*}
\MLDensity (X;(M,\bd,V)) = \frac{etr(VDM^T X))}{{}_0F_1(\frac{n}{2}, \frac{D^2}{4})}\;\mathbb{I}(M \in \SpaceM,\bd \in \SpaceD, V \in \SpaceV).
%\label{eq:MLDensity_MDV}
\end{equation*}
This parametrization ensures identifiability of all the parameters ($M$,$\bd$ and $V$). For notational convenience we omit the indicator function part and use the following form of the $\ML$ density for rest of the article
\begin{equation}
\MLDensity (X;(M,\bd,V)) = \frac{etr(VDM^T X))}{{}_0F_1( \frac{n}{2}, \frac{D^2}{4})},
\label{eq:MLDensity_MDV}
\end{equation}

with respect to the normalized Haar measure $[dX]$~\citep{Muirhead:2009}. From~\cite{Khatri:1977} (page 96) note that the normalizing constant can be simplified as follows --

\begin{equation*}
{}_0F_1\left(\frac{n}{2},\frac{F^TF}{4}\right)={}_0F_1\left(\frac{n}{2},\frac{D^2}{4}\right).
\end{equation*}

Thus ${}_0F_1(\cdot)$ only depends on the eigenvalues of the matrix $F^TF$, which are the diagonal elements of the matrix $D^2$. The parametrization with $M, D$ and $V$ enables us to represent the intractable hypergeometric function of matrix argument as a function of vector $\bd$, diagonal entries of $D$, paving a path for an efficient posterior inference. This makes posterior inference computationally tractable. Note that an alternative parametrization through polar decomposition with $M$ and $K$~\citep{Mardia:2009} may pose computational challenges since the elliptical part $K$ lies on a positive semi-definite cone and inference on positive semi-definite cone is not that straightforward~\citep{Hill:1987,Bhatia:2007,Schwartzman:2006}. In this article, we use  $M, D$ and $V$ parameters based representation for $\ML$ distribution for most part of our theory.

In the following subsection we study a few important properties of the hypergeometric function of matrix argument $\hypinline$, which are required for subsequent sections.

%%%%%%%%%%%%%%%%%%%%%%%%%%%%%%%%%%%%%
\subsection{Properties of $\hyp$}
\label{subsec:0F1_prop}

\begin{lemma}
\label{lem:0F1_upper_bound}
For any $p \times p$ diagonal matrix $D$ with positive elements, ${}_0F_1\left( \frac{n}{2}, \frac{D^2}{4}\right) \leq etr(D))$ when $n \geq p$.
\end{lemma}

{\bf Proof of Lemma~\ref{lem:0F1_upper_bound}. }

From Equation~\ref{eq:MLDensity_MDV}, we have
\begin{eqnarray}
&&\int_{\StiefelS} \MLDensity(X;(M,\bd,V))\,[dX] = 1 \nonumber \\
&\implies& \int_{\StiefelS} \frac{etr(VDM^T X))}{{}_0F_1( \frac{n}{2}, \frac{D^2}{4})}\,[dX] = 1 \nonumber \\
&\implies& {}_0F_1\left( \frac{n}{2}, \frac{D^2}{4}\right) = \int_{\StiefelS} etr(VDM^T X))\,[dX].
\label{eq:0F1_integral_form}
\end{eqnarray}
We know that $\MLDensity(X;(M,\bd,V))$ has the {\em{unique modal orientation}} $MV^T$ (page 32 in~\cite{Chikuse:2012}). Hence it follows from Equation~\ref{eq:0F1_integral_form} that
\begin{eqnarray}
{}_0F_1\left( \frac{n}{2}, \frac{D^2}{4}\right) &\leq& \int_{\StiefelS} etr(VDM^TMV^T))\,[dX] \nonumber \\
&=& etr(D))\int_{\StiefelS} [dX] = etr(D)),
\label{eq:0F1_upper_bound}
\end{eqnarray}
where $[dX]$ is the normalized Haar measure on $\StiefelS$.

\qedwhite
\newline
%%%%%%%%%%%%%%%

\begin{lemma}
\label{lem:diagonal_less_norm}
Let $A$ be a $n \times p$ real matrix with $n\geq p$. If $\normtwo{A} \leq \delta (< \delta)$ for some $\delta>0$ then   $\lvert A_{j,j} \rvert \leq \delta (< \delta)$ for $j =1, .., p$. Here $A_{j,j}$ denotes the $(j,j)$-th entry of the matrix $A$ and  $\normtwo{A}$ is the spectral norm of the matrix $A$.
\end{lemma}

{\bf Proof of Lemma~\ref{lem:diagonal_less_norm}.}

From the assumptions of the Lemma~\ref{lem:diagonal_less_norm} along with the definition of the spectral norm, it follows that  $l^T A^T A\; l \leq \delta^2 (< \delta^2)$ for all $l\in \mathbb{R}^p$ with $l^T l=1$. In particular, $e_j^T A^T A\; e_j \leq \delta^2 (< \delta^2)$ where $e_j\in \mathbb{R}^p$ such that its  $j$-{th} entry equals $1$ while rest  of its entries are $0$. Hence we have that $\sum_{k=1}^{n}A_{k,j}^2 \leq \delta^2 (< \delta)$ implying the fact that $|A_{j,j}| \leq \delta (< \delta).$

\qedwhite 
\newline
%%%%%%%%%%%%%%%%

\begin{lemma}
\label{lem:F_lowerbound} 
Let $D$ be a $p \times p$ diagonal matrix with positive diagonal elements $\bd = \{d_1, d_2, \cdots, d_p\}$. Then for any $\delta>0$ and $n \geq p$, there exists a positive constant, $\ConstOne_{n,p,\delta}$, depending on $n, p$ and $\delta$, such that 
$$_0F_1\left(\frac{n}{2}, \frac{D^2}{4}\right) > \ConstOne_{n,p,\delta} \;etr\left((1-\delta)D\right).$$
\end{lemma}

{\bf Proof of Lemma~\ref{lem:F_lowerbound}.}

Note that, $D$ is a $p\times p$ diagonal matrix with positive diagonal elements   $d_1, .., d_p$.  For the case $n \geq p$, define 

\begin{equation}
\widetilde{M} = \left[
\begin{array}{c}
\iMat\;\;\;\; \\
\mathbf{0}_{n-p,p} 
\end{array}
\right],
\widetilde{V} = \iMat
\;\;\mbox{and}\;\;
I^{\star}:=
\left[
\begin{array}{c}
\iMat \;\;\;\; \\
\mathbf{0}_{n-p,p} 
\end{array}
\right],
\label{eq:defn_M_tilde_V_tilde}
\end{equation}

where $\iMat$ denotes the $p\times p$ identity matrix and $\mathbf{0}_{n-p,p}$ represents the zero matrix of dimension $(n-p) \times p$. For arbitrary given positive constant $\delta>0$, consider
$$ B_{\delta}:=\left\{X \in \mathcal{V}_{n,p }, \text{ such that }   \normtwo{X- I^{\star}}  <\delta\right\},$$

where $\normtwo{\cdot}$ denotes the spectral norm of a matrix. Let $\HaarMu$ denotes the normalized Haar measure on the $\mathcal{V}_{n,p}$. Clearly,  $0<\HaarMu \left(B_{\delta}\right)<\infty $, as $B_{\delta}$ is a non-empty open subset of $\mathcal{V}_{n,p }$. 
%Define $\RestMu(A):=\frac{\HaarMu (A \cap B_{\delta})}{\HaarMu(B_{\delta})}$ for any $A \subset \mathcal{V}{n,p}$. Note that $\RestMu(\cdot)$ is normalized Haar measure  ( also a  probability measure)  restricted on the set $B_{\delta}$. Denote $X_{B_{\delta}}:=\int_{B_{\delta}} X \;\;\;d\RestMu(X) $, i.e. $X_{{\delta}}=\mathbb{E}_{\RestMu}(X)$. 
Now from Equation~\ref{eq:MLDensity_MDV} we have, 
\begin{eqnarray}
_0F_1\left(\frac{n}{2}, \frac{D^2}{4}\right)& = & \int_{\mathcal{V}_{n,p}} etr\left(\widetilde{V}D\widetilde{M}^T X\right)d\HaarMu(X).\nonumber\\
& \geq  & \int_{B_{\delta}} etr\left(\widetilde{V}D\widetilde{M}^T X\right)d\HaarMu(X).
%& =& \HaarMu(B_{\delta})  \int_{B_{\delta}} etr\left(\widetilde{V}D\widetilde{M}^T X\right)d\RestMu(X).\nonumber\\
%& \geq & \HaarMu(B_{\delta}) \;\; etr\left(\widetilde{V}D\widetilde{M}^T X_{{\delta}}\right),
\label{eq:ZeroFone_lower_1}
\end{eqnarray}
%where the last inequality follows from Jensen's inequality.
 
%Let $X$ be distributed as $\RestMu$. From the definition of the set $B_{\delta}$, it follows that for arbitrary $l\in \mathbb{R}^p$ with $l^Tl=1$, 
%\begin{eqnarray}
%\label{eq:ZeroFone_lower_2}
%& & \mathbb{E}\left(l^T (I^{\star}-X)^T (I^{\star}-X) l \right) <\delta^2 \nonumber \\
%&\implies & l^T (I^{\star}-X_{{\delta}})^T (I^{\star}-X_{{\delta}}) l+ l^T\,\mathbb{E}[(X-X_{{\delta}})^T (X-X_{{\delta}})]\,l <\delta^2 \nonumber\\
%&\implies & l^T (I^{\star}-X_{{\delta}})^T (I^{\star}-X_{{\delta}}) l <\delta^2,
%\end{eqnarray}
%where all the expectations are performed with respect to the probability measure $\RestMu$. 

%We have $\normtwo{X- I^{\star}} < \delta$ as $X \in B_{\delta}$. 
%It follows from Equation~\ref{eq:ZeroFone_lower_2} that $\normtwo{I^{\star}-X_{{\delta}}}< \delta.$ 
Using Lemma~\ref{lem:diagonal_less_norm} we know that $X_{{j,j}}>(1-\delta)$ for $j = 1, 2, \ldots, p$ where $X \in B_{\delta}$. Note that, $X_{{j,j}}$ denotes the $(j,j)$-th entry of the matrix $X$. Hence from Equation~\ref{eq:defn_M_tilde_V_tilde} and~\ref{eq:ZeroFone_lower_1} it follows that, 
\begin{eqnarray}
\label{eq:ZeroFone_lower_3}
_0F_1\left(\frac{n}{2}, \frac{D^2}{4}\right) 
& \geq  & \int_{B_{\delta}} \exp\left({\sum_{j=1}^p {X}_{j,j}\,d_j}\right) \,d\HaarMu(X), \nonumber \\
%& \geq & \HaarMu(B_{\delta}) \;\exp\left({\sum_{j=1}^p {X_{B_{\delta}}}_{j,j} d_j}\right),\nonumber\\
&>&  \HaarMu(B_{\delta}) \; etr\left((1-\delta)D\right),
\end{eqnarray}

where the last inequality uses the fact that $d_j>0$ for all $j =1, \ldots p.$ Finally we denote  $ \ConstOne_{n,p,\delta}:= \HaarMu(B_{\delta})>0$ as it depends on  $n,p$ along with $\delta$, to conclude that 

$$_0F_1\left(\frac{n}{2}, \frac{D^2}{4}\right) > \ConstOne_{n,p,\delta} \;etr\left((1-\delta)D\right).$$

\qedwhite
\newline

%%%%%%%%%%%%%%%%%%%%
\begin{lemma}
For any $p \times p$ diagonal matrix $D$ with positive elements $\bd \in \SpaceD$, the hypergeometric function of matrix argument denoted by $\hyp$ is log-convex with respect to $\bd$ where $n \geq p$.
 
\label{lem:0F1_log_convex}
\end{lemma}

{\bf Proof of Lemma~\ref{lem:0F1_log_convex}.}\\
From Equation~\ref{eq:MLDensity_MDV}, we have
\begin{equation}
\hyp = \int_{\StiefelS} etr(VDM^TX)\,[dX], 
%= \int_{\StiefelS} etr(DM^TX)\,[dX],
\label{eq:0F1_int_form}
\end{equation}
for arbitrary $M \in \SpaceM$ and $V \in \StiefelS$ where $n \geq p$. Without loss of generality, we can take $M = \widetilde{M} = \left[ \begin{array}{l}
\iMat \\
\bm{0}_{(n-p),p} \end{array} \right]$ 
and $V = \iMat$.

Let $D_1$ and $D_2$ be two $p \times p$ diagonal matrix with positive diagonal entries $\bd_1$ and $\bd_2$, respectively and $\bd_1 \neq \bd_2$. From Equation~\ref{eq:0F1_int_form}, we have
\begin{eqnarray}
{}_0F_1\left(\frac{n}{2},\frac{D_1^2}{4}\right) =  \int_{\StiefelS} etr(D_1\widetilde{M}^TX)\,[dX] \nonumber \\
{}_0F_1\left(\frac{n}{2},\frac{D_2^2}{4}\right) = \int_{\StiefelS} etr(D_2\widetilde{M}^TX)\,[dX].
\label{eq:0F1_D1_D2}
\end{eqnarray}

Let $\lambda \in [0,1]$ be any real number. We have

\begin{eqnarray}
%\exp\left(f(\lambda \bd_1 + (1-\lambda)\bd_2\right)) &=& 
&&{}_0F_1\left(\frac{n}{2},\frac{{\left(\lambda\,D_1 + (1-\lambda)\,D_2\right)}^2}{4}\right) \nonumber \\
&=&\int_{\StiefelS} etr((\lambda\,D_1 + (1-\lambda)\,D_2)\tilde{M}^TX)\,[dX] \nonumber \\
&=&\int_{\StiefelS} {\left(etr(D_1\tilde{M}^TX)\right)}^\lambda {\left(etr(D_2\tilde{M}^TX)\right)}^{1-\lambda}\,[dX] \nonumber \\
&<& {\left(\int_{\StiefelS}\,etr(D_1\tilde{M}^TX)\,[dX]\right)}^\lambda {\left(\int_{\StiefelS}\,etr(D_2\tilde{M}^TX)\,[dX]\right)}^{1-\lambda} \nonumber \\
&=&{\left({}_0F_1\left(\frac{n}{2},\frac{D_1^2}{4}\right)\right)}^\lambda\,{\left({}_0F_1\left(\frac{n}{2},\frac{D_2^2}{4}\right)\right)}^{1-\lambda}.
%&=&\exp\left(\lambda\,f(\bd_1)\right)\,\cdot\exp\left((1-\lambda)\,f(\bd_2)\right)\nonumber \\
%&=&\exp\left(\lambda\,f(\bd_1)+(1-\lambda)\,f(\bd_2)\right).
\label{eq:convexity_0F1}
\end{eqnarray}
Note that the inequality is due to H\"older~\citep{Hardy:1952} and note that in this case $\bd_1 \neq \bd_2$. 
%Now consider a real-valued function $\funch(\bd) := \log \hyp$ on $\mathbb{R}^p$. 
Therefore from Equation~\ref{eq:convexity_0F1} we have,
\begin{eqnarray}
\log\,{}_0F_1\left(\frac{n}{2}, \left(\lambda\,\frac{D_1^2}{4} + (1-\lambda)\, \frac{D_2^2}{4}\right)\right) &<& \lambda\,\log\,{}_0F_1\left(\frac{n}{2}, \frac{D_1^2}{4}\right) + \nonumber \\
&& \;\;(1-\lambda)\,\log\,{}_0F_1\left(\frac{n}{2}, \frac{D_2^2}{4}\right).
\end{eqnarray}
Hence $\log\,\hyp$ is a convex function or equivalently $\hyp$ is a log-convex function of the diagonal entries $\bd$ of matrix $D$.

\qedwhite
\newline 
%%%%%%%%%%%%%%%%%%%%%

\begin{lemma}
For any $p \times p$ ($p\geq 2$) diagonal matrix $D$ with positive elements $\bd \in \SpaceD$, then for $i=1, 2, \cdots, p$ we have 
$$ 0 < \; \partialhyp{i} < \hyp$$ 
where $n \geq p$.
\label{lem:0F1_prime}
\end{lemma}
%%%%%%%%%%%%%

{\bf Proof of Lemma~\ref{lem:0F1_prime}.}

\paragraph{Right hand side inequality:} Proceeding similar way as Lemma~\ref{lem:0F1_log_convex} we have
\begin{equation}
\hyp = \int_{\StiefelS} etr(D\widetilde{M}^TX)\,[dX], \;\;\mbox{where $\widetilde{M} = \left[ \begin{array}{l}
\iMat \\
\bm{0}_{(n-p),p} \end{array} \right]$}. 
\label{eq:0F1_int_form_tilde_M}
\end{equation}
From Equation~\ref{eq:0F1_int_form_tilde_M}, we have 
\begin{eqnarray}
\hyp &=& \int_{\StiefelS} \exp\left(\sum_{j=1}^p d_j\,X_{j,j}\right)\,[dX]
\label{eq:0F1_int_form_X_i_0}
\end{eqnarray}
Consider the set $\mathcal{V}_{0} := \left\{ X \in \StiefelS : X_{i,i}=1 \right\}.$ 
%It is easy to see that $\mathcal{V}_{0} := \left\{ X \in \StiefelS : X_{\cdot, i}=(,0,0,...,1,0,0,0) \right\} $. 
Note that $\mathcal{V}_{0}$ is  isomorphic  to the lower dimensional Stiefel manifold, $\mathcal{V}_{n,p-1}$. $\mathcal{V}_{0}$, being a lower dimensional subspace of $\StiefelS$, has measure zero i.e. $\int_{\StiefelS }\IndVzero{}[dX]=0$, where $\IndVzero{}$ is the indicator function for $X$ to be in the set $\mathcal{V}_0$. 
%Let $\widetilde{X} = \left(X_{1,1}, X_{2,2}, \cdots, X_{p,p}\right)$.
From Equation~\ref{eq:0F1_int_form_X_i_0}, we have 
\begin{eqnarray}
\hyp &=& \int_{\StiefelS} \exp\left(\sum_{j=1}^p d_j\,X_{j,j}\right)\,\IndVzero{c}\,[dX],
\label{eq:0F1_new_form}
\end{eqnarray}
where $\mathcal{V}^c_0$ is the complement of $\mathcal{V}_0$. Hence,

%%$$$$$%%%%%%%%%%

\begin{eqnarray}
\partialhyp{i} &=& \int_{\StiefelS} X_{i,i}\,\IndVzero{c}\,\exp\left(\sum_{j=1}^p d_j\,X_{j,j}\right)\,[dX].\nonumber\\ 
\label{eq:0F1_int_form_X_i_1}
\end{eqnarray}
Observe that, $\normtwo{X}=1$ on $\StiefelS$. Hence from Lemma~\ref{lem:diagonal_less_norm}  we have $\lvert  X_{i,i} \rvert  \leq 1$. Also,  $ X_{i,i} \neq 1$ when $X \in \mathcal{V}^{c}_{0}$. As a result, we conclude that $X_{i,i} < 1$ on ${\StiefelS \cap \mathcal{V}^{c}_{0} }$. Subsequently, it follows from Equations~\ref{eq:0F1_new_form} and \ref{eq:0F1_int_form_X_i_1} that, 
\begin{eqnarray}
\partialhyp{i}&\stackrel{}{<} & \int_{\StiefelS} \exp\left(\sum_{j=1}^p d_j\,X_{j,j}\right)\,\IndVzero{c}\,[dX] \nonumber \\
& = & \hyp. %\quad (\mbox{from equation~\ref{eq:0F1_new_form}}).
\label{eq:0F1_int_form_X_i_2}
\end{eqnarray}

%%$$$$$%%%%%%%%%%

\paragraph{Left hand side inequality:} 
Consider  $\StiefelS^{i,+}:=\left\{ X \in \StiefelS : X_{i,i}>0 \right\}$, $\StiefelS^{i,-}:=\left\{ X \in \StiefelS : X_{i,i}<0 \right\}$ and $\StiefelS^{i,0}:=\left\{ X \in \StiefelS : X_{i,i}=0 \right\}$. 
Clearly, $\StiefelS^{i,+} , \StiefelS^{i,0}$ and $\StiefelS^{i,-}$ forms a partition of $\StiefelS$. Hence from equation~\ref{eq:0F1_int_form_X_i_0} we have, 
\begin{eqnarray}
 & & \partialhyp{i} \nonumber\\
 &=& \int_{\StiefelS^{i,+}} X_{i,i} \exp\left(\sum_{j=1}^p d_j\,X_{j,j}\right)\,[dX]+ \int_{\StiefelS^{i,0}}  X_{i,i}\exp\left(\sum_{j=1}^p d_j\,X_{j,j}\right)\,[dX] \nonumber \\
 &&\hspace{2in}+\int_{\StiefelS^{i,+}} X_{i,i}\exp\left(\sum_{j=1}^p d_j\,X_{j,j}\right)\,[dX]\nonumber\\
 &=& \int_{\StiefelS^{i,+}} X_{i,i} \exp\left(\sum_{j=1}^p d_j\,X_{j,j}\right)\,[dX]+ \int_{\StiefelS^{i,-}}  X_{i,i}\exp\left(\sum_{j=1}^p d_j\,X_{j,j}\right)\,[dX].\nonumber \\
\label{eq:0F1_int_form_X_i_5}
\end{eqnarray}

Let $\Gamma_{}$ be the  $n \times n $ diagonal matrix such that $\Gamma_{j,j}=1$ for $j= 1, \ldots , n, j \neq i$ and $\Gamma_{i,i}=-1$.  $\Gamma_{}$ is an orthogonal matrix as $\Gamma^T\Gamma = \bf{I}_n$. It is easy to show that $\StiefelS^{i,+}= \left\{ \Gamma\, X : X \in \StiefelS^{i,-} \right\}$.

Consider the change of variable $Y:= \Gamma X$.  Using standard algebra we can show that $X_{i,i}=-Y_{i,i}$ and $X_{j,j}=Y_{j,j}$ for $j =1, \ldots p, j \neq i $. As the normalized Haar  measure on $\StiefelS$ is invariant under orthogonal transformation from Left  i.e. $[dX]=[dY]$ ~\cite{Chikuse:2012}, we get that  
\begin{eqnarray}
\int_{\StiefelS^{i,-}}  X_{i,i}\exp\left(\sum_{j=1}^p d_j\,X_{j,j}\right)\,[dX]
& =&  - \int_{\StiefelS^{i,+}}  Y_{i,i}\exp\left(-d_i\,Y_{i,i}+\sum_{j=1, j \neq i}^p d_j\,Y_{j,j}\right)\,[dY] \nonumber\\
 &=& - \int_{\StiefelS^{i,+}}  X_{i,i}\exp\left(-d_i\,X_{i,i}+\sum_{j=1, j \neq i}^p d_j\,X_{j,j}\right)\,[dX].\nonumber\\
\label{eq:0F1_int_form_X_i_5_1}
\end{eqnarray}

From Equations~\ref{eq:0F1_int_form_X_i_5}  and~\ref{eq:0F1_int_form_X_i_5_1} we have,

\begin{eqnarray}
& & \partialhyp{i} \nonumber\\
 %&=& \int_{\mathcal{V}_{+}} X_{i,i} \, \exp\left(\sum_{j=1}^p d_j\,X_{j,j}\right)\,[dX]- \int_{\mathcal{V}_{+}}  X_{i,i}\exp\left(-d_i\,X_{i,i}+\sum_{j=1}^p d_j\,X_{j,j}\right)\,[dX]\nonumber\\
&=& \int_{\StiefelS^{i,+}} X_{i,i} \, \exp\left(\sum_{j=1, j \neq i}^p d_j\,X_{j,j}\right) \Bigg(\exp\left( d_i\,X_{i,i}\right)- \exp\left(-d_i\,X_{i,i}\right)\Bigg)\,[dX]\nonumber\\
&=& \int_{\StiefelS^{i,+}} X_{i,i}  \exp\left( \sum_{j=1, j \neq i}^p d_j\,X_{j,j}\right) 2\, \sinh\left( d_i\,X_{i,i}\right) \,[dX]
\label{eq:0F1_int_form_X_i_6}
\end{eqnarray}

where $\sinh$ is the hyperbolic sin function. Note that $\sinh\left( d_i\,X_{i,i}\right)>0$ as  $ d_i>0$ and $X_{i,i}>0$ on $\StiefelS^{i,+}$ . Hence from Equation~\ref{eq:0F1_int_form_X_i_6} it follows that, 
\begin{eqnarray}
\partialhyp{i}  &\stackrel{}{>}& 0.
\label{eq:derivative_postivity}
\end{eqnarray}

From Equations~\ref{eq:0F1_int_form_X_i_2} and~\ref{eq:derivative_postivity}, we have the result.

\qedwhite
\newline
All five lemmas will be used for a theoretical development of a conjugate prior family for $\ML$ distributions, which we discuss next.

%%%%%%%%%%%%%

\section{Bayesian framework for $\ML$ distribution}
\label{sec:bayesian_framework_single_ML}
In this section we develop a comprehensive Bayesian framework related to $\ML$ distribution. We construct a novel class of conjugate priors and study their properties. We also derive the posterior form and comment on hyperparameter settings. 

\subsection{Prior construction}
\label{subsec:prior_construct}
In the context of the exponential family of distributions,~\cite{Diaconis:Ylvisaker:1979} (DY) provides a standard procedure to obtain a class of conjugate priors when the distribution is represented through natural parametrization~\cite{Casella:2002}. But we realize that for the $\ML$ distribution DY theorem could not be applied directly. We postpone the discussion on the DY theory later in Section~\ref{subsec:DY_inapplicable} since a direct application of their construction is not possible. Instead, we propose two different conjugate priors next aiming for scalable and flexible posterior inference. 

In this context, we would also like to mention that the construction of the class of priors in~\cite{Hornik:2013} is based on the direct application of DY, which is also not quite appropriate for $\ML$ distribution. The idea of constructing a conjugate prior on the natural parameter $F$ and using a transformation afterwards involves calculation of complicated Jacobean term~\cite{Hornik:2013}. Hence the corresponding class of prior obtained by this transformation would lack the interpretation of the corresponding hyperparameters. As the DY theorem is not directly applicable, an appropriate modification is required in order to use with $\ML$ distribution (see details in Section~\ref{subsec:DY_inapplicable}). In this section we construct a new class of conjugate prior for $\ML$ density. We then show that the hyperparameters of the constructed class of priors are easily interpretable from practitioners point of view. We further extend our investigation to study properties that are essential for the hyperparameter selection and posterior inference. In the following paragraphs we design both joint and independent prior structures for the parameters of the $\ML$ distribution.

\begin{defn}
\label{defn:joint_prior}
The probability density function of the joint conjugate prior with respect to the appropriate product measure $\prodMeasure$ on $\StiefelS \times \Rplus^p \times \SpaceV$ on the parameters $M, D$ and $V$ for $\ML$ distribution is proportional to 
\begin{eqnarray}
g(M,\bd,V \,;\,\nu,\priorXzero) = \frac{etr\left( \nu\,VDM^T\priorXzero\right)}{ \left[_0 F_1 (\frac{n}{2}, \frac{D^2}{4})\right]^{\nu}},
\end{eqnarray}
as long as $g(M,\bd,V \,;\,\nu,\priorXzero)$ can be integrable.
Here $\nu > 0$ and $\priorXzero \in \Rnp$.  
\end{defn}
%We call this class of distributions as joint modified Diaconis-Ylvisaker ($\JMDY$) class for subsequent discussions.

Although joint prior structure has some desirable properties (see Theorem~\ref{thm:DY_MDV_property} and Section~\ref{subsec:linearity_modal_param}), it sometimes difficult to incorporate strength of prior belief which could differ for different parameters. For example, if a practitioner has strong prior belief on 
$M$ but has very less knowledge about parameters $D$ and $V$, then $\JMDY$ may not be the optimal choice for prior structure. We design a class of conditional conjugate prior which would be better suited for this type of situation due to flexibility. Also, it is customary to come up with independent prior structure~\citep{Gelman:2014, Khare:Pal:Su:2017} for parameters of curved exponential family~\citep{Casella:2002}, where the parametrization differs from the natural parametrization. In order to develop conditional conjugate prior structure we assume independent priors on $M$, $\bd$ and $V$. It is easy to see that conditional conjugate priors for both $M$ and $V$ are $\ML$ distribution whereas the following definition is used to construct the conditional conjugate prior for $D$.

\begin{defn}
\label{defn:indep_prior}
The probability density function of the conditional conjugate prior for $D$ with respect to the Lebesgue measure on $\Rplus^p$ is proportional to
\begin{eqnarray}
g(\bd \,;\,\nu,\BoEta) = \frac{ \exp(\nu\,\BoEta^T\bd)}{\left[_{0}F_1 \left(  \frac{n}{2}, \frac{D^2}{4}\right)\right]^{\nu}},
\end{eqnarray}
as long as $g(\bd \,;\,\nu,\BoEta)$ can be integrable. Here $\nu > 0$, $\BoEta \in \mathbb{R}^p$ and $n\geq p$.
\end{defn}
Note that, $g(\bd \,;\,\nu,\BoEta)$ is a function of $n$ as well, however we do not vary $n$ anywhere in our construction and thus we omit the symbol $n$ from the notation of $g(\bd \,;\,\nu,\BoEta)$.

We refer this particular class of distributions defined in Definition~\ref{defn:joint_prior} and Definition~\ref{defn:indep_prior} as joint modified Diaconis-Ylvisaker ($\JMDY$) and independent modified Diaconis-Ylvisaker ($\IMDY$) class, respectively for subsequent discussions.

Theorem~\ref{thm:DY_joint_prior} and Theorem~\ref{thm:DY_indep_prior} provides conditions on $\nu, \priorXzero$ and $\BoEta$ so that $g(M,\bd,V \,;\,\nu,\priorXzero)$ and $g(M,\bd,V \,;\,\nu,\BoEta)$ are integrable, respectively. We state and prove the following lemma which is necessary to prove these theorems.  
%%%%%%%%%%%%%%%%%%%%%%%%%%%%%%%%%

\begin{lemma}
\label{lem:prior_DY1}
Let $\priorXzero \in \Rnp$ and  $D$ be  a diagonal matrix with positive diagonal entries. If $\normtwo{\priorXzero} < 1$, then for arbitrary  $M\in \StiefelS, V \in \SpaceV$,  
\begin{eqnarray}
\frac{etr\left( VDM^T\priorXzero\right)}{ _0 F_1 (\frac{n}{2}, \frac{D^2}{4})} < \frac{etr(-\epsilon_0\,D)}{\ConstOne_{n,p,\epsilon_0}},
\end{eqnarray}
where $\epsilon_0=\frac{1}{2}\left(1-\normtwo{\priorXzero}\right)$ and $\ConstOne_{n,p,\epsilon_0} >0 $ is a constant depending on $n,p$ and $\epsilon_0$.
\end{lemma}

{\bf Proof of Lemma~\ref{lem:prior_DY1}.}

Note that, $0<\epsilon_0<\frac{1}{2}$ as $\normtwo{\priorXzero} < 1$. Assume $Y_0= M^T \priorXzero V \in \Rpp$. For arbitrary $l\in \mathbb{R}^p$ with $\vecnorm{l}=1$, we have
\begin{eqnarray}
l^T  Y_0^T Y_0 l &=&  (V\,l)^T  \priorXzero^T \priorXzero (V\,l) -l^T  V^T \priorXzero^T ({\bf{I}}_{n}-M M^T) \priorXzero V l \nonumber \\
&\leq&( 1-2 \epsilon_0)^2.
\label{eq:YnormIsLess}
\end{eqnarray}
The last inequality follows as $\normtwo{\priorXzero} = 1-2\epsilon_0$ and $({\bf{I}}_{n}-MM^T) $ is a non-negative definite matrix. From Equation~\ref{eq:YnormIsLess} it follows that $\normtwo{Y_0} \leq 1-2\epsilon_0$. Hence, we can apply Lemma~\ref{lem:diagonal_less_norm} we obtain that $\lvert{Y_0}_{j,j}\rvert < 1-2\epsilon_0$ for $j=1,\cdots, p$, where $Y_{0,j}$ is the $j$-{th} diagonal element of the matrix $Y_0$. Now applying Lemma~\ref{lem:F_lowerbound} we have,
\begin{eqnarray}
\frac{etr\left( VDM^T \priorXzero\right)}{ _0 F_1 (\frac{n}{2}, \frac{D^2}{4})} &<& \frac{etr(D Y_0-(1-\epsilon_0)D)}{\ConstOne_{n,p,\epsilon_0}}< \frac{etr(-\epsilon_0\,D)}{\ConstOne_{n,p,\epsilon_0}}. \nonumber
\end{eqnarray}
\qedwhite
\newline

\begin{theorem}
\label{thm:DY_joint_prior}
Let  $M\in \StiefelS$ and $V \in \SpaceV$  and $D$ be a diagonal matrix with positive diagonal elements $\bd\in \Rplus^p$. Let $\priorXzero \in \Rnp$ with $n\geq p$, then for any $\nu>0$,
\begin{enumerate}[(a)]
\item if $\normtwo{\priorXzero}<1$, we have
\begin{eqnarray*}
\int_{\StiefelS}\int_{\SpaceV} \int_{\Rplus^p} g(M,\bd,V \,;\,\nu, \priorXzero)\; d\bd\; d\HaarMu(V)\; d\HaarMu(M)<\infty,
\end{eqnarray*}

\item if $\normtwo{\priorXzero}>1$, we have
\begin{eqnarray*}
\int_{\StiefelS}\int_{\SpaceV} \int_{\Rplus^p} g(M,\bd,V \,;\,\nu, \priorXzero)\; d\bd\; d\HaarMu(V)\; d\HaarMu(M) = \infty,
\end{eqnarray*}

\end{enumerate}
where $g(M,\bd,V ; \nu, \priorXzero)$ is defined in Definition~\ref{defn:joint_prior}.
\end{theorem}
 
{\bf Proof of Theorem~\ref{thm:DY_joint_prior}.}

\paragraph{$(a)$} When $\normtwo{\priorXzero}<1$: \\
The function $g(M,\bd,V\,;\,\nu, \priorXzero)$ can be normalized to construct a probability density function with respect to the product measure $\prodMeasure$. Consider that 
\begin{eqnarray*}
& & \int_{\StiefelS}\int_{\SpaceV} \int_{\Rplus^p} g(M,\bd,V\,;\,\nu, \priorXzero) \;d\bd\; d\HaarMu(V)\; d\HaarMu(M)\nonumber\\
& =& \int_{\StiefelS}\int_{\SpaceV} \int_{\Rplus^p}  \frac{etr\left(\nu VDM^T \priorXzero\right)}{ \left[_0 F_1 (\frac{n}{2}, \frac{D^2}{4})\right]^{\nu}} \;d\bd\; d\HaarMu(V)\; d\HaarMu(M)\nonumber\\
& \stackrel{(i)}{<}& \int_{\StiefelS}\int_{\SpaceV} \int_{\Rplus^p}  \frac{etr(-\nu\epsilon_0D )}{{(\ConstOne_{n,p,\epsilon_0})}^{\nu}} \;d\bd\; d\HaarMu(V)\; d\HaarMu(M)\nonumber\\
&=& \int_{\StiefelS} d\HaarMu(M)\int_{\SpaceV}  d\HaarMu(V)\; \int_{\Rplus^p}  \frac{etr(-\nu\epsilon_0D )}{ {(\ConstOne_{n,p,\epsilon_0})}^{\nu}} \;d\bd\;\nonumber\\
& \stackrel{(ii)}{=} & \frac{1}{{\ConstOne^{\nu}_{n,p,\epsilon_0}}} \prod_{j=1}^{p}  \int_{\mathbb{R}_{+}}  \exp(-\nu\epsilon_0d_j)\;dd_j \\
&<&\infty,
\end{eqnarray*}
where the inequality $(i)$ is due to Lemma~\ref{lem:prior_DY1} while $(ii)$ follows as $\HaarMu$ is the normalized Haar measure.  Note that, here we write $[dV] = d\HaarMu(V)$ and $[dM] = d\HaarMu(M)$.

\paragraph{$(b)$} When $\normtwo{\priorXzero}>1$:\\
Let $\priorXzero:=M_{\priorXzero} D_{\priorXzero}  V^T_{\priorXzero}$ be the the unique SVD~\citep{Chikuse:2012} decomposition for the matrix $\priorXzero$. Note that, using sub-multiplicativity 
$$
\normtwo{\priorXzero} \leq \normtwo{M_{\priorXzero}} \normtwo{D_{\priorXzero}} \normtwo{V^T_{\priorXzero}} = \normtwo{D_{\priorXzero}} = D_{\priorXzero, 1}.
$$ 
Hence there exists an $\epsilon_0>0$ such that,  $ D_{\priorXzero, 1} >  (1+\epsilon_0)$ where $D_{\priorXzero, 1}$ denotes the first diagonal element of the diagonal matrix $D_{\priorXzero}$. Now consider the fact that 
\begin{eqnarray}
& & \int_{\StiefelS}\int_{\SpaceV} \int_{\Rplus^p} g(M,\bd,V\,;\,\nu, \priorXzero) d\bd\; d\HaarMu(V)\; d\HaarMu(M)\nonumber\\
&\geq& \int_{\StiefelS}\int_{\SpaceV} \int_{\SpaceD} g(M,\bd,V\,;\,\nu, \priorXzero) d\bd\; d\HaarMu(V)\; d\HaarMu(M)\nonumber\\
& =& \int_{\StiefelS}\int_{\SpaceV} \int_{\SpaceD}  \frac{etr\left(\nu\, VDM^T \priorXzero\right)}{ \left[_0 F_1 (\frac{n}{2}, \frac{D^2}{4})\right]^{\nu}} d\bd\; d\HaarMu(V)\; d\HaarMu(M)\nonumber\\
& =& \int_{\StiefelS}\int_{\SpaceV} \int_{\SpaceD}  \frac{etr\left(\nu\, DM^T M_{\priorXzero} D_{\priorXzero}  V^T_{\priorXzero} V\right)}{ \left[_0 F_1 (\frac{n}{2}, \frac{D^2}{4})\right]^{\nu}} d\bd\; d\HaarMu(V)\; d\HaarMu(M).
\label{eq:Thm1IntegralInfinite1}
\end{eqnarray}

Consider the change of variable via the following orthogonal transformations

\[
\diam{M}=
\left[\begin{array}{c}M_{\priorXzero} \;\;,  \;\;\overline{M}_{\priorXzero}  \end{array}\right] \; M,
\hspace{.5 in }
\diam{V}= V^{T}_{\priorXzero} V,
\]
where $\overline{M}_{\priorXzero}  $  is matrix containing the bases for the orthogonal complement of the column space of ${M}_{\priorXzero} $.   Note that  $\left[\begin{array}{c}M_{\priorXzero} \;\;,  \;\;\overline{M}_{\priorXzero}  \end{array}\right]^T M_{\priorXzero}= {(I^{\star})}^T  $ where $
 I^{\star}:=\left[\begin{array}{c} \iMat \;\;, \;\; \mathbf{0}_{n-p,p} \end{array}\right]^T$.  As the Haar measure on the Stiefel manifold is invariant under the orthogonal transformations~\citep{Chikuse:2012}, from Equation~\ref{eq:Thm1IntegralInfinite1} we get that, 
 
\begin{eqnarray}
& & \int_{\StiefelS}\int_{\SpaceV} \int_{\Rplus^p} g(M,\bd,V\,;\,\nu, \priorXzero) \,d\bd\; d\HaarMu(V)\; d\HaarMu(M)\nonumber\\
& \geq& \int_{\StiefelS}\int_{\SpaceV} \int_{\SpaceD}  \frac{etr\left(\nu\, D \diam{M}^T I^{\star}  D_{\priorXzero} \diam{V} \right)}{ \left[_0 F_1 (\frac{n}{2}, \frac{D^2}{4})\right]^{\nu}} \,d\bd\; d\HaarMu(\diam{V})\; d\HaarMu(\diam{M}).
\label{eq:Thm1IntegralInfinite2}
\end{eqnarray}
 
Consider 
$$
\mathcal{V}_{n,p}^{\dagger}:=\left\{M \in \StiefelS : \normtwo{I^{\star}-M}<\frac{\delta_0}{2}  \right\}; \;\; \mathcal{V}_{p,p}^{\dagger}:=\left\{V \in \SpaceV : \normtwo{\iMat-V}<\frac{\delta_0}{2}  \right\},
$$ 
where $\delta_0={\epsilon_0}/({2\;\normtwo{D_{\priorXzero}}})$. Note that  $\delta_0 > 0$ as $0 < \normtwo{D_{\priorXzero}} < \infty$. Clearly $\mathcal{V}_{n,p}^{\dagger}$ and $\mathcal{V}_{p,p}^{\dagger}$ are  open subsets of $\StiefelS$ and $\SpaceV$ respectively. Hence, $\HaarMu(\mathcal{V}_{n,p}^{\dagger}) > 0$ and $\HaarMu(\mathcal{V}_{p,p}^{\dagger}) > 0$.

If $M \in \mathcal{V}_{n,p}^{\dagger}$ and $V \in \mathcal{V}_{p,p}^{\dagger}$ then using sub-multiplicativity of $\normtwo{\cdot}$~\citep{Conway:1990} and triangle inequality, we get
\begin{eqnarray}
\normtwo{{M}^T I^{\star} D_{\priorXzero} {V} - D_{\priorXzero}} 
&\leq &\normtwo{{M}^T I^{\star} D_{\priorXzero} {V} - D_{\priorXzero}V}+ \normtwo{D_{\priorXzero} {V} - D_{\priorXzero}} \nonumber \\
&\leq &\normtwo{{M}^T I^{\star} - \iMat}\; \normtwo{D_{\priorXzero}V}+ \normtwo{D_{\priorXzero}}\normtwo{ {V} -\iMat} \nonumber \\
&= &\normtwo{({M} - I^{\star})^TI^{\star} }\; \normtwo{D_{\priorXzero}V}+ \normtwo{D_{\priorXzero}}\normtwo{ {V} -\iMat} \nonumber \\
&\leq &\normtwo{({M} - I^{\star})^T} \normtwo{I^{\star} }\; \normtwo{D_{\priorXzero}}\normtwo{V}+ \normtwo{D_{\priorXzero}}\normtwo{ {V} -\iMat} \nonumber \\
&\leq &\normtwo{({M} - I^{\star})^T}\;\normtwo{D_{\priorXzero}}+ \normtwo{D_{\priorXzero}}\normtwo{ {V} -\iMat} \nonumber \\
&\leq & \delta_0 \; \normtwo{D_{\priorXzero} } \nonumber \\
&= & \frac{\epsilon_0}{2}.
\label{eq:Thm1IntegralInfinite3}
\end{eqnarray}
Let $\lambda_{1}, \ldots, \lambda_{p} $ be diagonal elements of the matrix ${M}^T I^{\star} D_{\priorXzero} {V}$. From Lemma ~\ref{lem:diagonal_less_norm} we get that $\vert\lambda_j- D_{\priorXzero,j}\rvert \leq \epsilon_0/2$ for $j =1, \ldots, p$. Here $D_{\priorXzero,j}$ denotes the $j$-th diagonal element of the matrix $D_{\priorXzero}$. Hence  for arbitrary $M \in \mathcal{V}_{n,p}^{\dagger}$ and $V \in \mathcal{V}_{n,p}^{\dagger}$, we have 
\begin{eqnarray}
tr\left( {M}^T I^{\star} D_{\priorXzero} {V}\right)=\sum_{j=1}^{p} \lambda_j \geq  \sum_{j=1}^{p} \left(D_{\priorXzero,j}-  \; \frac{\epsilon_0}{2} \right),
\label{eq:Thm1IntegralInfinite4}
\end{eqnarray} 
as $\lambda_j \geq  \left(D_{\priorXzero,j}-  \; \frac{\epsilon_0}{2} \right)$ for all $j =1, 2, \cdots, p$.

Now from Equation~\ref{eq:Thm1IntegralInfinite2}, we have 
\begin{eqnarray}
& & \int_{\StiefelS}\int_{\SpaceV} \int_{\Rplus^p} g(M,\bd,V\,;\,\nu, \priorXzero) d\bd\; d\HaarMu(V)\; d\HaarMu(M)\nonumber\\
& \geq & \int_{ \mathcal{V}_{n,p}^{\dagger}}\int_{ \mathcal{V}_{p,p}^{\dagger}} \int_{\SpaceD}  \frac{etr\left(\nu\, D \diam{M}^T I^{\star}  D_{\priorXzero} \diam{V} \right)}{ \left[_0 F_1 (\frac{n}{2}, \frac{D^2}{4})\right]^{\nu}} d\bd\; d\HaarMu(\diam{V})\; d\HaarMu(\diam{M})\nonumber\\
&\stackrel{(iii)}{\geq} & \int_{ \mathcal{V}_{n,p}^{\dagger}}\int_{ \mathcal{V}_{p,p}^{\dagger}} \int_{\SpaceD}  \frac{\exp\left( \nu\,  \sum_{j=1}^{p} d_j\left(D_{\priorXzero,j}-  \; \frac{\epsilon_0}{2} \right) \right)}{ \left[_0 F_1 (\frac{n}{2}, \frac{D^2}{4})\right]^{\nu}} d\bd\; d\HaarMu(\diam{V})\; d\HaarMu(\diam{M}),\nonumber\\
&\stackrel{(iv)}{\geq}& \int_{ \mathcal{V}_{n,p}^{\dagger}}\int_{ \mathcal{V}_{p,p}^{\dagger}} \int_{\SpaceD}  \frac{\exp\left( \nu\,  \sum_{j=1}^{p} d_j\left(D_{\priorXzero,j}-  \; \frac{\epsilon_0}{2} \right) \right)}{ \left[ etr(D) \right]^{\nu}} d\bd\; d\HaarMu(\diam{V})\; d\HaarMu(\diam{M}),\nonumber\\
& \geq & \HaarMu(\mathcal{V}_{n,p}^{\dagger})\; \HaarMu(\mathcal{V}_{p,p}^{\dagger})\, \int_{\SpaceD}  \exp\left( \nu\,  \sum_{j=1}^{p} d_j\left(D_{\priorXzero,j}-1-  \; \frac{\epsilon_0}{2} \right) \right) d\bd , \nonumber\\
& \stackrel{(v)}{\geq} & \HaarMu(\mathcal{V}_{n,p}^{\dagger})\; \HaarMu(\mathcal{V}_{p,p}^{\dagger})\, \int_{\SpaceD}  \exp\left( \nu\, \frac{\epsilon_0}{2} d_1 \right)  \prod\limits_{j=2}^p \exp\left( \nu\, d_j\left(D_{\priorXzero,j}-1- \; \frac{\epsilon_0}{2} \right) \right) d\bd, \nonumber\\
& =& \infty,
\label{eq:Thm1IntegralInfinite5}
\end{eqnarray}
where $(iii)$ and $(iv)$ follow from Equation~\ref{eq:Thm1IntegralInfinite4} and Lemma~\ref{lem:0F1_upper_bound}, respectively. Finally, $(v)$ follows as $D_{\priorXzero,1}> (1 +
\epsilon_0)$.

\qedwhite
\newline

\paragraph{Remark for Theorem~\ref{thm:DY_joint_prior}.} 
One could notice that the conditions mentioned in this theorem is not entirely necessary and sufficient conditions. We have not addressed the case where $\normtwo{\priorXzero} = 1$. This scenarios could be broken into two cases (a) all the eigenvalues of $\priorXzero$ are equal to $1$ and (b) only a few eigenvalues are equal to $1$ and rest are strictly less than $1$. In both the cases, it seems that the problem is more involved than the current one and we have not investigated the finiteness of the corresponding integral in detail for those cases. For now, we leave those for future work.

\begin{theorem}
\label{thm:DY_indep_prior}
Let $D$ be  diagonal matrix with diagonal elements $\bd \in \mathbb{R}_{+}^p$. Let  $\BoEta=\left(\eta_1, \ldots, \eta_p\right)\in \mathbb{R}^p$ and $n$ be any integer with $n\geq p$. Then for any $\nu > 0$, 
\begin{eqnarray}
\int_{\mathbb{R}_{+}^{p}} g(\bd ; \nu, \BoEta, n)\;d\bd  < \infty, \nonumber
\end{eqnarray}
if and only if $\max\limits_{1 \leq j \leq p} \eta_j < 1 $, where $g(\bd ; \nu, \BoEta, n)$ is defined in Definition~\ref{defn:indep_prior}.
\end{theorem}

{\bf Proof of Theorem~\ref{thm:DY_indep_prior}.} 

\paragraph{Sufficient condition:}
For any $\BoEta:=\left(\eta_1, \ldots, \eta_p\right) \in \mathbb{R}^{p} $, define $\BoEta^{+}:=\left(\eta^{+}_{1}, \ldots , \eta^{+}_{p} \right)$ where $ \eta^{+}_{j}$ equals $\eta_j$ when $\eta_j>0$ and zero otherwise. Define $D_{\BoEta}$ to be the diagonal matrix with diagonal elements $ \BoEta^{+}$. Let us consider the following matrices 
\[
\priorXzero = 
\left[
\begin{array}{c}
D_{\BoEta} \;\;\;\; \\
\mathbf{0}_{n-p,p} 
\end{array}
\right], \;\;
M^{\star}=
\left[
\begin{array}{c}
{\bf I}_{p,p} \;\;\;\; \\
\mathbf{0}_{n-p,p} 
\end{array}
\right]\;\text{and }
V^{\star}=\iMat.
\]

Note that $\widetilde{M} \in \SpaceM$, $\widetilde{V} \in \SpaceV$  and $ D_{\BoEta}={\widetilde{M}}^T  \priorXzero  \widetilde{V} $. Now from Definition~\ref{defn:indep_prior} we get that   

\begin{eqnarray}\label{eq:Lemma2}
\int_{\mathbb{R}_{+}^{p}} g(\bd ; \nu, \BoEta, n)\;d\bd  
 & =& \int_{\mathbb{R}_{+}^{p}} \frac{\exp(\nu\,\sum_{j=1}^{p} \eta_j d_j)}{\left[ _0 F_1 (\frac{n}{2}, \frac{D^2}{4})\right]^{\nu}}\;d\bd\nonumber\\ 
&\stackrel{}{\leq} & \int_{\mathbb{R}_{+}^{p}} \frac{\exp(\nu\sum_{j=1}^{p} \eta^{+}_{j} d_j)}{\left[ _0 F_1 (\frac{n}{2}, \frac{D^2}{4})\right]^{\nu}}\;d\bd \nonumber \\
&= &   \int_{\mathbb{R}_{+}^{p}} \frac{etr\left( \nu \,D D_{\BoEta}\right)}{\left[ _0 F_1 (\frac{n}{2}, \frac{D^2}{4})\right]^{\nu}}\;d\bd\nonumber\\
&\stackrel{}{=} &  \int_{\mathbb{R}_{+}^{p}} \frac{etr\left(\nu{\widetilde{V} D {\widetilde{M}}^T \priorXzero}\right)}{\left[ _0 F_1 (\frac{n}{2}, \frac{D^2}{4})\right]^{\nu}}\;d\bd \nonumber\\
&\stackrel{(vi)}{<} & \int_{\mathbb{R}_{+}^{p}} \frac{etr(-\nu\epsilon_0 D)}{{(\ConstOne_{n,p,\epsilon_0})}^{\nu}} \;d\bd \nonumber\\
&\stackrel{}{=} &  \frac{1}{{(\ConstOne_{n,p,\epsilon_0})}^{\nu}} \prod_{j=1}^{p}  \int_{\mathbb{R}_{+}} \exp(-\nu\epsilon_0d_j)\;dd_j \nonumber\\
& <& \infty, 
\end{eqnarray}
where the inequality at step  $(vi)$  follows from Lemma~\ref{lem:prior_DY1} with appropriate $\epsilon_0>0$.

\paragraph{Necessary condition:}
Let $\BoEta \in \mathbb{R}^p$ be such that $\max\limits_{j=1,\ldots p} \eta_j\geq 1$. There exist  at least one $j \in \{1, \ldots p\}$ such that $\eta_j\geq 1$. Without loss of generality, let us assume that $\eta_1\geq 1$. From Definition~\ref{defn:indep_prior}, we have

\begin{eqnarray}\label{eq:onlyIfIndepPrior1}
& & \int_{\mathbb{R}_{+}^{p}} g(\bd \,; \,\nu, \BoEta,n)\;d\bd  \nonumber\\
& =& \int_{\mathbb{R}_{+}^p} \frac{\exp(\nu\sum_{j=1}^{p} \eta_j d_j)}{\left[ _0 F_1 (\frac{n}{2}, \frac{D^2}{4})\right]^{\nu}}\;d\bd 
\nonumber\\
&\stackrel{}{\geq } & \int_{\mathbb{R}_{+}^p} \frac{\exp(\nu\,\sum_{j=1}^{p} \eta_{j} d_j)}{etr(\nu D)}\;d\bd \nonumber \\
&= &  \prod_{j=1}^p\int_{\mathbb{R}_{+}} \exp\left(\nu (\eta_{j}-1) d_j\right)\;d d_j \nonumber \\
&= &  \int_{\mathbb{R}_{+}} \exp\left(\nu (\eta_{1}-1) d_1\right)d d_1\prod_{j=2}^p\int_{\mathbb{R}_{+}} \exp\left(\nu (\eta_{j}-1) d_j\right)\;d d_j \nonumber \\
& =& \infty, \nonumber  
 \end{eqnarray}
where the inequality is due to Lemma~\ref{lem:0F1_upper_bound}.
\qedwhite
\newline
%%%%%%%%%%%%%%%%%%%%%
\paragraph{Remark for Theorem~\ref{thm:DY_indep_prior}.} 
We could alternatively parametrize $\IMDY$ in the following way
$g(\bd \,;\, \nu, \BoEta) \propto {\exp\left(\sum_{j=1}^{p} \eta_j d_j\right)}/{\left[ _0 F_1 (\frac{n}{2}, \frac{D^2}{4})\right]^{\nu}}$
when $\max\limits_{1 \leq j \leq p} \eta_{j} < \nu$.
In this parametrization if we set $\nu = 0$ and $\boldsymbol{\beta}:=-\BoEta$ then $g(\bd \,;\, \nu, \BoEta)$ refers to the {\it{Exponential}} distribution with parameter $\bbeta$.
%%%%%%%%%%%%%%%%%%%%%%%%%%%%%%%%%

\subsection{Properties of $\IMDY$ and $\JMDY$ class of distributions}
\label{subsec:prior_prop}
%\paragraph{Characterization for concentration parameter ($\nu$) for  unimodal class of probability density functions}

%In subsequent paragraphs whenever we use the ``concentration", it means probability concentration around the unique mode unless stated otherwise.

The following lemmas are essential to study theoretical properties of the conjugate prior mentioned in Section~\ref{subsec:prior_construct}.

%\subsubsection{Unimodality of prior density function of $D$ [only diagonal] [{\attn{Need to proofread}}]}

\begin{lemma}
The probability density function for the prior distribution of $\bd \sim \mbox{\IMDY}(\bd;\nu,\bEta)$ denoted by $g(\bd;\nu,\BoEta):= {\exp(\nu\,\bEta^T\bd)}/{{\left[{}_0F_1\left(\frac{n}{2},\frac{D^2}{4}\right)\right]}^{\nu}}$, is log-concave as a function of $\bd$ where $D$ is the diagonal matrix with diagonal elements $\bd$, $\underset{1\leq j\leq p}{\max} \eta_j < 1$, $\nu > 0$ and $n \geq p$.
\label{lem:density_D_unimodal}
\end{lemma}

{\bf Proof of Lemma~\ref{lem:density_D_unimodal}.}

From Definition~\ref{defn:indep_prior} we have,
\begin{eqnarray}
g(\bd;\nu,\bEta)&:=& \DYlang, \nonumber\\
\implies \log g(\bd;\nu,\bEta)&:=& \logDYlang
\label{eq:prior_density_d}
\end{eqnarray}

From Lemma~\ref{lem:0F1_log_convex}, it follows that $-\nu\,\lhyp$ is concave function of $\bd$. Also, $\nu\,\bEta^T\bd$ is a linear function of $\bd$. Therefore from Equation~\ref{eq:prior_density_d} it is clear that $\log g(\bd;\nu,\bEta)$ is a concave function of $\bd$. 

\qedwhite
\newline

\begin{lemma}
The distribution of $\bd$ is unimodal if $0 <\eta_j < 1$ for all $j=1, 2, \cdots, p$. The mode of the distribution is characterized by the parameter $\BoEta$ and it does not dependent on the parameter $\nu$.
\label{lem:unique_mode_independent_of_nu}
\end{lemma}

{\bf Proof of Lemma~\ref{lem:unique_mode_independent_of_nu}.}

Let $l(\variableX,\nu,\BoEta)= \log(g(\variableX;\nu, \BoEta))$.
If $\widehat{\variableX}$ is the mode of the distribution then 

\begin{eqnarray}
\label{eq:priorModeEquation}
& & \frac{\partial}{\partial \variableX} \,l(\variableX,\nu,\BoEta) \bigg\rvert_{\variableX=\widehat{\variableX}} \; \; =0, \nonumber\\
& \implies &  \nu \BoEta -\nu \frac{\partial }{\partial \variableX}  \log\left( _0F_1 \left(\frac{n}{2}, \frac{D^2}{4}\right)\right)\bigg\rvert_{\variableX=\widehat{\variableX}} \;\;=0, \nonumber\\
& \implies &  \frac{\partial }{\partial \variableX}  \log\left( _0F_1 \left(\frac{n}{2}, \frac{D^2}{4} \right)\right)\bigg\rvert_{\variableX=\widehat{\variableX}} =\BoEta, \nonumber\\
& \implies & h(\widehat{\variableX})=\BoEta, 
\end{eqnarray}

where  $h(\bd):= \left(h_1(\bd), h_2(\bd),\cdots ,h_p(\bd)\right)$ with $h_j(\bd) := {\left(\frac{\partial }{\partial d_j}\,{}_0F_1 \left(\frac{n}{2}, \frac{D^2}{4} \right)\right)}/{_0F_1 \left(\frac{n}{2}, \frac{D^2}{4} \right)}$ for $j=1, 2,\cdots, p$. The function $h_j(\bd)$ is strictly increasing as the function $_0F_1 \left(\frac{n}{2}, \frac{D^2}{4} \right)$ is log-convex (see Lemma~\ref{lem:0F1_log_convex}). Also, it follows from Lemma~\ref{lem:0F1_prime} that $0< h_j(\bd) < 1$ for all $\bd \in \SpaceD$. Hence the Equation~\ref{eq:priorModeEquation} has a unique solution when $0< \eta_j < 1$ for all $j=1, 2, \cdots, p$. Also it is clear that the solution does not depend on $\nu$. On the other hand, given any $\widehat{\bd} \in \SpaceD$ we can always find a $\BoEta$ satisfying Equation~\ref{eq:priorModeEquation} such that $0<\max\limits_{1 \leq j \leq p} \eta_j<1$.
\qedwhite
\paragraph{Remark:} In the case of $\eta_j \leq 0$, the density defined in~\ref{defn:indep_prior} is decreasing as a function of $d_j$ on the set $\Rplus$. Therefore, mode does not exist.
\newline

In order to introduce the notion of ``concentration" for $\IMDY$ class of distributions we require the concept of level set. 
%%%%%%%%%%%%%%%%%%%%
Let unnormalized probability density function for $\IMDY$ class of distributions, $g(\bx;\nu,\BoEta)$, achieves the maximum value at $\m$ and let 
$$
{\SetWithModePrime}_l = \left\{ \bx \in \Rplus^p: g(\bx;1,\BoEta)/g(\m;1,\BoEta) > l \right\}
$$ be the level set of order $l$ containing the mode $\m$ where $0 \leq l < 1$.
Note that, to define the level set we could have used any fixed value of $\nu_0 > 0$ in $g(\bx;\nu_0,\BoEta)$ instead of $g(\bx;1,\BoEta)$, however without loss of generality we choose $\nu_0 = 1$.

%%%%%%%%%%%%%%%%%%%%%%%%%%%%
\begin{lemma}
Let $\BoEta \in \mathbb{R}^p$ be a fixed vector such that $0<\max\limits_{1 \leq j \leq p} \eta_j<1$. Whenever $\bd \sim \mbox{\IMDY}(\bd;\nu,\bEta)$, we have

\begin{enumerate}[(a)]
\item  $P_\nu(\SetWithMode_l)$ is an increasing function of $\nu$.
\item  For any open set $\SetWithMode \subset \Rplus^p$ containing $\m$, $P_\nu(\bd \in \SetWithMode)$ goes to $1$ as $\nu \to \infty$,
\end{enumerate}
\label{lem:cn}	
where $P_{\nu}(\cdot)$ denotes the probability distribution corresponding to $\bd \sim \mbox{\IMDY}(\bd;\nu,\bEta)$.
\end{lemma}
	
{\bf{Proof of Lemma~\ref{lem:cn}}.}
\paragraph{$(a)$ }
Note that, from definitions of unimodality and level set we have
\begin{equation}
\left[\frac{g(\by;\nu,\BoEta)}{g(\bx;\nu,\BoEta)}\right] >  1 \;\;\mbox{for all $\by \in {\SetWithModePrime}$ and for all $\bx \in {\SetWithModePrime}^c$}.
\label{eq:g_1_y_g_1_x_ge_1}
\end{equation}  
Consider the  function 
\begin{equation}
r(\nu,\bx) := \int _{\SetWithModePrime}  \frac{g(\by;\nu,\BoEta)} {g (\bx;\nu,\BoEta)}\,d\by
 =  \int _{\SetWithModePrime}  {\left[\frac{g(\by;1,\BoEta)} {g (\bx;1,\BoEta)}\right]}^\nu\,d\by,
\label{eq:g_nu_y_x_ratio}
\end{equation}
where $\bx \in {\SetWithModePrime}^c$.  
Using equation~\ref{eq:g_1_y_g_1_x_ge_1} it is easy to see that ${\left[\frac{g(\by;1,\BoEta)} {g (\bx;1,\BoEta)}\right]}^\nu$ is monotonically increasing in $\nu$ for all $\by \in \SetWithModePrime$.  Hence $r(\nu,\bx)$ is increasing function in $\nu$ for any $\bx \in {\SetWithModePrime}^c$.
  
Note that,
\begin{equation}
\frac{ P_\nu(\bd \in {\SetWithModePrime}^c)}{P_\nu(\bd \in \SetWithModePrime)} = \frac{\int_{{\SetWithModePrime}^c} g(\bx;\nu,\BoEta)\,d\bx}  {\int_{\SetWithModePrime} g(\by;\nu,\BoEta)\,d\by } =  \int_{{\SetWithModePrime}^c } \frac{1}  {\int_{\SetWithModePrime} \frac{g(\by;\nu,\BoEta)}{g(\bx;\nu,\BoEta)} \,d\by} \,d\bx = \int _{{\SetWithModePrime}^c} \frac{1}{r(\nu,\bx)}\,d\bx.
\label{eq:P_nu_d}
\end{equation}
Hence ${ P_\nu(\bd \in {\SetWithModePrime}^c)}/{P_\nu(\bd \in \SetWithModePrime)}$ is a decreasing function of $\nu$ as $\frac{1}{r(\nu,\bx)}$ is a decreasing function in $\nu$ for every $\bx \in {\SetWithModePrime}^c$ or equivalently ${P_\nu(\bd \in \SetWithModePrime)}$ increasing function in $\nu$.

\qedwhite
\newline

\paragraph{$(b)$ } 
Let $\bd \sim \mbox{\IMDY}(\cdot ; \nu, \BoEta)$ with $0<\eta_j<1$ for $j = 1, \dots p$. Let $\m $ be the mode the distribution. Note that the value of $\m$ only depends on the parameter $\BoEta$ and does not depend on the parameter $\nu$. Let $f(\variableX; \nu, \BoEta)$ be the corresponding probability density function. Hence for the class of distribution function defined in Definition~\ref{defn:indep_prior}, it follows that, 
\begin{eqnarray}
f(\variableX; \nu, \BoEta)=\frac{1}{K_{\nu, \BoEta}}\frac{ \exp(\nu\;  \BoEta^T \variableX ) }{\left[_{0} F_1 \left(  \frac{n}{2}, \frac{D^2}{4}\right)\right]^{\nu}},
\end{eqnarray}
where $K_{\nu, \BoEta}$ is the appropriate normalizing constant.

Let us define the function $g(\bd; \BoEta)={ \exp(\;  \BoEta^T \variableX ) }/{_{0} F_1 \left(  \frac{n}{2}, \frac{D^2}{4}\right)}$. 
Let $\SetWithMode$ be any open set containing $\m$, the mode of the density function $f(\variableX; \nu, \BoEta)$. Consider,  the set $\TargetComp:= \left\{\bd : g(\bd; \BoEta) \leq \zeta \right\}$, where $\zeta = \sup\limits_{\bd \in S^{c} } g(\bd; \BoEta)$. It is easy to show that $ S^{c} \subseteq \TargetComp$.
 
Consider the fact that ,   
\begin{eqnarray}
& & \frac{g(\bd; \BoEta)}{g(\lambda \m + (1-\lambda)\bd; \BoEta)} \quad \mbox{where $\lambda \in (0,1)$} \nonumber\\
&=&  \frac{ \exp(\;  \BoEta^T \variableX ) }{_{0} F_1 \left(  \frac{n}{2}, \frac{D^2}{4}\right)}
 \frac{_{0} F_1 \left(  \frac{n}{2}, \frac{ \left[ \lambda D_{m} + (1-\lambda) D\right]^2}{4}\right)}{ \exp(\;  \BoEta^T (\lambda \m + (1-\lambda)\bd )) } \nonumber\\
 &\stackrel{(vii)}{\leq} &  \frac{ \exp(\;  \BoEta^T \variableX ) }{_{0} F_1 \left(  \frac{n}{2}, \frac{D^2}{4}\right)}
 \frac{ \left[_{0} F_1 \left(  \frac{n}{2}, \frac{  D_{m}^2}{4}\right)\right]^{\lambda}\; 
 \left[ _{0} F_1 \left(  \frac{n}{2}, \frac{  D^2}{4}\right)\right]^{1-\lambda}
 }{ \exp(\;  \BoEta^T (\lambda \m + (1-\lambda)\bd )) } \nonumber\\
 &\leq &  \left[\frac{ \exp(\;  \BoEta^T \variableX ) }{_{0} F_1 \left(  \frac{n}{2}, \frac{D^2}{4}\right)} \right]^{\lambda}
 \left[  \frac{ _{0} F_1 \left(  \frac{n}{2}, \frac{  D_{m}^2}{4}\right)\; 
 }{ \exp(\;  \BoEta^T \m ) }\right]^{\lambda} \nonumber\\
 & =& \left[\frac{g(\bd; \BoEta)}{g( \m ; \BoEta)}\right]^{\lambda},
\end{eqnarray}
where $D_m$ is the diagonal matrix with diagonal $\m$. Note that, inequality $(vii)$ follows from the fact that ${}_0F_1(\cdot)$ is a log-convex function.

Hence we have, 
\begin{eqnarray}
\frac{f(\bd\; ; \nu,\BoEta)}{f\left(\lambda \m+ (1-\lambda)\bd ; \nu,\BoEta\right)}
=\left[\frac{g(\bd \; ; \BoEta)}{g\left(\lambda \m+ (1-\lambda)\bd ; \BoEta\right)}\right]^{\nu}
\leq  \left[  \frac{g(\bd;\BoEta)}{g(\m;\BoEta)}\right]^{\nu \;\lambda}.
\end{eqnarray}

\begin{eqnarray}
P_{\nu}(\TargetComp) & = &  \int_{\TargetComp}f(\bd\;;\;\nu,\BoEta) \;d\,\bd  \nonumber\\
& =&  \int_{\TargetComp}\frac{f(\bd\;;\;\nu,\BoEta)}{ f(\lambda \m+ (1-\lambda)\bd\;;\;\nu,\BoEta)} \; f(\lambda \m+ (1-\lambda)\bd\;;\;\nu,\BoEta)  \;d\,\bd \nonumber\\
& \leq&  \int_{\TargetComp}  \left[  \frac{g(\bd;\BoEta)}{g(\m;\BoEta)}\right]^{\nu\lambda} \; f(\lambda \m+ (1-\lambda)\bd\;;\;\nu,\BoEta)  \;d\,\bd \nonumber\\
& \leq&  \int_{\TargetComp}  \left[  \frac{\zeta}{g(\m;\BoEta)}\right]^{\nu \lambda} \; f(\lambda \m+ (1-\lambda)\bd\;;\;\nu,\BoEta)  \;d\,\bd\nonumber\\
 &=&\left[\frac{\zeta}{g(\m;\BoEta)}\right]^{\nu \lambda} \int_{\TargetComp} \; f(\lambda \m+ (1-\lambda)\bd\;;\;\nu,\BoEta)  \;d\,\bd \nonumber\\
 &\leq& \left[  \frac{\zeta}{g(\m;\BoEta)}\right]^{\nu \lambda}.
\end{eqnarray}
Hence we have,
$$
\lim_{\nu\to\infty} P_{\nu}(\SetWithMode) \geq 1 - \lim_{\nu\to\infty} P_{\nu}(\TargetComp)\geq 1 - \lim_{\nu\to\infty}  \left[  \frac{\zeta}{g(\m;\BoEta)}\right]^{\nu \lambda}=1
$$
as $\zeta< g(\m;\BoEta)$. 
%because $\TargetComp$ is an open set. 
\qedwhite
\newline

The following two theorems establishes few important properties of  $\IMDY$ and $\JMDY$ class of distributions.

\begin{theorem}
\label{thm:DY_D_property}
Let $\variableX \sim \mbox{\IMDY}(\cdot ; \nu, \BoEta)$ for some $\nu>0$ and $\max\limits_{1 \leq j \leq p} \eta_j<1$ where $\BoEta=\left(\eta_1,\ldots, \eta_p \right)$. Then  
\begin{enumerate}[(a)]
\item The distribution of $\bd$ is log-concave.
\item The distribution of $\bd$ is unimodal if $\eta_j > 0$ for all $j=1, 2, \cdots, p$. The mode of the distribution is characterized by the parameter $\BoEta$ and it does not dependent on the parameter $\nu$.
\item The parameter $\nu$ relates to the concentration of the probability around mode of the distribution. Larger values of $\nu$ implies larger concentration of probability near the mode of the distribution.
%\item the distribution have exponentially light (thin?) right tail %\Pal{Need to define, cite appropriate reference and prove}
\end{enumerate}
\end{theorem}

{\bf Proof of Theorem~\ref{thm:DY_D_property}.}\\
%Let  the  random variable $\variableX \sim  \mbox{\IMDY}(\cdot ; \nu, \BoEta)$, then the density is proportional to 
%\begin{eqnarray}
%g(\variableX; \nu, \BoEta)=\frac{ \exp(\nu\;  \BoEta^T \variableX ) }{\left[_{0} F_1 \left(  \frac{n}{2}, \frac{D^2}{4}\right)\right]^{\nu}},  
%\end{eqnarray}
%where $D$ is the diagonal matrix with diagonal elements $\variableX$,  $\max\limits_{1 \leq j \leq p} \eta_j<1$ , $\nu>0$ and  $n\geq p$. Following are the properties of the distribution.

%\paragraph{(a):}
Proof of part $(a), (b)$ and $(c)$ follow from Lemma~\ref{lem:density_D_unimodal},~\ref{lem:unique_mode_independent_of_nu} and~\ref{lem:cn}, respectively.

\qedwhite
\newline

We call the parameter $\BoEta$ as modal parameter and $\nu$ as Concentration parameter.

\begin{defn}
\label{defn:modal_def}
The parameter $\eta$ in the distribution that belongs to the class of distributions $\IMDY$ is defined as ``modal parameter". 
\end{defn}

\begin{defn}
\label{defn:concentration_def}
The scalar parameter $\nu$ in the distribution that belongs to the class of distributions $\IMDY$ is defined as ``concentration parameter". 
\end{defn}

%%%%%%%%%%%%%%%%%%%%%%%%%%%

\begin{theorem}
\label{thm:DY_MDV_property}
Let $(M,\variableX, V) \sim  \mbox{\JMDY}(\cdot ; \nu, \priorXzero)$ for some $\nu>0$ and $\normtwo{\priorXzero}<1$. Then  
\begin{enumerate}[(a)]
\item The distribution has unique mode. The mode is characterized by the parameter $\priorXzero$ and it does not dependent on the parameter $\nu$.
%\item The parameter $\nu$ relates to the concentration of the probability around mode of the distribution. Larger values of $\nu$ implies larger concentration of probability near mode of the distribution.
\item Conditional distribution of $M$ given $(\bd,V)$ and $V$ given $(M,\bd)$ are $\ML$ distributions whereas conditional distribution of $\bd$ given $(M,V)$ is $\IMDY$ class of distribution. 
\end{enumerate}
\end{theorem}

{\bf Proof of Theorem~\ref{thm:DY_MDV_property}.}\\
%Let  the  random variable $(M,\variableX,V) \sim  \mbox{\JMDY}(\cdot ; \nu, \priorXzero)$, then 
The joint density is proportional to 
\begin{eqnarray}
g(M, \bd, V; \nu, \priorXzero)=\frac{etr(\nu\,V D M^T \priorXzero)}{\left[_{0} F_1 \left(  \frac{n}{2}, \frac{D^2}{4}\right)\right]^{\nu}},  
\label{eq:JMDY_unnormalized_density}
\end{eqnarray}

%In the $\JMDY$ class, again $\nu$ is the ``Concentration parameter" while $\priorXzero$ is the ``modal parameter". 

\paragraph{$(a)$ }
%%%% follow Chikuse proof

Let us write the SVD~\citep{Chikuse:2012} of $\priorXzero = M_{\priorXzero}D_{\priorXzero}V_{\priorXzero}^T$. 
%It is easy to verify that the density attains the unique maximum value at $M = M_{\priorXzero}$ and $V = V_{\priorXzero})$. Now, for $D$ we can put back the value of $M$ and $V$ and following  Lemma~\ref{lem:unique_mode_independent_of_nu}, we could solve for $D$ for which density would be maximum. Neither of the solution depends on $\nu$.
We have,
\begin{eqnarray}
etr(\nu\,V D M^T \priorXzero) &=& etr(\nu\,D M^T M_{\priorXzero}D_{\priorXzero}V_{\priorXzero}^TV) \nonumber \\
&=& etr(\nu\,V_{\priorXzero}^TV D \, U_M D_M V_M^T \,D_{\priorXzero}) \nonumber \\
&=& etr(\nu\,V_1 D\, U_M D_M V_M^T \, D_{\priorXzero})
\label{eq:unique_mode_JMDY}
\end{eqnarray}
where SVD of is written as $M^T M_{\priorXzero} = U_MD_MV_M^T$ and $V_1 = V_{\priorXzero}^TV$ is an orthogonal matrix.
 Therefore we have,
\begin{eqnarray}
etr(\nu\,V D M^T \priorXzero) &=& etr(\nu\,V_1 D\; U_M D_M V_M^T\; D_{\priorXzero}) \nonumber \\
&\stackrel{(viii)}{\leq}& etr(\nu\,DD_MD_{\priorXzero}),
\label{eq:ApplicationKristof}
\end{eqnarray}
where the inequality $(viii)$ follows from~\cite{Kristof:1969} (see Theorem on page $5$) as $V_1, U_M$ and $V_M$ are orthogonal matrices while $D$, $D_M$ and $D_{\priorXzero}$ are diagonal matrices with nonnegative diagonal entries. Note that, using sub-multiplicativity of the $\normtwo{\cdot}$~\citep{Conway:1990}, we have
\begin{equation}
\normtwo{D_M} = \normtwo{U_M^TM^T M_{\priorXzero}V_M} \leq \normtwo{U_M^T} \normtwo{M^T}\normtwo{M_{\priorXzero}} \normtwo{V_M} \leq 1.\nonumber
\end{equation}
Therefore, using Lemma~\ref{lem:diagonal_less_norm}, we infer that all the diagonal entries of $D_M$ is less than or equal to $1$. Hence from Equation \ref{eq:ApplicationKristof}, we get that
\begin{eqnarray}
etr(\nu\,V D M^T \priorXzero) &\leq& etr(\nu\,DD_{\priorXzero}).
\end{eqnarray}

Therefore, it follows from~\cite{Kristof:1969} that $M=M_\priorXzero$ and $V = V_\priorXzero$ are unique maximizers when $M_\priorXzero \in \SpaceM$ and $V_\priorXzero \in \SpaceV$. 
Note that, this does not depend on the choice of $\nu$.

Now putting back the value of $M$ and $V$, we write the expression given in the Equation~\ref{eq:JMDY_unnormalized_density} which can now be seen as ${etr(\nu\,D D_\priorXzero)}/{\left[_{0} F_1 \left(\frac{n}{2}, \frac{D^2}{4}\right)\right]^{\nu}}$.  
Note that, the diagonal elements of $D_\priorXzero$ is between $0$ and $1$ as $\normtwo{\priorXzero} < 1$. 
Hence using part (b) of Theorem~\ref{thm:DY_D_property} we know 
that ${etr(\nu\,D D_\priorXzero)}/{\left[_{0} F_1 \left(\frac{n}{2}, \frac{D^2}{4}\right)\right]^{\nu}}$ has a unique maximizer which also does not depend on the choice of $\nu$.

\qedwhite
\newline

\paragraph{$(b)$ }
For $\JMDY$ prior structure, the conditional distribution of $M$ given $(\bd,V)$ is proportional to
\begin{equation*}
etr\left( \nu\,{(\priorXzero V D)}^T M)\right).
\end{equation*}
This distribution is an $\ML$ distribution with parameters $M_\priorXzero^M, D_\priorXzero^M, V_\priorXzero^M$ where SVD decomposition~\citep{Chikuse:2012} of $\nu\,{(\priorXzero V D)} = M_\priorXzero^M D_\priorXzero ^M {(V_\priorXzero^M)}^T$.

Similarly, the conditional distribution of $V$ given $M$ and $\bd$  is proportional to 
\begin{equation*}
etr\left( \nu\,{(\priorXzero^T M D)}^T V)\right).
\end{equation*}
Therefore, it is another $\ML$ distribution with parameters $M_\priorXzero^V, D_\priorXzero^V, V_\priorXzero^V$ where SVD decomposition of $\nu\,{(\priorXzero^T M D)} = M_\priorXzero^V D_\priorXzero^V {(V_\priorXzero^V)}^T$.

Finally, the conditional distribution of $\bd$ given $(M,V)$ is a distribution that belongs to $\IMDY$ class of distributions with parameters $\nu$ and $\BoEtapsiD$, where $\BoEtapsiD = \{ {\etapsiD}_1,  {\etapsiD}_2, \cdots, {\etapsiD}_p \}$ and ${\etapsiD}_j$ is the $j$-th diagonal element of the matrix $M^T\priorXzero V$.

\qedwhite
\newline

In next subsection (Section~\ref{subsec:linearity_modal_param}) we show that the posterior ``modal parameter" is a linear combination of the prior ``modal parameter" and a function of sample mean.

%%%%%%%%%%%%%%%%%%%%%%%

The following lemmas are useful from the practitioner viewpoint. The result will help to truncate the right tail of the distribution at an appropriate point according to a criteria involving only the unnormalized density function. 

%%%%%%%%%%%%%%%%%%%%
\begin{lemma}
Let $\bd \sim \IMDY(\cdot ; \nu, \BoEta)$ for some $\nu>0$ and $\max\limits_{1 \leq j \leq p} \eta_j < 1$ where $\BoEta = \left(\eta_1,\ldots, \eta_p \right)$. 
Let $m$ be the mode of the conditional  distribution, $g_1 (\cdot) := g(\cdot\,;\,\nu,\BoEta\,\mid\,(d_2, \ldots , d_p))$, of the variable $d_1$ given $(d_2, \ldots , d_p)$. Then
$Q(d_1) = g_1(d_1+b)/g_1(d_1)$ is strictly decreasing when $b > 0$ and $d_1 > m$ where $m$ is the mode of the density function given in Definition~\ref{defn:indep_prior}.
\label{lem:Q_d_1}
\end{lemma}

{\bf{Proof of Lemma~\ref{lem:Q_d_1}}.}\\
We have,
\begin{eqnarray}
\log(g_1(d_1)) &=& \nu\,\eta_1\,d_1 - \log \left(\hyp\right) \nonumber\\
\implies \frac{\partial^2}{\partial d_1^2}\left(\log g_1(d_1)\right) &=& - \frac{\partial^2 }{\partial d_1^2} \left(\log \left(\hyp\right) \right) < 0,
\end{eqnarray}
as $\log \left(\hyp\right)$ is a strictly convex function (from Lemma~\ref{lem:0F1_log_convex}). Therefore $\frac{\partial}{\partial d_1}\left(\log g_1(d_1)\right) = g_1^{\prime}(d_1)/g_1(d_1)$ is a strictly decreasing function in $d_1$. 

\begin{eqnarray*}
\log(Q(d_1)) &=& \log(g_1(d_1+b)) - \log(g_1(d_1)) \nonumber \\ 
\implies \frac{\partial}{\partial d_1}\left(\log Q(d_1) \right) &=& \frac{g_1^{\prime}(d_1+b)}{g(d_1+b)} - \frac{g_1^{\prime}(d_1)}{g_1(d_1)} < 0,
\end{eqnarray*}
as $g_1^{\prime}(d_1)/g_1(d_1)$ is a strictly decreasing function. Therefore, $Q(d_1)$ is also a strictly decreasing function in $d_1$.
\qedwhite
\newline

\begin{lemma}
\label{lem:right_tail_prob_bound}
Let $\bd \sim \IMDY(\cdot ; \nu, \BoEta)$ for some $\nu>0$ and $\max\limits_{1 \leq j \leq p} \eta_j<1$ where $\BoEta=\left(\eta_1,\ldots, \eta_p \right)$. 
Let $m$ be the mode of the conditional  distribution, $g_1(\cdot) := g(\cdot\,;\,\nu,\BoEta\,\mid\,(d_2, \ldots , d_p))$, of the variable $d_1$ given $(d_2, \ldots , d_p)$. 
Let  $B>m$, be such that  $\frac{g_1(B)}{g_1(m)}<\epsilon$ for some $\epsilon>0$, then $P(d_1 > B \mid d_2, \ldots, d_p)< \epsilon$.
\end{lemma}

{\bf Proof of Lemma~\ref{lem:right_tail_prob_bound}.}\\
The unnormalized conditional density of the random variable $d_{1}$ is proportional to 
\begin{eqnarray*}
g_1(d_1)=\frac{exp(\nu\;\eta_1\,d_1)}{\hyp^{\nu}}.
\end{eqnarray*}

Let $f(d_1\,;\,\nu,\BoEta \mid (d_2,\ldots, d_p))$ be the density function for the conditional distribution of $d_{1}$ given $\left(d_{2}, \ldots, d_{p}\right)$. For notational convenience, for rest of this lemma we use $f_1(\cdot)$ as the conditional probability density function. Hence we have,
\begin{eqnarray*}
f_1(d_1)= \frac{1}{\ConstCondDen^1} \frac{ \exp{(\nu\,\eta_1 d_1)}  }{\hyp^{\nu}},
\end{eqnarray*}
where $\ConstCondDen^1$ is an appropriate normalizing constant. From Lemma~\ref{lem:Q_d_1}, it follows that $f_1(B+x)/f_1(m+x)$ is a decreasing function of $x$ when $B>m$. Hence for all $x>0$, 
\begin{eqnarray*}
\frac{f_1(B+x)}{f_1(m+x)} = \frac{g_1(B+x)}{g_1(m+x)} < \frac{g_1(B)}{g_1(m)} \stackrel{(viii)}{<} \epsilon,
\end{eqnarray*}
where the inequality at $(viii)$ follows due to the assumption of the lemma.  Therefore, 
\begin{eqnarray*}
P(d_1 > B \mid  \left(d_{2}, \ldots, d_{p}\right)) & = & \int_{B}^{\infty} f_1(y) dy  \nonumber \\
&=&  \int_{0}^{\infty} \frac{f_1(B+x)}{f_1(m+x)} f_1(m+x)\; dx \nonumber \\
 &<& \epsilon \, P\left(d_1>m \,\mid (d_2,\ldots,d_p) \right) \nonumber \\
 &<& \epsilon. \nonumber
\end{eqnarray*}

\qedwhite
\newline
%%%%%%%%%%%%%%%%%%%%%
\subsection{Linearity for posterior modal parameter} 
\label{subsec:linearity_modal_param}
Let $W_i$ for $i=1, 2, \cdots, N$ be the samples drawn from $\ML$ distribution with parameters $M, \bd, V$. If we consider a Bayesian analysis with the prior class $\JMDY$ with parameters $\nu$ and $\priorXzero$, then the probability density for the joint posterior distribution of $M, \bd$ and $V$ given ${\{W_i\}}_{i=1}^N$ is proportional to 

\begin{eqnarray}
&&g(M,\bd,V \,;\,\nu,\priorXzero) \times \prod_{i=1}^N \frac{etr(VDM^T W_i)}{{}_0F_1( \frac{n}{2}, \frac{D^2}{4})} \nonumber \\
&=&
\frac{etr\left( \nu\,VDM^T\priorXzero\right)}{ \left[_0 F_1 (\frac{n}{2}, \frac{D^2}{4})\right]^{\nu}} \times \prod_{i=1}^N \frac{etr(VDM^T W_i)}{{}_0F_1( \frac{n}{2}, \frac{D^2}{4})} \nonumber \\
&=& \frac{etr\left((\nu+N)\,VDM^T\left(\frac{\nu}{\nu+N}\priorXzero + \frac{N}{\nu+N}\overline{W}\right)\right)}{ \left[_0 F_1 (\frac{n}{2}, \frac{D^2}{4})\right]^{\nu+N}},
\end{eqnarray}
where $\overline{W} = \sum_{i=1}^N W_i/N$ and $N$ is the number of data points. Observe that, the posterior distribution is also in $\JMDY$ class with concentration parameter $(\nu + N)$ and modal parameter $\left(\frac{\nu}{\nu+N}\priorXzero + \frac{N}{\nu+N} \overline{W}\right)$.

On the other hand, when we consider a Bayesian analysis with the prior class $\IMDY$ with parameters $\nu$ and $\BoEta$, then the conditional probability density for posterior distribution of $\bd$ given $M$, $V$, ${\{W_i\}}_{i=1}^N$ is proportional to 
\begin{eqnarray}
&&g(\variableX\;; \nu, \BoEta) \times \prod_{i=1}^N \frac{etr(VDM^T W_i)}{{}_0F_1( \frac{n}{2}, \frac{D^2}{4})} \nonumber \\
&=& \frac{ \exp(\nu\;\BoEta^T \variableX ) }{\left[_{0} F_1 \left(  \frac{n}{2}, \frac{D^2}{4}\right)\right]^{\nu}} \times \prod_{i=1}^N \frac{etr(VDM^T W_i)}{{}_0F_1( \frac{n}{2}, \frac{D^2}{4})} \nonumber \\
&=& \frac{ \exp(\nu\;\BoEta^T \variableX ) }{\left[_{0} F_1 \left(  \frac{n}{2}, \frac{D^2}{4}\right)\right]^{\nu}} \times \frac{etr(DM^T N \overline{W} V)}{{\left[{}_0F_1( \frac{n}{2}, \frac{D^2}{4})\right]}^N} \nonumber \;\;\mbox{ where $\overline{W} = {\sum_{i=1}^N W_i}/{N}$}\\
&=&\frac{ \exp(\nu\;\BoEta^T \variableX ) }{\left[_{0} F_1 \left(  \frac{n}{2}, \frac{D^2}{4}\right)\right]^{\nu+N}} \times {\exp(N\sum_{j=1}^{p} d_jY_{j,j})}\;\;\mbox{ where $Y = M^T\overline{W} V \in \Rpp$} \nonumber \\
&=& \frac{ \exp(\nu\;\BoEta^T \variableX )\,\exp(N\sum_{j=1}^{p} d_jY_{j,j}) }{\left[_{0} F_1 \left(  \frac{n}{2}, \frac{D^2}{4}\right)\right]^{\nu+N}} \nonumber \\
&=& \frac{ \exp\left((\nu+N)\;{\left(\frac{\nu}{\nu+N}\BoEta + \frac{N}{\nu+N}\BoEta_{Y}\right)}^T\bd \right)}{\left[_{0} F_1 \left(  \frac{n}{2}, \frac{D^2}{4}\right)\right]^{\nu+N}} \;\;\mbox{ where $\BoEta_{Y} = \left(Y_{1,1},\cdots, Y_{p,p}\right)$} \nonumber \\
\label{eq:posterior_cond_IMDY}
\end{eqnarray}
Here the conditional posterior distribution of $\bd$ is in $\IMDY$ class with concentration parameter $(\nu + N)$ and modal parameter ${\left(\frac{\nu}{\nu+N}\BoEta + \frac{N}{\nu+N}\BoEta_{Y}\right)}$.
\newline

%%%%%%%%%%%%%%%%%%%%%
Finally, in the following subsection we talk about the several reasons for not being able to use DY theorem directly in our case.
%%%%%%%%%%%%%%%%%%%%
\subsection{Inapplicability of DY theorem to construct prior for $\ML$ distribution}
\label{subsec:DY_inapplicable}
According to the assumption of DY, for a $d$-dimensional exponential family distribution, $\mu$ be the measure defined on the Borel sets of $\mathbb{R}^d$. In the context of th $\ML$ distribution $\mu$ is the measure defined on the Stiefel manifold. The symbol $\mathcal{X}$ is used to denote the interior of the support of the measure $\mu$. As showed in~\cite{Hornik:2013} $\mathcal{X}:=\left\{ X: \normtwo{X}<1\right\}$. According to the assumptions of DY $\int_{\mathcal{X}}dP_{\theta}(X)=1$ (See the paragraph after equation (2.1) on page 271 in~\cite{Diaconis:Ylvisaker:1979}). On the contrary for matrix Langevin distribution 
$$ \int_{\mathcal{X}}dP_{\theta}(X) =\int_{\mathcal{X}}f_{\ML}\left( X\right)[dX]=0.$$

During the proof of Theorem $1$ in~\cite{Diaconis:Ylvisaker:1979} Dy constructs a probability measure restricted on set $A$ as follows.
$$\mu_A(B)=\frac{\mu(A\cap B)}{\mu(A)}, \mbox{ where }\mu(A)>0. $$
Also, $x_A = \int Z \;d\mu_A(Z)$. In the context of the proof of Theorem 1 in~\cite{Diaconis:Ylvisaker:1979} uses the crucial fact that $x_A$ are dense in $supp(\mu)$ (See the line after Equation (2.4) on page 272 in~\cite{Diaconis:Ylvisaker:1979}). 

\indent In the context of the $\ML$ distribution $supp(\mu)$ is the Stiefel manifold. It can be shown that similar construction in the case of $\ML$ distribution would lead to $x_A$ where $x_A$ does not belong to the Stiefel manifold i.e. $x_A \not\in supp(\mu)$. Hence $x_A$ will not be dense $supp(\mu)$. As a result, Theorem $1$ in~\citep{Diaconis:Ylvisaker:1979} is not applicable for $\ML$ distribution.
Note that a modified DY construction can be formulated that would enable us constructing prior on $F$. However, our parametrization is different than the natural parametrization, therefore we require a new approach to construct the prior distribution on $M,\bd$ and $V$.

%\item  {\attn{NTT about $\nu<0$ case.}} 

%%%%%%%%%%%%%%%%%%%%%

\begin{figure}
\centering
\begin{tabular}{cc}
\includegraphics[scale=0.23]{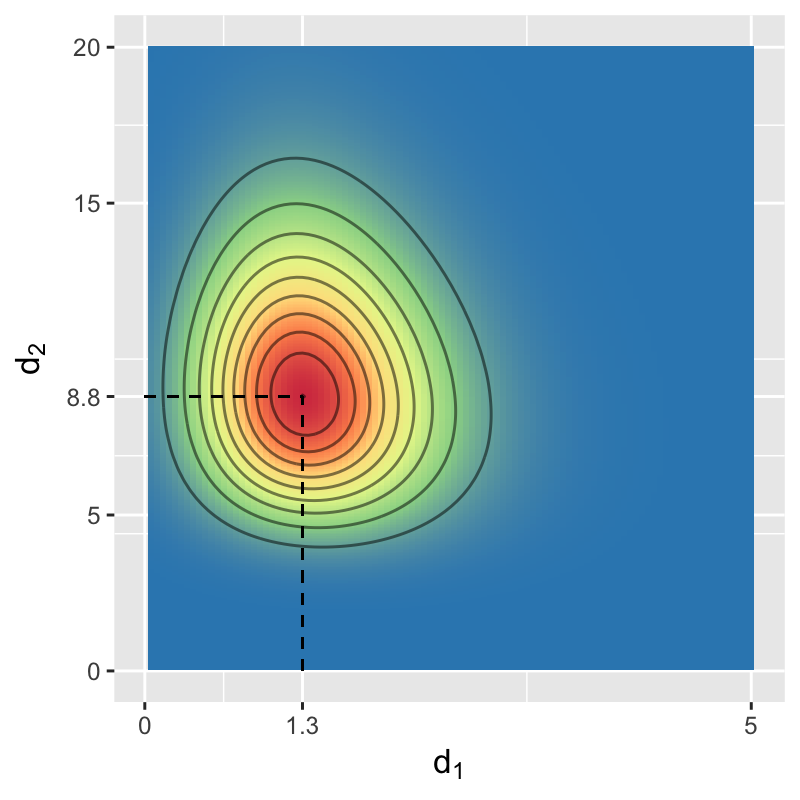}& 
\includegraphics[scale=0.23]{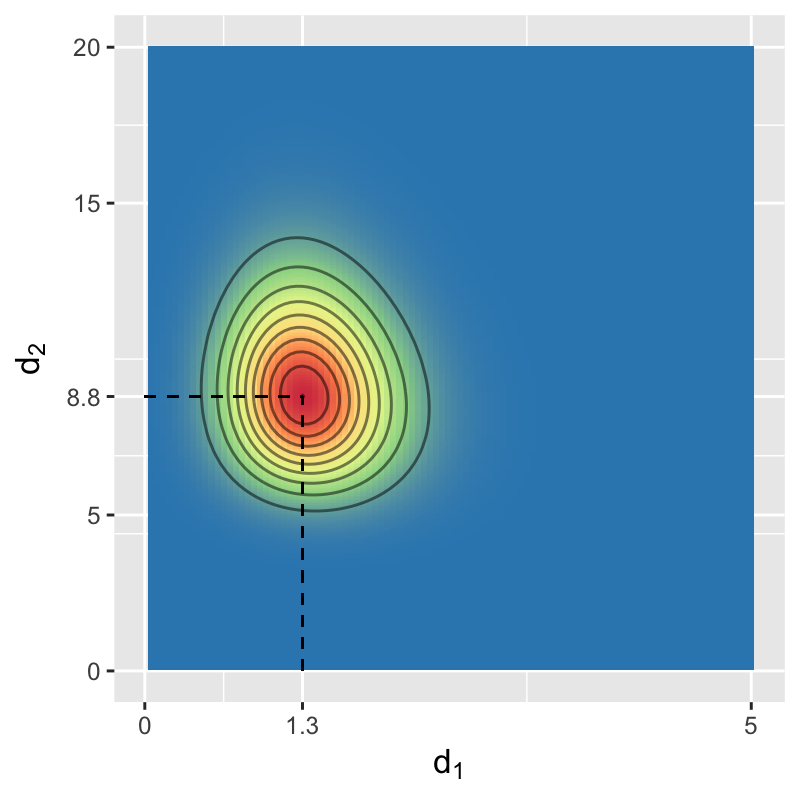}\\
(a) $\nu = 10, \BoEta = [0.50,0.89]$ &
(b) $\nu = 20, \BoEta = [0.50,0.89]$ \\
\includegraphics[scale=0.23]{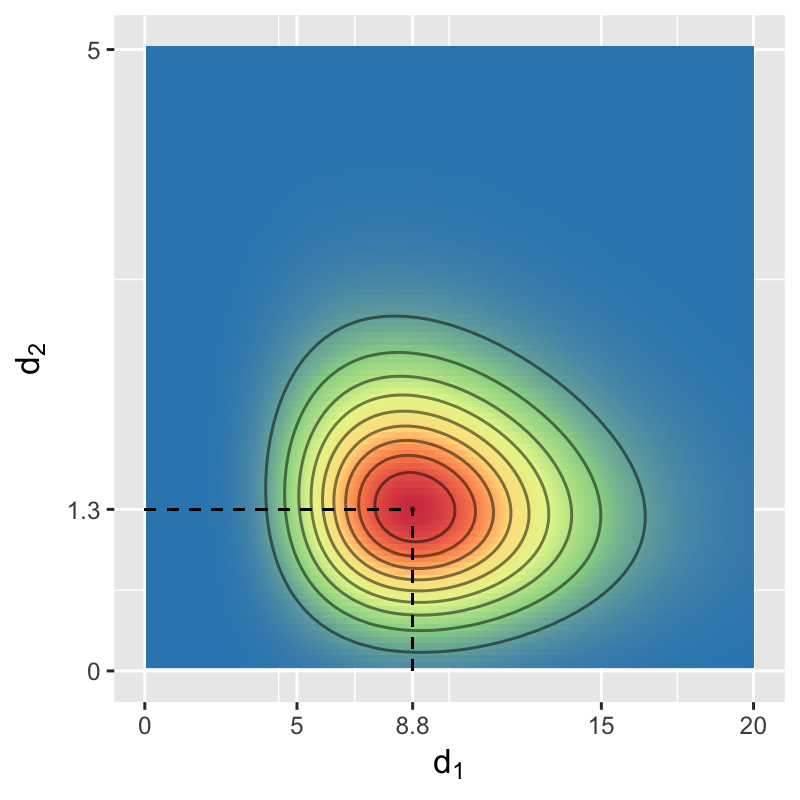}& 
\includegraphics[scale=0.23]{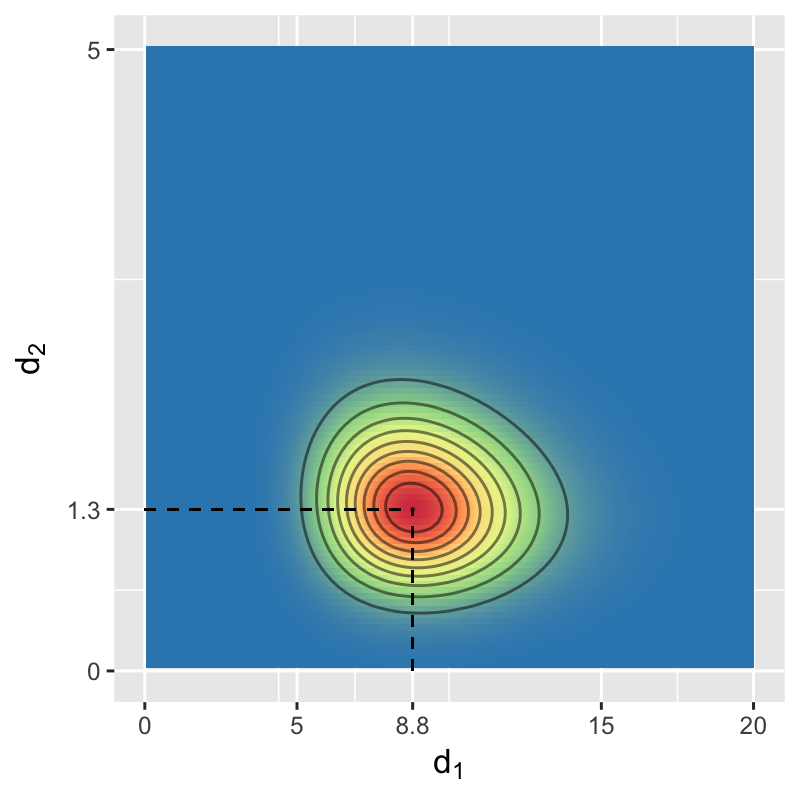}\\
(c) $\nu = 10, \BoEta = [0.89,0.50]$ &
(d) $\nu = 20, \BoEta = [0.89,0.50]$ \\
\includegraphics[scale=0.23]{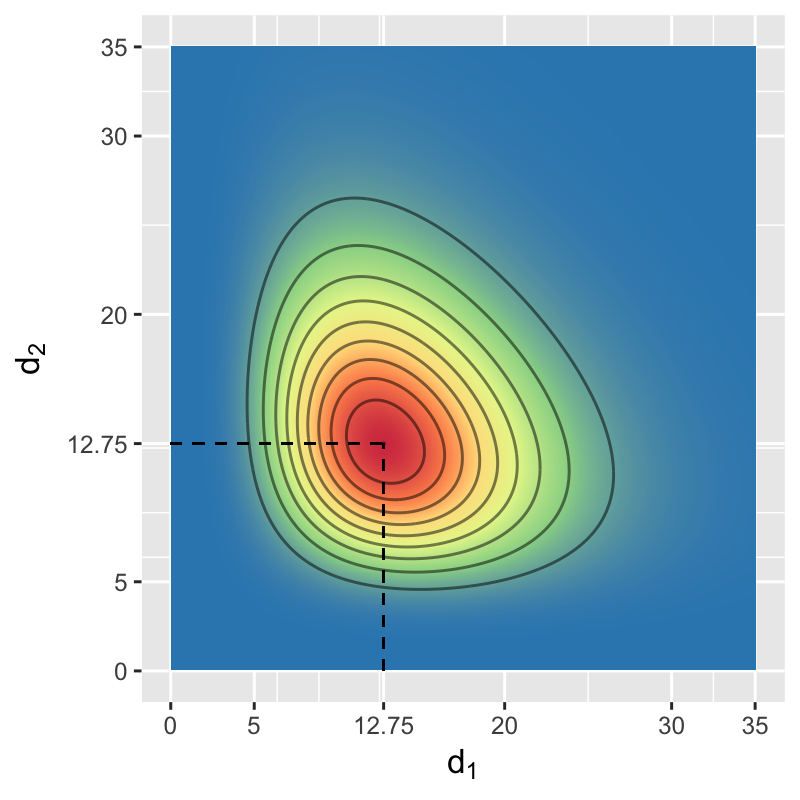}& 
\includegraphics[scale=0.23]{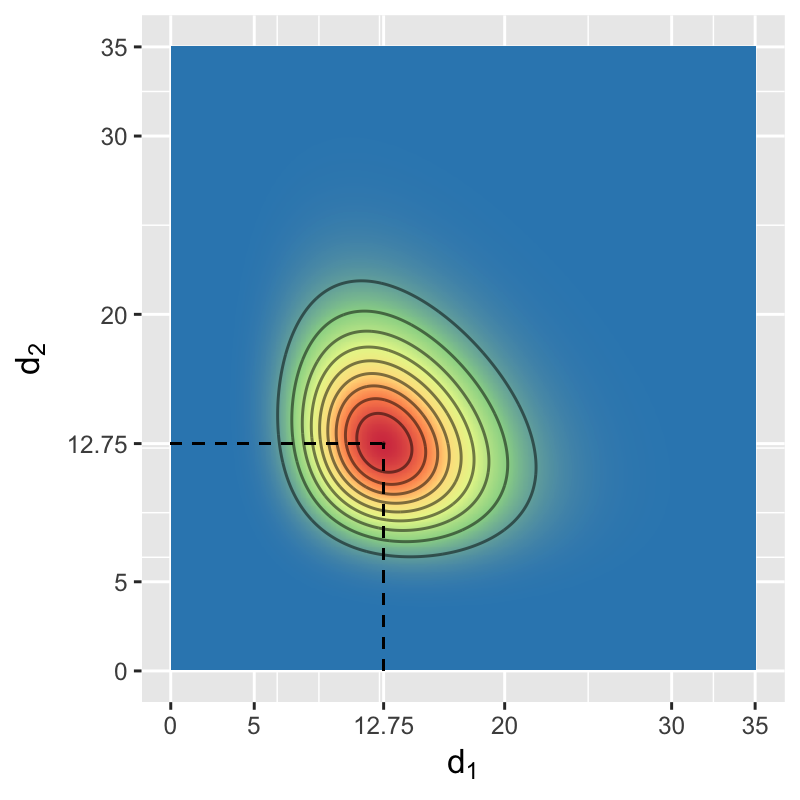}\\
(e) $\nu = 10, \BoEta = [0.94,0.94]$ &
(f) $\nu = 20, \BoEta = [0.94,0.94]$ 
\end{tabular}
\caption{Prior density plots for different values of $\nu$ and $\BoEta$}
\label{fig:prior_plots}
\end{figure}

%%%%%%%%%%%%%%%%%%%%%
\paragraph{Plots for conditional prior of $\bd$ given $M$ and $V$}
Figure~\ref{fig:prior_plots} shows plots for prior densities for different values of $\nu$ and $\BoEta$. Note that, with the same value of $\BoEta$ the location of the mode remain the same for different values of $\nu$ (see each row of Figure~\ref{fig:prior_plots}). As $\nu$ increases, the probability concentration around the mode of the distribution increases.
 
%%%%%%%%%% PLOTS %%%%%%%%%%%%%
%\begin{comment}
%\end{comment}

%%%%%%%%%%%%%

\begin{figure}
\centering
\begin{tabular}{cc}
\includegraphics[scale=0.23]{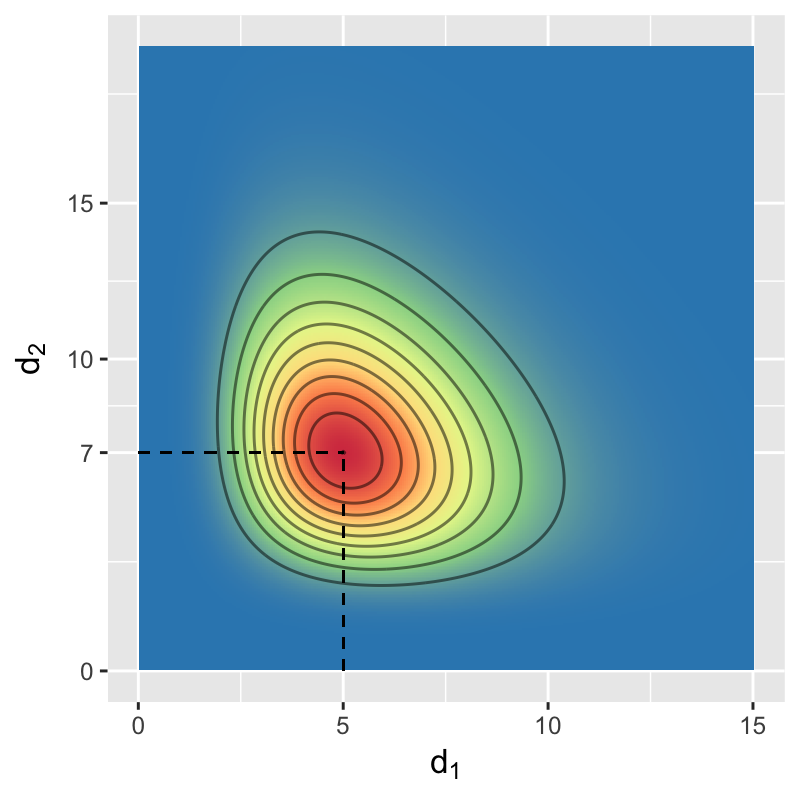}& 
\includegraphics[scale=0.23]{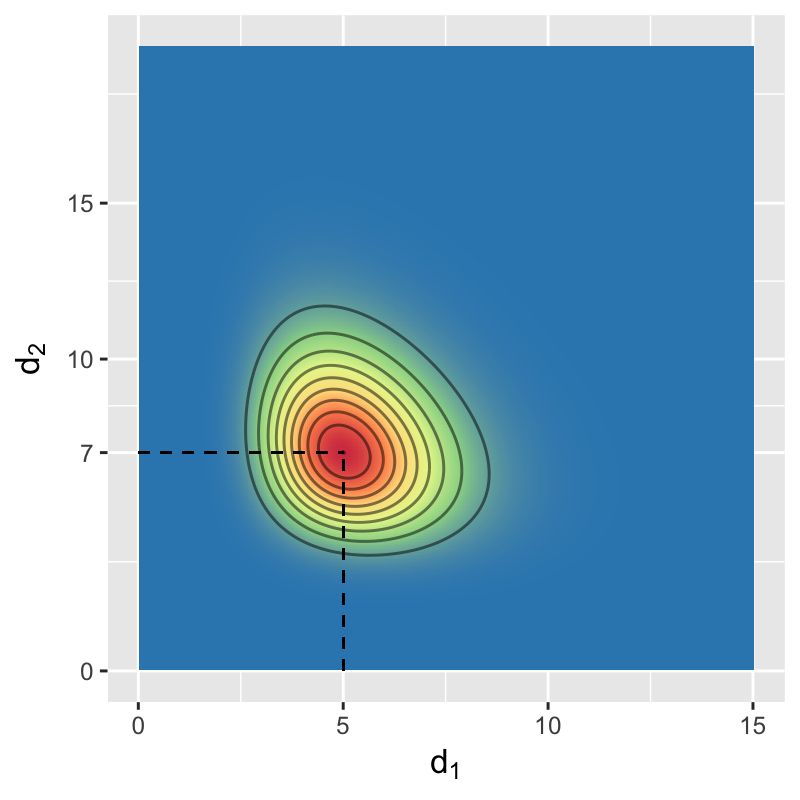}\\
(a) $\nu = 10, \BoEta = [0.85,0.88]$ &
(b) $\nu = 20, \BoEta = [0.85,0.88]$ 
\end{tabular}
\caption{Find appropriate $\BoEta$ from given mode of the distribution $\bd_{mode}$.}
\label{fig:eta_from_mode}
\end{figure}

\paragraph{Finding the modal parameter from the mode}
We have given an example when the practitioner wants to set a particular mode denoted by $\bd_{mode}$. We solve for the corresponding $\BoEta_{mode}$ from Equation~\ref{eq:priorModeEquation}. For example, let us denote the mode by $(5,7)$ and after solving for $\BoEta_{mode}$, we have $\BoEta_{mode} = (0.85, 0.88)$. In the Figure~\ref{fig:eta_from_mode}, we see that the mode is shown by $(5,7)$ for two different setting of $\nu$ which incorporates the strength of the belief in the value of the mode. Here we take $\nu = 10$ and $\nu = 20$.
%(0.8499639, 0.8824125)
%\begin{comment}
%\end{comment}

\subsection{Hyperparameter selection procedure}
\label{subsec:hyperparameter_singleML}
%\begin{itemize}
%\item 
%Uniform improper prior - both joint and indep. 
For both $\JMDY$ and $\IMDY$ class of distributions, we have uniform prior over respective parameters whenever the probability density function is proportional to $1$. For $\JMDY$, it can be achieved by setting $\nu = 0$ in Definition~\ref{defn:joint_prior}. For $\IMDY$, $\nu = 0$ provides the uniform prior on parameter $\bd$.
The resulting priors would be improper as in this case, the integral over the entire space becomes infinite. However, in this case, it is necessary to check the propriety of posterior distributions.

%CS inequality for posterior proprity (improper prior)

%\paragraph{}
%how to incorporate prior belief (example):
In order to incorporate the prior belief for $\IMDY$ class of distributions, one can find the appropriate value of hyperparameter $\BoEta$ from Equation~\ref{eq:priorModeEquation} once mode of $\bd$ (denoted by $\bd_{mode}$) is given. Note that, we get a feasible $\BoEta$ for every real $\bd_{mode} \in \SpaceD$. The other parameter $\nu$ sets the strength of one's prior belief. It is important to realize that there is a strong relationship between $\nu$ and number of data samples. 
For setting the hyperparameters of the prior distribution for $M$ and $V$, one can use $M_{mode}$ and $V_{mode}$, respectively with the appropriate parameters for $\ML$ distribution.
 
%$\nu$ strength of prior belief which can be brought into the model via the number of data samples. According to $\ML$ distribution hyperparams for given $M_{mode}$ and $V_{mode}$ 

On the other hand for $\JMDY$ class of distribution, we set appropriate value of hyperparameter $\BoEta$ from Equation~\ref{eq:priorModeEquation} when mode of $\bd$ is given. Next, we construct a diagonal matrix, $D_{\BoEta}$ with the diagonal entries $\BoEta$. The hyperparameter $\priorXzero$ can be constructed in the following way, $\priorXzero = M_{mode} D_{\BoEta} V^T_{mode}$ where $M_{mode}$ and $V_{mode}$ are the choices for the modes of their respective distributions.
 
%Empirical prior -- joint/indep. 
In order to setup an empirical prior framework, one could obtain the maximum likelihood estimator (MLE) using the technique described in~\cite{Chikuse:2012}. We could set the hyperparameters in such a way that the mode of the prior distribution is same as MLE. Also note that, the ``Empirical Bayesian" procedure~\citep{Robbins:1985, Casella:1985} is out of scope of this study.

%\end{itemize}

%%%%%%%%%%%%%%%%%%%%%%
\section{Bayesian framework for Mixture of $\ML$ distributions}
\label{sec:bayesian_mixture_model}
In this section, we develop a framework for a finite mixture of $\ML$ distributions. We talk about posterior form and consistency. We also elaborate on sampling technique. 
%Generative model for observed data, \\
%Hypergeometric function and its properties \\ 
%For unique representation of parameters on Stiefel we put a restriction on each column of 3x2 orthogonal matrix by adjusting sign appropriately

%\subsection{Remark}
%(**) Note that, one could have a prior on $F_k$ directly but that would not control the direction and the scale of the parameters separately. Mardia's book reference for interpretability of $M_k$.
%Note that, we could have put a prior on $F$ directly. But a single parameter does not help to control direction and scale individually. Instead singular value decomposition (SVD) of $F$ gives us $F = MDV^t$, where $MV^t$ is the mode of the distribution, so the directional component actually comes from this part and scale parameters is governed by the eigenvalues of $D$ matrix. Thus the interpretation and roles of the parameters becomes easier to understand. 

\subsection{Mixture model}
\label{subsec:mix_model}
Cluster analysis helps to determine the internal structure of data in an unsupervised way when no information other than the observed values of data is available~\citep{Picard:2007}. Finite mixture model allows us to cluster data points by assuming that each component of the mixture comes from a suitable parametric distribution and the mixture distribution is constructed by a convex combination of a number of individual component distributions. This number of components is typically specified initially.

We describe our framework as a finite mixture of $\ML$ distribution with a fixed number of mixture component $C$. Details on the selection number of mixture component is described in Section~\ref{subsec:model_sel}. One of the popular techniques of clustering data is to model the data by a mixture of appropriate distributions. For example Gaussian mixture model is one of the most popular methods which has been used in numerous application spanning from computer vision to computational neuroscience~\citep{Stauffer:1999, Mckenna:1999, Kaewtrakulpong:2002, Lewicki:1998, Wood:2004}, in the context of directional data mixture of Von Mises~\citep{Mcgraw:2006,Mardia:2007,Tang:2009,Bangert:2010,Reisinger:2010,Hornik:2014} or mixture of $\ML$ distributions used in~\cite{Lin:2017}.

Consider a product parameter space denoted by
$\ProdSpace := \SpaceM^C \times \SpaceD^C \times \SpaceV^C$. Let $\btheta := {\{\theta_c\}}_{c=1}^C = {\{M_c, \bd_c, V_c\}}_{c=1}^C$ denote any point in $\ProdSpace$. Let $\SpacePi := \{ \langle \pi_1, \pi_2,\cdots, \pi_C \rangle\in (0,1)^C :\sum_{c=1}^C \pi_c = 1 \}$ be the $C$-Simplex, and $\bpi\in\SpacePi$ be any point in it. Let us also denote $\ThetaAndPi:=\ProdSpace \times \SpacePi$. 

Now consider a class of finite mixture of $\ML$ densities denoted by $\ClassML := \{ f(X;(\btheta,\bpi)) = \sum_{c=1}^C \pi_c\, \MLDensity(X;\theta_c): (\btheta,\bpi) \in \Xi\}$. Let $X_i \in \StiefelS$ ($i=1, 2, \cdots, N$) be the observed data from mixture of $\ML$ distributions. 

\begin{equation*}
f(X;(\btheta,\bpi)) = \sum_{c=1}^C \pi_c\, \MLDensity(X;\theta_c)\;\;\mbox{as $f \in \ClassML$}.
\end{equation*}
For convenience it is customary to introduce latent cluster assignment variable to make sampling easier~\citep{Mclachlan:2004, Bishop:2006}. Therefore this mixture model can be described as the following
\begin{equation}
X_i \,\mid\, (Z_i = c) \sim \MLDensity(X_i\,;\, \theta_c) \;\;\mbox{with}\; P(Z_i = c) = \pi_c \;\mbox{for $c=1,\cdots,C$}, 
\label{eq:mix_model}
\end{equation}
where $\pi_c > 0 \mbox{  and  }\sum_{c=1}^C \pi_c= 1$ and $Z_i$ is the latent cluster assignment for $i$-th data point, $X_i$. The likelihood function for the parameter $\btheta$ is given by
\begin{equation}
L(\btheta) = \prod_{i=1}^N \prod_{c=1}^C {\left[ \pi_c\,\MLDensity(X_i \,\mid \,\theta_c)\right]}^{\mathbb{I}(Z_i = c)} 
\label{eq:likelihood_fn}
\end{equation}

%\subsection{Inference}
%\label{sec:inference}
%Prior setup \\
%Hyperparameter setting - (1) weakly informative/flat (2) empirical \\
%Posterior distribution, Proper/Improper posterior condition (constraint mixture model)\\
%Gibbs steps\\
%sampling technique for D\\
%fast computation\\
%posterior maximization as an alternative to MLE\\
%convexity in M step\\
%convergence criterion \\

%\subsection{Posterior distribution}
%\label{subsec:posterior}
In Section~\ref{sec:bayesian_framework_single_ML} we talk about the prior structure and its properties in detail. We assume two different class of prior structures. In the first one, we have 
\begin{eqnarray}
(M_c, D_c,V_c) &\sim & \mbox{\JMDY}(\cdot\,;\,\nu_c, \priorXzero_c)\nonumber \\ 
\bpi &\sim & \mbox{Dir}(\cdot;\,(\alpha_1, \alpha_2, \cdots, \alpha_C)),
\label{eq:joint_prior_model}
\end{eqnarray}
while in the second one, we have
\begin{eqnarray}
M_c &\sim & {\ML}(\cdot\,;\,\hyparam{M}{M}, \hyparam{M}{D}, \hyparam{M}{V})\nonumber \\ 
D_c &\sim & \mbox{\IMDY}(\cdot\,;\,\nu_c, \BoEta_c)\nonumber \\ 
V_c &\sim & \ML(\cdot\,;\,\hyparam{V}{M}, \hyparam{V}{D}, \hyparam{V}{V})\nonumber \\
\bpi &\sim & \mbox{Dir}(\cdot;\,(\alpha_1, \alpha_2, \cdots, \alpha_C)).
\label{eq:indep_prior_model}
\end{eqnarray}

For both the prior structures the conditional posterior distributions of the parameters would be similar. Therefore, we choose to use independent prior structure given in Equation~\ref{eq:indep_prior_model} to demonstrate the posterior computation described in Section~\ref{subsec:sampling}.

The posterior density of $(\btheta,\Bpi, Z_i)$ given ${\{X_i\}}_{i=1}^N$ is proportional to
\begin{equation}
\left\{\prod_{i=1}^N \prod_{c=1}^C {\left[ \pi_c\,\MLDensity(X_i \,\mid \,\theta_c)\right]}^{\mathbb{I}(Z_i = c)} \right\}\,f_{prior}(\bpi,\btheta) 
\label{eq:post_density_1}
\end{equation}
From Equation~\ref{eq:post_density_1} it follows that the posterior density is proportional to
\begin{eqnarray}
&&\prod_{c=1}^C \Bigg\{{\pi_c}^{(\alpha_c+N_c-1)}\, \frac{etr\left(\left(V_c\,D_c\,M_c^T\right)\,N_c\,\overline{X}_c + G_c^0\,M_c + H_c^0\,V_c\right)} {{}_0F_1(\frac{n}{2};D_c^2/4)^{\nu_c + N_c}}  \nonumber \\ 
&&\hspace{2.5in} \exp(\nu_c\,\BoEta_c^T{\bd}_c)\,\mathbb{I}({\bd}_{c} \in \SpaceD)\Bigg\},\nonumber\\
\label{eq:post_mix}
\end{eqnarray}
where $N_c = \sum_{i=1}^N \mathbb{I}(Z_i=c)$ and $\overline{X}_c =\frac{1}{N_c}\,\sum_{i=1}^N X_i \mathbb{I}(Z_i=c)$ for $c=1, \cdots, C$. Also we have,
\begin{eqnarray}
G_c^0 = \hyparam{M}{V}\,\hyparam{M}{D}\,{(\hyparam{M}{M})}^T 
\;\;\mbox{and}\;\; H_c^0 = \hyparam{V}{V}\,\hyparam{V}{D}\,{(\hyparam{V}{M})}^T.
\label{prior_svd}
\end{eqnarray}

\subsection{Hyperparameter selection for mixture model}
\label{subsec:hype_sel}
The class of prior distributions specified in Equations~\ref{eq:indep_prior_model} and~\ref{eq:joint_prior_model} are flexible in the sense, empirical information and/or prior knowledge about any parameters can be incorporated in the model via appropriate hyper-parameter choices. On the other hand, in the absence of prior knowledge, one can specify  hyper-parameters values such that the corresponding prior distributions becomes weakly informative or vague.  In the following section we note down two  specific procedures to select the value of hyper-parameters focusing independent prior structure in Equation~\ref{eq:indep_prior_model} in mind. Similar procedure can easily be developed to select hyper-parameters for the joint prior structure  described in Equation~\ref{eq:joint_prior_model}.

\paragraph{Weakly informative prior}
If the prior probability density function is proportional to $1$ then we refer the corresponding prior as uniform prior. We can construct uniform prior using the prior structure defined in Equation~\ref{eq:indep_prior_model} by choosing  $\alpha_c =1$, $\nu_{c}=0$,   $\hyparam{M}{D}={\boldsymbol 0}_{p,p}$ and $\hyparam{M}{D}={\boldsymbol 0}_{p,p}$ for $c=1,\ldots, C$. Here ${\boldsymbol 0}_{p,p}$ denotes the zero matrix of dimension $p \times p$. Note that, the other hyperparameters,  $\BoEta_c, \hyparam{M}{M}, \hyparam{M}{V}, \hyparam{V}{M}, \hyparam{V}{D}$, are not required to be specified in this case. Note that the uniform prior designed here is improper in nature and the improper priors are not allowed for mixture models as it leads to invalid posterior. As a remedy one may construct ``constrained mixture model"~\citep{DIEBOLT:ROBERT:1994} %Pal:Kang:Guo:2017} 
by introducing some additional constraint to ensure propriety for corresponding posterior. As it is tangential to the current discussion, we avoid the detailed construction on `constrained mixture model' for the current model in this article. Without going into the additional complexity, one may construct weakly informative,  proper prior by choosing $\BoEta_c$ to be very close to zero (such as $0.01$) instead of zero.

\paragraph{Empirical prior}
We first gather the empirical information by fitting a EM based  algorithm to the data to obtain the maximum likelihood estimator of the parameters (see Section~\ref{subsec:EM_part}). Once we have a basic basic estimates of the cluster assignments,  we compute number of points, $n_c^{\dagger}$,  assigned in each clusters and  rough estimates of the cluster specific  parameters,  $M^{\dagger}_c, \bd^{\dagger}_c, V^{\dagger}_c$,  for $c \in  \left\{1, \ldots , C\right\}$. The idea is to choose appropriate hyper-parameter values in Equation~\ref{eq:indep_prior_model}, so that the corresponding prior distributions have modes at the values $M^{\dagger}_c, D^{\dagger}_c, V^{\dagger}_c$. For the prior distribution of $\bd_c$ use the procedure described in Section~\ref{subsec:hyperparameter_singleML} to set appropriate value of ${\BoEta}_c$. For $c=1\ldots C$, we set $\hyparam{M}{M}=M^{\dagger}_c,\hyparam{M}{V}=\iMat $  and $\hyparam{V}{M}=V^{\dagger}_c,\hyparam{V}{V}=\iMat $. The choice for $\nu_c$ and $\hyparam{M}{D}, \hyparam{M}{D}$ are crucial and it may not desired to set very high values for these parameters. We set $\nu_c= {n_{c}^{\dagger}}/{K^{\dagger}} $ and the values for  $\hyparam{M}{D}, \hyparam{M}{D}$ to be close to ${n_{c}^{\dagger}}/{K^{\dagger}}$. Here $K^{\dagger}$ determines the relative strength of the prior distribution appropriately. To select hyper parameters from the parameter $\bpi$ we set  $\alpha_c={n_{c}^{\dagger}}/{K^{\dagger}}$ for $c = 1, \ldots C.$

%{\attn{CUT}}
%\begin{itemize}
	%\item Posterior mode finding for conditional distribution (NR method) - derivation of a computationally tractable form for the derivative of log conditional density
	%\item Uniqueness of posterior mode - log concavity of density function : If the support is $R^+$, then then density has to increase and then has to decrease to 0 asymptotically or it has to decrease monotonically to 0 asymptotically. (for $R$, it must also start from 0). In all the cases, there must be one maxima which would occur in between  or at the left starting point, respectively (or in between).
	
	%\item infinite support is reduced to a compact support (0 to pseudoinfinity) such that probability beyond pseudoinfinity can be made arbitrarily small 
	%\item as long as mode of conditional posterior distribution of $d_1$ and $d_2$ is not large, i.e. $<80$,the previous discrete sampler works reasonably well. On the contrary if $S_{11}$ is close to N ($S_{11}$ is $> N*0.95$), then mode of $d_1$ is large resulting a blow up in hypergeometric function (of a matrix argument). However, with the help of the approximation of ${}_0F_1$ for large argument provided by Mardia, we could construct an efficient sampler for $d_1$,$d_2$.
	%\item Good initial value generation scheme for MCMC chain applying Chikuse's MLE idea and hierarchical clustering ( or EM algorithm) 
%\end{itemize}

\medskip 

In any Bayesian model, consistency of the posterior distribution is a desirable property. In the following subsection we establish posterior consistency for our mixture model.
%\subsection{POSTERIOR CONSISTENCY}
%%%%%%%%%%%%%%%%%%%%%%%%%
%%%% Consistency part %%%%%%%%%%%%
%%%%%%%%%%%%%%%%%%%%%%%%%
%%%%%%%%%%%%%%%%%%%%%%%%%
%%%% Consistency part %%%%%%%%%%%%
%%%%%%%%%%%%%%%%%%%%%%%%%
\subsection{Weak and Strong Posterior Consistency} 
\label{subsec:consistency}

Consider a product parameter space denoted by
$\ProdSpace := \SpaceM^C \times \SpaceD^C \times \SpaceV^C$. Let $\btheta :=
{\{\theta_c\}}_{c=1}^C = {\{M_c, \bd_c, V_c\}}_{c=1}^C$ denote any point in
$\ProdSpace$, and $\btheta_0 := {\{M_c^0, \bd_c^0, V_c^0\}}_{c=1}^C \in
\ProdSpace$ a particular point.
%${\left\{\underbrace{M_c, \bd_c, V_c}_{\theta_c}\right\}}_{c=1}^C$
Let $\SpacePi := \{ \left( \pi_1, \pi_2,\cdots, \pi_C \right) \in (0,1)^C
:\sum_{c=1}^C \pi_c = 1 \}$ be the $C$-Simplex, and $\bpi\in\SpacePi$
be any point in it.

%\footnote{since, not directly used, we can skip it. Note that this specific method of combining spaces, i.e product and the metric itself is not playing any role in any subsequent  proofs in current text, implying any well-defined  metric on any combined space would do.However, this specific definition may be used to rigorize some continuity arguments specifically to rigorize some continuity arguments (which o.w we may well take for granted :))used in (blue parts) in Lemma 2 and a prior mass line in theorem 3. Lin et al skipped the metric definition and hence their `` piecewise continuity" arguments are a little incomplete!} 
Consider the distance metric $d(\cdot,\cdot)$ on the parameter space $\ThetaAndPi:=\ProdSpace \times \SpacePi$ constructed from appropriate distance metrics in the respective parameter spaces:
\begin{eqnarray}
d(\btheta_1,\btheta_2) &:=& \sqrt{\sum_{c=1}^{C}\left[ d^2_{St}(M_c^1,M_c^2) +
	d^2_{Eu}(\bd_c^1,\bd_c^2) + d^2_{St}(V_c^1,V_c^2)\right]}\nonumber \\
d((\btheta_1,\bpi_1),(\btheta_2,\bpi_2)) &:=& \sqrt{ d^2_{Eu}(\bpi_{1},\bpi_{2}) + d^2(\btheta_1,\btheta_2)}
\label{eq:metric_theta_pi}
\end{eqnarray}
and, likewise,
\begin{equation}
d((X_1, \btheta_1,\bpi_1),(X_2, \btheta_2,\bpi_2)) := \sqrt{ d^2_{St}(X_{1},X_{2}) + d^2_{Eu}(\bpi_{1},\bpi_{2}) + d^2(\btheta_1,\btheta_2)}
\label{eq:prod_dist_metric}
\end{equation}
where $d_{Eu}$ is the Euclidean distance and $d_{St}$ is the geodesic
distance on the Stiefel manifold. 
Also consider a class of finite mixture of $\ML$ densities denoted by $\ClassML := \{ f(X;(\btheta,\bpi)) = \sum_{c=1}^C \pi_c\, \MLDensity(X;\theta_c): (\btheta,\bpi) \in \Xi\}$.

We alternatively
denote $f(X; (\btheta,\bpi))$ by $\fthetapi{}(X)$ when we wish to emphasize the
parametrization. $\fthetapi{}: \StiefelS \rightarrow \mathbb{R}^+$ is a family
of probability density functions with respect to the normalized Haar measure
$[dX]$ on $\StiefelS$. Observe that $\ThetaAndPi$ and $\StiefelS$ are complete
separable metric spaces and that $(\btheta, \bpi) \to \fthetapi{}$ is
one-to-one and $(X,\btheta,\bpi) \to f(X;(\btheta,\bpi))$ is measurable.

%%%%%
The prior $\Pi$ is defined on $\ThetaAndPi$.
Let $X_1, X_2, \cdots,X_N$ be independent and identically distributed with
probability density function $\fthetapi{0}$. The posterior distribution
$\Pi(A \mid X_1, X_2, \cdots, X_N)$ for any measurable subset $A$ of
$\ThetaAndPi$ is given by

\begin{equation}
\Pi(A \mid X_1, X_2, \cdots, X_N) =
\frac
{\int_A R_n((\btheta,\bpi))d\Pi((\btheta,\bpi))}
{\int_{\ThetaAndPi} R_n((\btheta,\bpi))d\Pi((\btheta,\bpi))}
\label{eq:induced_measure_posterior}
\end{equation}
where
\begin{equation}
R_n((\btheta,\bpi))=
\prod_{i=1}^N \frac{f(X_i;(\btheta,\bpi))}{f(X_i;(\btheta_0,\bpi_0))}.
\label{eq:induced_measure_posterior2}
\end{equation}

In our model, $\Pi((\btheta,\bpi))$ in
Equation~\ref{eq:induced_measure_posterior} is defined with respect to the appropriate product measure 
%normalized Haar measure 
on $\ProdSpace$ and the Lebesgue measure on $\SpacePi$. 
The prior $\Pi$ is given by the Equation~\ref{eq:indep_prior_model}.
%The prior $\Pi$ has the following structure
%\begin{eqnarray}
%\{\pi_1, \pi_2, \cdots, \pi_C\} &\sim& \mbox{Dir}(\alpha_1, \alpha_2, \cdots, \alpha_C)\nonumber \\
%M_c &\sim& {\ML}(G_0) \nonumber\\
%V_c &\sim& {\ML}(H_0) \nonumber\\
%D_c &\sim& DY_{ML}(\balpha,\bbeta,\nu),
%\label{eq:prior_struc}
%\end{eqnarray}
%where $DY_{ML}(\balpha,\bbeta,\nu) := \exp\left(\sum_{j=1}^p d_j \beta_j\right) \prod_{j=1}^p \frac{d_j^{\alpha_j-1}}{{[{}_0F_1(n/2,D^2/4]}^{\nu}}$.

For $\epsilon>0$ define, respectively, a neighborhood in parameter space, a Kullback-Leibler (KL) neighborhood, a weak neighborhood,  and a Hellinger neighborhood 
%\footnote{pulled this up here, as per Subhadip} 
of $(\btheta_0,\bpi_0)$ (corresponding to the true
density $\fthetapi{0}$) in $\ThetaAndPi$ as

\begin{eqnarray*}
N_{\epsilon}((\btheta_0,\bpi_0)) &=& \left\{ (\btheta,\bpi)\in \ThetaAndPi :
d\left((\btheta,\bpi),(\btheta_0,\bpi_0)\right) < \epsilon \right\}, \\
{KL}_{\epsilon}((\btheta_0,\bpi_0)) &=& \left\{ (\btheta,\bpi)\in \ThetaAndPi :
\int_{\StiefelS} \fthetapi{0}(X) \log \frac{\fthetapi{0}(X)}{\fthetapi{}(X)}
[dX] < \epsilon \right\}, \\
\mathcal{U}_{\epsilon}((\btheta_0,\bpi_0)) &=& \left\{ (\btheta,\bpi)\in \ThetaAndPi :
\bigg\lvert \int_{\StiefelS} g(X)\fthetapi{0}(X)[dX] - \int_{\StiefelS} g(X)\fthetapi{}(X)[dX] \bigg\rvert < \epsilon \right\}, \\
\mathcal{W}_{\epsilon}((\btheta_0,\bpi_0)) &=& \left\{ (\btheta,\bpi)\in \ThetaAndPi :
\left( \int_{\StiefelS}\left(\sqrt{\fthetapi{0}(X)} -
\sqrt{\fthetapi{}(X)}\right)^2 [dX] \right)^{1/2}  < \epsilon \right\}.
\end{eqnarray*}
The weak neighborhood definition holds if the corresponding equation is satisfied for all bounded and continuous functions $g$ on $\StiefelS$.

%%%%%%%%%%%%%%%%%%%%%
\begin{lemma}
\label{lem:ml_prop}
A finite mixture of $\ML$ densities is strictly positive, bounded away from zero and bounded from above.
\end{lemma}
{\bf Proof of Lemma~\ref{lem:ml_prop}.}

Let $f(X; (\btheta,\bpi)) \in \ClassML$ be a density function that is a
$C$-component mixture of $\ML$ distributions parametrized by $(\btheta,\bpi)
\in\Xi$, that is,
\begin{equation}
f(X; (\btheta,\bpi)) = \sum_{c=1}^C \pi_c\, \MLDensity(X; \theta_c).
\label{eq:mixture_with_theta}
\end{equation}

Since the density function $\MLDensity\left(\cdot ; \theta_c\right): \StiefelS
\to \mathbb{R}^+$ is continuous on the compact manifold $\StiefelS$, 
%\footnote{Just directly from here, we can argue that the mixture also is continuous and hence reaches +ve  max and min from extreme value theorem. Of course to get explicit bounds, the rest of the arguments are needed} 
the extreme value theorem~\citep{Rudin:1964} 
%\footnote{Seems this is as common as Taylor's theorem and like, so we may even skip the reference} 
dictates that $\MLDensity\left(\cdot ; \theta_c\right)$
is bounded and attains at least one minima and maxima. In particular,
$\MLDensity(X; \theta_c)$ has the {\em{unique modal orientation}} $M_cV_c^T$
(page 32 in~\cite{Chikuse:2012}) where $\theta_c =\left(M_c,\bd_c,V_c \right)$.
Likewise, it is easy to see that the minimum value of the density
function occurs at $-M_cV_c^T$. Hence for any $X\in \StiefelS $, we have 

\begin{eqnarray}
& & \frac{etr\left(V_cD_cM_c^T\; (M_cV_c^T)\right)}{{}_0F_1(n/2; D_c^2/4)} \geq \MLDensity(X; \theta_c) \geq \frac{etr\left(-V_cD_cM_c^T\; (M_cV_c^T)\right)}{{}_0F_1(n/2; D_c^2/4)}, \nonumber\\
&\implies& \frac{etr\left(D_c\right)}{{}_0F_1(n/2; D_c^2/4)} \geq \MLDensity(X; \theta_c) \geq \frac{etr\left(-D_c\right)}{{}_0F_1(n/2; D_c^2/4)}, \nonumber\\
 & \implies& \frac{\exp (\sum_{i=1}^p d_{ic})}{{}_0F_1(n/2; D_c^2/4)} \geq \MLDensity(X; \theta_c) \geq \frac{\exp (-\sum_{i=1}^p d_{ic})}{{}_0F_1(n/2; D_c^2/4)}>0.
\label{eq:lower_bound_density_ML}
\end{eqnarray}
%On putting back the value of the mode in equation~\ref{eq:MLDensity}, we have 
%\begin{eqnarray*}
%\exp(trace(VDM^TX)) &\leq& = \exp(trace(VDM^TMV^T)) \\
%&=& \exp(trace(VDV^T)) = \exp(trace(V^TVD)) \\
%&=& \exp (trace(D)) = \exp\left(\sum_{i=1}^p d_i \right).
%\end{eqnarray*}
%The lower bound attained by $\MLDensity$ is non-negative (as it is a density function). It is easy to show that the minimum value of the density occurs at $-MV^T$ and the corresponding density value would be $\exp (-\sum_{i=1}^p d_i)/({}_0F_1(n/2; D^2/4))$.
%Therefore the density value of $\ML$ distribution for any $X \in \StiefelS$ has the following equation 
%\begin{equation}
%UB_c = \frac{\exp (\sum_{i=1}^p d_{ic})}{{}_0F_1(n/2; D_c^2/4)} \geq \MLDensity(X; \theta_c) \geq \frac{\exp (-\sum_{i=1}^p d_{ic})}{{}_0F_1(n/2; D_c^2/4)} > 0.
%\label{eq:lower_bound_density_ML}
%\end{equation}
%We later use equation~\ref{eq:lower_bound_density_ML} to claim that our density kernel is bounded from above and below as well. It is also clear that the density kernel is strictly positive. 
\noindent
Let $UB = \max\limits_{1\leq c\leq C}\frac{\exp (\sum_{i=1}^p d_{ic})}{{}_0F_1(n/2; D_c^2/4)}$ and $LB = \min\limits_{1\leq c\leq C}\frac{\exp (\sum_{i=1}^p -d_{ic})}{{}_0F_1(n/2; D_c^2/4)}$. Using Equations~\ref{eq:mixture_with_theta} and~\ref{eq:lower_bound_density_ML}, we get
\begin{equation}
0 < LB \leq f(X; (\btheta,\bpi)) \leq UB < \infty.
\label{eq:lower_bound_mixture_density_ML}
\end{equation}

\qedwhite
\newline

\begin{lemma}
\label{lem:theta_to_density}
Let $(\btheta,\bpi),(\btheta_0,\bpi_0) \in \Xi$. Then for any $\epsilon > 0$, there exists a $\delta > 0$ such that
$$
d((\btheta,\bpi),(\btheta_{0},\bpi_0)) < \delta \implies \sup_{X \in \StiefelS} \bigg\lvert \log \frac{f(X;(\btheta_0,\bpi_0))}{f(X;(\btheta,\bpi))} \bigg\rvert < {\epsilon},
$$
where $f(X;(\btheta,\bpi)), f(X;(\btheta_0,\bpi_0)) \in \ClassML$.
\end{lemma}
{\bf Proof of Lemma~\ref{lem:theta_to_density}.}
%Proof follows from the fact that the density function is continuous w.r.t the parameter $\theta$ and the random variable $X$ and sample space of $X$ is $\StiefelS$, which is compact. We also know that $f(X;\theta)$ and $f(X;\theta_0)$ strictly positive and bounded from above (from lemma~\ref{lem:ml_prop}). Using all the properties, for any $\epsilon>0$, there exists a $\delta >0$, such that

%f0(X) being strictly positive and bounded away from 0. Also, bounded from above.
%and delta can be chosen such that if d(theta0, theta) < delta then for all such f(X)
%f(X) is strictly positive and there exists a strictly positive bound such that all f(X) > that bound. Also, there exists a bound such that all f(X) smaller than that bound.
%we cleanly claim that |log f0/f| < epsilon

%\begin{eqnarray*}
%d(\theta,\theta_0) < \delta &\implies& \underset{X\in\StiefelS}{\mbox{sup}} \lvert \log f_0(X) - \log f(X) \rvert < \epsilon,\\
%&\implies& d(\theta_c,\theta_{0c}) < \delta \quad\forall c=1, 2, \cdots, C\qquad \mbox{is it immediate ??}\\
%&\implies& \underset{X\in\StiefelS}{\mbox{sup}} \bigg\lvert \log \frac{f_0(X)}{f(X)} \bigg\rvert < \epsilon.
%\end{eqnarray*}

%Let us assume that $\theta_0$ is the true value of the parameter. 
Let $ f(X; (\btheta,\bpi)) \in \ClassML $, that is, let
$f(X; (\btheta, \bpi)) = \sum_{c=1}^C \pi_c\, \MLDensity(X;\theta_c)$. Note
that for all $c=1, \ldots, C, $ the function $\MLDensity(X;\theta_c)$ is
continuous in $X$ and $\theta_c $. Since $f(X; (\btheta, \bpi))$ is a linear
combination of functions $\left\{ \MLDensity(X ; \theta_c) \right\}_{c=1}^{C}$
with weights $\{\pi_c\}_{c=1}^{C}$, it too is continuous in $X \in \StiefelS$
and $(\btheta,\bpi) \in \ThetaAndPi$. Moreover, from Lemma~\ref{lem:ml_prop},
$f(X; (\btheta, \bpi))$ is bounded away from $0$ and $\infty$.
Hence $\log\,f(X; (\btheta,\bpi))$  is continuous in $X \in \StiefelS$ and $(\btheta,\bpi) \in \ThetaAndPi$, since $\log$ is continuous and well defined over the range.

Let $B_{(\btheta_0,\bpi_0)}\subset \ThetaAndPi$ be a compact ball around $(\btheta_0,\bpi_0)$ with strictly positive, bounded radius. %\footnote{did read Subhadip's comment, but seems current text is still ok. any resultant $\delta$ would be  ball in the overall space..Maybe something that requires an argument here here could be the fact that we can create such a compact ball at all. This will present us another scope of utilizing the specific  Euclidean metric definition \sss -- did not change anything \sse}

Now, consider the  function
$\log\,f(X;(\btheta,\bpi))$, restricted to the domain
$\StiefelS \times B_{(\theta_0,\bpi_0)}$.
Within  both  $\StiefelS$ and $B_{(\theta_0,\bpi_0)}$, $f$ is continuous in each argument $\theta$, $\pi$ and $X$ 
%\footnote{needs a proof, however short/trivial. Note that following the trail of argument in this para would require building up continuity in $\theta$ from each individual piece of $\theta$ i.e (M, D,V).  Lin et al spends some time even for each such piece!  which makes it seem a non-trivial result \sss -- need to discuss\sse}. 
The compactness of  both these spaces ensures uniform continuity individually in each argument.The latter ensures uniform continuity  of the joint function 
%\footnote{ a trivial triangular argument is sufficient inserted for this last statement \sss -- triangular ? \sse} 
within the joint space  $\StiefelS \times B_{(\theta_0,\bpi_0)}$.

%\footnote{ Either the above slowly-building argument for joint continuity, or a one-stroke Arunabha da’s   argument for composition of functions along with equivalence of $ R^n$and Stiefel topologies.. in which case we need to spell that out clearly  with references}  \ech

Now, since $\StiefelS \times B_{(\btheta_0,\bpi_0)}$ is compact, the function restricted to this domain is {\em uniformly} continuous. Therefore for any
$\epsilon>0$, there exists a $\delta>0$, such that 
\begin{equation*}
d\left( \;(X_1,\btheta_0,\bpi_0),(X_2,\btheta,\bpi)\; \right) < \delta \implies \lvert \log\,f(X_1;(\btheta_0,\bpi_0)) -  \log f(X_2;(\btheta,\bpi)) \rvert < \epsilon,
\end{equation*}
for arbitrary $ (X_1,\btheta_0,\bpi_0) , (X_2,\btheta,\bpi)\in \StiefelS \times B_{(\btheta_0,\bpi_0)}$.
In particular, setting $X_1=X_2=X$, and using the fact that $d\left(\;(\btheta,\bpi),(\btheta_0,\bpi_0)\; \right) = d\left( \;(X,\btheta_0,\bpi_0),(X,\btheta,\bpi)\; \right) < \delta$ for all $X$ (clear from Equations~\ref{eq:metric_theta_pi} and~\ref{eq:prod_dist_metric}), we have 
\begin{eqnarray}
d\left(\;(\btheta,\bpi),(\btheta_0,\bpi_0)\; \right) < \delta &\implies& \sup_{X \in \StiefelS} \lvert \log f(X;(\btheta_0,\bpi_0)) - \log f(X;(\btheta,\bpi)) \rvert < \epsilon, \nonumber \\
&\implies& \sup_{X \in \StiefelS} \bigg\lvert \log \frac{f(X;(\btheta_0,\bpi_0))}{f(X;(\btheta,\bpi))}  \bigg\rvert < \epsilon.
%&\implies& \sup_{X \in \StiefelS} \bigg\lvert \log \frac{f_0(X)}{f(X)}  \bigg\rvert < \epsilon.
\end{eqnarray}

\qedwhite
\newline

\begin{theorem}{\bf (Weak Consistency)}
\label{thm:weak_consis}
For our prior $\Pi$ (defined in Definition~\ref{defn:indep_prior} in Section~\ref{sec:bayesian_framework_single_ML})
$$
\Pi(\mathcal{U}^c \mid X_1, X_2, \cdots, X_N) \to 0 \;\;\; a.s. \;\;\;  F_{\btheta_0,\bpi_0}^\infty,
$$
for any weak neighborhood $\mathcal{U}$ of $(\btheta_0,\bpi_0)$.
$F_{\btheta_0,\bpi_0}$ is the probability distribution function corresponding to the density $\fthetapi{0}$, and $F_{\btheta_0,\bpi_0}^\infty$ is the corresponding infinite product measure.
\end{theorem}

{\bf Proof of Theorem~\ref{thm:weak_consis}.}

For every $\epsilon>0$ there exists a $\delta>0$ such that
\begin{eqnarray*}
d\left( \;(\btheta,\bpi),(\btheta_0,\bpi_0)\; \right) < \delta &\implies& \sup_{X \in \StiefelS} \bigg\lvert \log \frac{f(X;(\btheta_0,\bpi_0))}{f(X;(\btheta,\bpi))}  \bigg\rvert < \epsilon,  \;\; \mbox{[from Lemma~\ref{lem:theta_to_density}]},\\
&\implies& \int_{\StiefelS} f(X;(\btheta_0,\bpi_0)) \bigg\lvert
\log\frac{f(X;(\btheta_0,\bpi_0))}{f(X;(\btheta,\bpi))} \bigg \rvert [dX] < \epsilon,\\
&\implies& \int_{\StiefelS} f(X;(\btheta_0,\bpi_0))
\log\frac{f(X;(\btheta_0,\bpi_0))}{f(X;(\btheta,\bpi))} [dX] < \epsilon.
\end{eqnarray*}

Hence, $N_{\delta}\left((\btheta_0,\bpi_0)\right) \subset
{KL}_{\epsilon}\left((\btheta_0,\bpi_0)\right)$.  It is easy to see that $\Pi$ puts strictly positive measure on 
$N_{\delta}\left((\btheta_0,\bpi_0)\right)$ for all $\delta>0$, because by the definition of distance metric given in Equation~\ref{eq:metric_theta_pi}, any neighborhood around $(\btheta_0,\bpi_0)$ has a positive measure.
%\bch maybe add a line that from the definition of our metric, any such Euclidean ball has a positive measure \ech. 
It follows that $\Pi$ puts strictly positive measure on
${KL}_{\epsilon}\left((\btheta_0,\bpi_0)\right)$ for all $\epsilon>0$.
The theorem then follows from ~\cite{Schwartz:1965}.	

\qedwhite
%\vspace{3mm}
%
%For $\epsilon>0$ define a Hellinger neighborhood of $(\btheta_0,\bpi_0)$
%(corresponding to the true density $\fthetapi{0}$) in $\ThetaAndPi$ as
%
%\begin{equation*}
%\mathcal{W}_{\epsilon}((\btheta_0,\bpi_0)) = \left\{ (\btheta,\bpi)\in \ThetaAndPi :
%	\left( \int_{\StiefelS}\left(\sqrt{\fthetapi{0}(X)} -
%	\sqrt{\fthetapi{}(X)}\right)^2 [dX] \right)^{1/2}  < \epsilon \right\}.
%\end{equation*}
%

\begin{theorem}{\bf (Strong/Hellinger Consistency)}
\label{thm:strong_consis}
For our prior $\Pi$
$$
\Pi(\mathcal{W}^c \mid X_1, X_2, \cdots, X_N) \to 0 \;\;\; a.s. \;\;\;  F_{\btheta_0,\bpi_0}^\infty,
$$
for any Hellinger neighborhood $\mathcal{W}$ of $(\btheta_0,\bpi_0)$.
\end{theorem}

{\bf Proof of Theorem~\ref{thm:strong_consis}.}

For any $\epsilon>0$ consider the weak neighborhood
$$
\mathcal{U}_{\epsilon^2}((\btheta_0,\bpi_0)) = \left\{ (\btheta,\bpi)\in \ThetaAndPi :
\bigg\lvert \int_{\StiefelS} g(X)\fthetapi{0}(X)[dX] - \int_{\StiefelS} g(X)\fthetapi{}(X)[dX] \bigg\rvert < \epsilon^2 \right\}.
$$ for all bounded and continuous functions $g$ on $\StiefelS$.

For each $(\btheta,\bpi)\in \mathcal{U}_{\epsilon^2}((\btheta_0,\bpi_0))$ choose
$$
g(X) := \left(\frac{\sqrt{\fthetapi{0}(X)}-\sqrt{\fthetapi{}(X)}}{\sqrt{\fthetapi{0}(X)}+\sqrt{\fthetapi{}(X)}} \right)
$$
Now from Lemma~\ref{lem:ml_prop}, the functions $\fthetapi{}$  are  bounded away from $0$, ensuring a positive lower bound for the denominator. The upper bound for the denominator is guaranteed from the upper bound  property that follows from the same Lemma.  A  similar argument holds for the numerator as well. This  ensures  boundedness of the function $g(X)$ and the continuity follows from the continuity of $f(X)$ (from Lemma~\ref{lem:ml_prop}). Thus $g(X)$ is a bounded and continuous function. Hence, 

\begin{eqnarray}
& & \bigg\lvert \int \left(\frac{\sqrt{\fthetapi{0}(X)}-\sqrt{\fthetapi{}(X)}}{\sqrt{\fthetapi{0}(X)}+\sqrt{\fthetapi{}(X)}} \right) \fthetapi{0}(X)\,[dX]- \nonumber \\
&&\hspace{1in} \int \left(\frac{\sqrt{\fthetapi{0}(X)}-\sqrt{\fthetapi{}(X)}}{\sqrt{\fthetapi{0}(X)}+\sqrt{\fthetapi{}(X)}} \right) \fthetapi{}(X)\,[dX] \bigg\rvert <\epsilon^2, \nonumber\\
& \implies &  \bigg \lvert \int {\left(\sqrt{\fthetapi{0}(X)}- \sqrt{\fthetapi{}(X)}\right)}^2\,[dX] \bigg\rvert < \epsilon^2, \nonumber\\
& \implies &  \left( \int {\left(\sqrt{\fthetapi{0}(X)}- \sqrt{\fthetapi{}(X)}\right)}^2\,[dX] \right)^{1/2} < \epsilon.
\end{eqnarray} 

Hence $\mathcal{U}_{\epsilon^2}((\btheta_0,\bpi_0))\subset
\mathcal{W}_{\epsilon}((\btheta_0,\bpi_0))$.
The Theorem now follows from an application of Theorem~\ref{thm:weak_consis}.
\qedwhite

\subsection{Sampling procedure}
\label{subsec:sampling}

%\begin{itemize}
%\item Gibbs full conditional ( main $D$ sampling; NR method; unimodality; approximate large value)
%\item computation of 0F1 - (2x2 special form by Bessel) 
%\item Initial cluster selection 
%\item MCMC convergence 
%\end{itemize}
In order to perform Bayesian inference, it is important to compute statistics related to the posterior distribution e.g. the posterior mean or posterior quantiles. The posterior density, defined in Equation~\ref{eq:post_mix}, is intractable in the sense that it is not possible to compute these quantities analytically by performing integration or to generate i.i.d. samples from the posterior distribution.

However we can design a Gibbs sampling Markov chain to generate samples from the posterior distribution. It is known that the Markov chain corresponding to Gibbs samplers would converge to the desired stationary distribution. 

In order to implement the Gibbs sampler, we sample cluster specific parameters along with the latent indicator $Z_i$ for cluster assignment for each data point $X_i$ where $i=1, 2, \cdots, N$. The conditional distribution of $Z_i$ given all other parameters follows 
\begin{equation}
P(Z_i=c \mid M_c, \bd_c,V_c, \Bpi, {\{X_i\}}_i^N) = \frac{\pi_c \, \MLDensity(X_i \mid M_c, \bd_c,V_c)}{ \sum_{c=1}^C\pi_c\,\MLDensity(X_i \mid M_c, \bd_c,V_c)}
\label{eq:conditional_Z}
\end{equation}
for $c=1, 2,\cdots, C$. The conditional posterior distribution of $\Bpi$ is given as
\begin{equation}
\Bpi \mid {\{M_c, \bd_c,V_c\}}_{c=1}^C, {\{X_i,Z_i\}}_i^N \sim \mbox{Dir} (\alpha_1+N_1, \alpha_1 + N_2, \cdots, \alpha_C + N_C),
\label{eq:conditional_pi}
\end{equation}
where $N_c = \sum_{i=1}^N \mathbb{I}(Z_i=c)$ for $c=1, \cdots, C$.
Given the latent cluster assignments, the conditional posterior distribution of cluster specific parameters are independent. Due to conditional functional conjugacy for $M_c$ and $V_c$, it is straightforward to show that the full conditional of the corresponding posterior would belong to the $\ML$ class of distribution. In particular,  
\begin{eqnarray}
M_c \mid (\bd_c,V_c,{\{X_i,Z_i\}}_i^N,\Bpi) &\sim& \ML\left(\cdot\,;\,(S^M_G, S^D_G, S^V_G)\right) \\
V_c \mid (M_c,\bd_c,{\{X_i,Z_i\}}_i^N, \Bpi) &\sim& \ML\left(\cdot\,;\,(S^M_H, S^D_H, S^V_H)\right),
\label{eq:conditional_M_V}
\end{eqnarray}  
where  ($S^M_G, S^D_G, S^V_G$) and $S^M_H, S^D_H, S^V_H)$ are SVD decompositions of matrices $(D_cV_c^T\,N_c\overline{X}_c^T + G_c^0)$ and $(D_cV_c^T\,N_c\overline{X}_c^T + H_c^0)$, respectively. Observe that 
$$
\overline{X}_c = \frac{1}{N_c}\sum_{i=1}^N X_i \,\mathbb{I}(Z_i=c).
$$
Efficient sampling from $\ML$ distribution is done using algorithm developed in~\cite{Hoff:2009}. 

%\prod_{c=1}^C \left\{{\pi_c}^{(\alpha_c+N_c-1)}\, \frac{etr\left(\left(V_c\,D_c\,M_c^T\right) \overline{X}_c + G_c^0\,M_c + H_c^0\,V_c\right)} {{}_0F_1(\frac{n}{2};D_c^2/4)^{\nu_c + N_c}} \exp(\nu_c\,\BoEta_c^T{\bd}_c)\,\mathbb{I}({\bd}_{c} \in \SpaceD)\right\},\nonumber\\

The conditional posterior distribution for $\bd_c$ given other parameters has the following density --
\begin{eqnarray}	
f(\bd_c \mid M_c, V_c, {\{X_i,Z_i\}}_{i=1}^N,\Bpi) &\propto&  \frac{\exp\left((\nu_c\,\BoEta_c^T+N_c\,\phi_c^T)\,{\bd}_c\right)}{{}_0F_1(\frac{n}{2};D_c^2/4)^{\nu_c + N_c}}\,\mathbb{I}({\bd}_{c} \in \SpaceD) \nonumber \\
\bd_c \mid (M_c, V_c, {\{X_i,Z_i\}}_{i=1}^N,\Bpi) &\sim & \mbox{\IMDY}\left(\cdot\,;\,(\nu_c+N_c), \frac{\nu_c\,\BoEta_c}{\nu_c+N_c}+ \frac{N_c\,\bphi_c}{\nu_c+N_c}\right), \nonumber \\
\label{eq:D_c_post}
\end{eqnarray}
where $\bphi_c = \{\phi_{c1}, \phi_{c2}, \cdots, \phi_{cp}\}$ with $\phi_{cj}$ is the $j$-th diagonal element of the matrix $M_c^T \overline{X}_c V_c$ for $j=1, 2, \cdots, p$. Note that, this can also be verified from the Equation~\ref{eq:posterior_cond_IMDY} in Section~\ref{subsec:linearity_modal_param}. Due to non-standard form of the posterior distribution given in Equation~\ref{eq:D_c_post}, sampling of ${\bd}_c$ is challenging. 

%\end{eqnarray*} 
The density corresponding to the full conditional distribution of $d_{cj}$, the $j$-th diagonal entry of $D_c$ for $j=1, 2, \cdots, p$, is given below,
\begin{eqnarray}
&&f\left(d_{cj} \mid \bd_{cj}^{-},M_c,V_c,{\{X_i,Z_i\}}_i^{N},\Bpi \right) \propto \nonumber \\
&& \hspace{0.6in} \frac{\exp(d_{cj}\,(\nu_c\,\eta_{cj} + N_c\, \phi_{cj}))}{{}_0F_1(n/2,D_c^2/4)^{\nu_c+N_c}} \,\mathbb{I}\left(d_{c(j+1)} < d_{cj} < d_{c(j-1)}\right).
\label{eq:dc_full_cond}
\end{eqnarray}
Also let $F_{cj}(\cdot)$ is the corresponding distribution function of conditional distribution of $d_{cj}$. We describe the detailed implementation of sampling from this conditional distribution in the following paragraph after this subsection.  
For a generic representation of posterior distribution for $d_{cj}$ for all $j=1, \cdots, p$, we define $d_{c0} = \infty$ and $d_{c(p+1)} = 0$. We also write 
\begin{equation}
\bd_{cj}^{-} := \left\{ d_{c1},\cdots, d_{c(j-1)}, d_{c(j+1)}, \cdots, d_{cp} \right\}.
\end{equation}

We have designed an efficient sampling scheme to sample $\bd_c$ using the set of $p$ distributions given in Equation~\ref{eq:dc_full_cond}. Observe that support of the distribution for $d_{c1}$ is $[ d_{c2},\infty)$ while that of the others are bounded. Note that the posterior distribution of $d_{c1}$ is unimodal (see Theorem~\ref{thm:DY_D_property}) and we exploit that fact to design an efficient sampler for ${\bd}_c$.
A description of the sampling steps is given in Algorithm~\ref{alg:gibbs} below.

Note that, because of log-concavity nature of the conditional distribution function for $d_{cj}$, we could have implemented adaptive rejection sampler (ARS) for it. However, the standard ARS algorithm can not be immediately implemented in this context because of involved computation with ${}_0F_1(\cdot)$ function. So we reserved this development for our future work.
%\PalCmnt{Discuss the distribution of each diagonal entries of D given all the other parameters is in fact  log concave hence ARS is a possibility. We need to investigate if we can actually apply ARS? The problem of truncation of tail still may be applicable as the ARS requires finite support for the underlying distribution to sample from. }

\paragraph{Gibbs algorithm}

The following algorithm outlines the steps of the Gibbs sampling algorithm which shows the full conditional distribution of the parameters at $k$-th step based on the samples drawn at $(k-1)$-th step and data. 

\begin{algorithm}[H]
\caption{Algorithm for MCMC method}
\small
\label{alg:gibbs}
\begin{algorithmic}
\vspace{2mm}
\For{$c=1,2,\cdots, C$}
	\State initialize the MCMC chain with $M_c^{(0)}, D_c^{(0)}, V_c^{(0)}$ and ${\Bpi}^{(0)}$
\EndFor
\vspace{2mm}
\State $k = 0$
\vspace{2mm}
\Repeat
\vspace{2mm}
\State $k = k+1$
\vspace{2mm}
\For{$i=1,2,\cdots, N$}
%\State  $Z_i^{(k)} \mid \left(M_c^{(k-1)} D_c^{(k-1)},V_c^{(k-1)}, \Bpi^{(k-1)}\right) \sim f_{multinom}(1,\Bpi^{(k-1)})$
\State  $Z_i^{(k)}  \sim \mbox{Categorical}\left(\cdot \,;\, \{1, 2, \cdots,C\} ,\Bpi^{(k-1)}, {\{M_c^{(k)}, D_c^{(k)},V_c^{(k)}\}}_{c=1}^C\right)$
\EndFor
\vspace{2mm}

\For{$c=1,2,\cdots, C$} 
\begin{eqnarray*}
N_c^{(k)} &=&  \sum_{i=1}^N \mathbb{I}(Z_i^{(k)}=c) \\
\overline{X}_c^{(k)} &=& \frac{1}{N_c}\,\sum_{i=1}^N X_i \, \mathbb{I}(Z_i^{(k)}=c)\\
M_c^{(k)} &\sim& \mbox{$\ML$} \left(\cdot \,;\, \left({(S^M_G)}^{(k-1)}, {(S^D_G)}^{(k-1)}, {(S^V_G)}^{(k-1)}\right)\right)\\ 
V_c^{(k)} &\sim& \mbox{$\ML$} \left(\cdot \,;\, \left({(S^M_H)}^{(k)}, {(S^D_H)}^{(k-1)}, {(S^V_H)}^{(k-1)}\right)\right)\\
d_{cj}^{(k)} &\sim& F_{cj}\left(\cdot \,;\, {(\bd_{cj}^{-})}^{(k)},M_c^{(k)},V_c^{(k)},{\{X_i,Z_i\}}_i^{N},\Bpi^{(k-1)} \right)\;\; \mbox{for all $j=1, 2, \cdots, p$}
\end{eqnarray*} 
 
\EndFor

\vspace{2mm}

\State  $\Bpi^{(k)} \sim \mbox{Dir} \left(\cdot \,;\, \alpha_1+N_1^{(k)}, \cdots, \alpha_C + N_C^{(k)}\right)$

\Until convergence
\vspace{2mm}
\end{algorithmic}
\end{algorithm}
Note that, ${(\bd_{cj}^{-})}^{(k)}$ is given by the following set
\begin{equation*}
{(\bd_{cj}^{-})}^{(k)} := \left\{ {(d_{c1})}^{(k)}, \cdots, {(d_{c(j-1)})}^{(k)},{(d_{c(j+1)})}^{(k-1)}, \cdots, {(d_{cp})}^{(k-1)} \right\}
\end{equation*}

%where form of $f_j(\cdot)$ is given in equation~\ref{eq:dc_full_cond}.
%\paragraph{Initial value selection}

The stationary distribution of the Gibbs sampling Markov chain is the posterior distribution corresponding to Equation~\ref{eq:post_mix}. Convergence to this stationary distribution does not on the choice of the initial point. However, in order to run the MCMC method it is required to initialize Algorithm~\ref{alg:gibbs} with certain values (e.g. $M_c^{(0)}, D_c^{(0)}, V_c^{(0)}$ and ${\Bpi}^{(0)}$). In practice, specifically in the case of large-scale dataset, it is often seen that bad choice of initial value might lead to slow convergence of the MCMC method. In order to come up with a reasonable choice of initial value, we first run a hierarchical clustering~\citep{Lattin:2003,Rokach:2005} on the entire dataset  with a fixed number of clusters, $C$ (for selection of optimal $C$ see Section~\ref{subsec:model_sel}) to get a initial cluster assignments for the data points. Based on the initial assignment, we adopt a maximum likelihood based technique described in~\cite{Chikuse:2012} to obtain the initial value of the cluster specific parameters. This initial point selection procedure has worked well for our simulated dataset.
We notice that the selection of initial point may not be crucial for small datasets. However, for large dataset choice of suitable initial point could save significant amount of time by reducing number of burn-in steps.
%\begin{eqnarray*}
%\mathbb{E}(g(d_{c1}^t\mid d_{c1}^{t-1})) = \int g(d_{c1}^t)dP(d_{c1}^t|d_{c1}^{t-1}) \leq \lambda\,g(d_{c1}^{t-1}) + C
%\end{eqnarray*}
%where $\lambda$ is a constant less than $1$ and $g$ is any function such that the sets $\{ x: g(x) \leq C\}$, are compact for all $C$. For example, $g(x) = \frac{1}{x^\delta}+x^\delta$ for $\delta > 0$.
%\begin{eqnarray*}
%log(f(D)) &\propto& -N \log({}_0F_1(n/2,D^2/4)) + (\alpha-1) (\log d_1 + \log d_2) + \left(d_1\,(S_{11}-\beta) + d_2\,(S_{22}-\beta)\right)  \,\mathbb{I} \{d_1 \geq d_2 \} 
%\end{eqnarray*}

%For a special case with $n=3 ,\; p=2 \;\;[\mbox{ where } d_1 > d_2 > 0]$ the above equation becomes --
%\begin{eqnarray*}
%P(D_c \mid M_c,V_c,{\{X_i\}}_i^N)&\propto&  \frac{1}{{}_0F_1(n/2,D_c^2/4)^{N_c}}  \times d_{c1}^{\alpha-1} \, d_{c2}^{\alpha-1} \exp\left(d_{c1}\,(S_c^{11}-\beta) + d_{2c}\,(S_c^{22}-\beta)\right) \,\mathbb{I} \{d_{c1} \geq d_{2c} \}
%\end{eqnarray*}

%%%%%%%%%%%%%%
\paragraph{Efficient Rejection Sampler} 
In this section we describe the rejection sampling procedure from the conditional distribution of $(d_1\,\mid\,(d_2, \cdots, d_p))$ when
$\bd \sim \IMDY(\cdot ; \nu, \BoEta)$ for some $\nu>0$ and $\max\limits_{1 \leq j \leq p} \eta_j<1$. Here $\BoEta=\left(\eta_1,\ldots, \eta_p \right)$. 
Let $m$ be the mode of the conditional  distribution, $g_1(\cdot) := g(\cdot\,;\,\nu,\BoEta\,\mid\,(d_2, \ldots , d_p))$, of the variable $d_1$ given $(d_2, \ldots , d_p)$ when $\eta_1 > 0$. In case, $\eta_1 < 0$, we explicitly set $m$ to be $0$.

Using  property of the conditional distribution described Lemma~\ref{lem:right_tail_prob_bound} the we compute a critical point $RT_{cric}$ so that $P\left(d_{1} > RT_{crit} \,\mid \,(d_2, \cdots, d_p), \{{X_j\}}_{j=1}^N \right)< \epsilon$ with the choice of $\epsilon=0.0001$. 

%We draw a Bernoulli distributed variable with  probability of head being $\epsilon$ and for practical purpose, we end up with tail  
%[ Theoretical justification; we draw a Bernoulli distributed variable with  probability $\epsilon$ ; For the practical purpose this ]. 
We restrict the support of the conditional posterior distribution for  $d_{c1}$ to the bounded interval $(0, RT_{crit}]$. We employ a efficient rejection sampling scheme to sample from the desired distribution in the following way.

Let $\delta = {RT_{crit}}/{N_{bin}}$ where $N_{bin}$ is the total number of partitions for the interval $(0, RT_{crit}]$. Consider, $k=(\left[ {m}/{\delta}\right]+1)$ where $\left[{m}/{\delta} \right]$ denotes the greatest integers less that or equal to $ {m}/{\delta}$. Now define the function 
\begin{eqnarray*}
\overline{g}_1(x) &:=& \sum_{j=1}^{k-1} g_1(j\,\delta)\; \mathbb{I}_{\left( (j-1) \delta, j \delta]\right)} (x)  + g_1(m) \mathbb{I}_{\left( (k-1) \delta, k \delta]\right)} (x) \\ 
&&\qquad +\sum_{j=k+1}^{N_{bin}} g_1((j-1)\,\delta)\;  \mathbb{I}_{\left( ( (j-1) \delta, j \delta]  \right)} (x).
\end{eqnarray*} 
Note that  $ \overline{g}_1(x) \geq g_1(x)$ for all $x \in (0, RT_{crit}]$ as $g_1(\cdot)$ unimodal log-concave function with maxima $m$. To sample from  for the distribution with density corresponding to the function $\overline{g}_1(\cdot)$ we consider,
$p_j = {q_j}/{\sum_{j=1}^{N_{bin}}q_j}$ for $j = 1, 2, \cdots, N_{bin}$ where,
$$
q_j = \threepartdef
{{g_1(j\delta)}{}}      {1\leq j< \left[\frac{m}{\delta}\right]+1,}
{{g_1(m)}{}}      { j= \left[\frac{m}{\delta}\right]+1,}
{{g_1((j-1)\delta)}{}} {\left[\frac{m}{\delta}\right]+1< j\leq M.}
$$
The steps of the rejection samplers are given below
\begin{itemize}
\renewcommand\labelitemi{--}
\item Sample $Z$ from the discrete distribution with the support  $\{ 1, 2, \ldots, N_{bin}\}$ corresponding probability $\{p_j \}_{j=1}^{M}$. 
\item Sample $y \sim Uniform\left(  (Z-1)\,\delta, Z\delta \right)$.
\item Sample $ U \sim Uniform (0,1) $.
\item Accept $y$ if  $U\leq \frac{g_1(y)}{\overline{g}_1(y)}   $.
\end{itemize}

Note that the efficiency of the sampler increases when we choose larger values for $N_{bin}$. 

%%%%%%%%%%%%%%%

%why F instead of M,D,V for summary ?
\paragraph{Posterior summary}
There are multiple ways to summarize the posterior distribution for the estimates of the parameters. We choose to use the parametrization given in Equation~\ref{eq:MLDensity_F} for posterior summary. This parametrization enables us to report the error in more interpretable way. Using $M, \bd, V$, it is challenging to report the error as $M$ and $V$ lie on a non-euclidean space. Generating summary of results for different parameters on $\StiefelS$ is not straightforward. Some generalized version of mean like Karcher mean could be investigated. Note that we can directly compare true $F$ and $\hat{F}$ as there is no constraint on the elements of $F$. This direct comparison is not immediately possible for $M, \bd, V$ parametrization given in Equation~\ref{eq:MLDensity_MDV} which is mainly done to achieve computational tractability.

\subsection{Model selection}
\label{subsec:model_sel}
In order to identify the optimum number of cluster we use Deviance Information Criteria  for Bayesian model selection ($DIC$)~\citep{Book:Gelman:Rubin:2003,Spiegelhalter:2002}. It has been successfully used as a model selection criteria in in various Bayesian models~\citep{Berg:Meyer:Yu:2004, Franccois:Laval:2011,Khare:Pal:Su:2017}. To explain the DIC criterion in the context of the current model, let  $\thetabf^{(C)} = \left\{M_c, \bd_c, V_c\right\}_{c=1}^{C} $ denote all the parameter vectors and the {\it deviance function} is defined as $Dev(\thetabf^{(C)}):= -2 log L(\thetabf^{(C)})$ where $L(\cdot)$ is the likelihood function defined in Equation~\ref{eq:likelihood_fn}.
%In order to identify the optimum number of cluster we considered DIC~\citep{Book:Gelman:Rubin:2003,Spiegelhalter:2002}, a popular technique in Bayesian model selection. To explain the standard DIC criterion, let  $\thetabf^{(C)} = \left\{M_c, D_c, V_c\right\}_{c=1}^{C} $ denote all the parameter vectors and the {\it deviance function} is defined as $Dev(\thetabf^{(C)}):= -2 log L(\thetabf^{(C)})$ where $L(\cdot)$ is the likelihood function defined in equation~\ref{eq:likelihood_fn}. 
Let $\{\thetabf^{(C,i)}\}_{i=1}^{S}$ be $S$ values of the parameters, sampled  from the appropriate posterior distribution in Equation~\ref{eq:post_mix}. The $DIC$ score with a given choice for $C$, is computed as $DIC^{(C)}:=\overline{Dev}^{(C)}+\sum_{i=1}^{S} \left (Dev(\thetabf^{(C,i)})-\overline{Dev}^{(C)}\right)^2/(2(S-1))$ where $\overline{Dev}^{(C)}=\sum_{i=1}^{S}Dev(\thetabf^{(C,i)})/S$(~\cite{Book:Gelman:Rubin:2003}, page 185). To infer the number of clusters, samples are generated from different Markov chain assuming different values of $C$. The optimum number of cluster is given by $C_{opt} = \mbox{argmax}_C \; {DIC}^{(C)}$. For detailed discussion on DIC see~\cite{DeIorio:Robert:2002, Book:Gelman:Rubin:2003, Robert:2006}. Specifically, in the context of the mixture model, \cite{DeIorio:Robert:2002} described possible limitations for the standard DIC criterion. Following the alternative criteria proposed in~\cite{Robert:2006}, we considered several score functions (i.e. ${DIC}_2$, ${DIC}_3$, ${DIC}_4$, ${DIC}_5$, ${DIC}_6$, ${DIC}_7$, ${DIC}_8$  as defined in~\cite{Robert:2006}). We conducted an extensive numerical study with several simulated data sets. We found that the score function ${DIC}_5$ outperforms other alternative criteria in terms of efficiency for the model. Also, the computation of ${DIC}_5$ takes significantly less time than that of standard ${DIC}$. Therefore one may use ${DIC}_5$ instead of standard ${DIC}$ whenever computation of standard ${DIC}$ takes significantly longer time particularly for any large dataset.
Additional details along with a table comparing the performance of different ${DIC}$ scores in our simulation study is given in Section~\ref{sec:sim}) where we observe that ${DIC}_5$ score can identify the correct number of clusters in most of the cases in our model.     

\subsection{Iterative method to find posterior mode}
\label{subsec:EM_part}
In this section, we develop an iterative optimization technique to obtain point estimator for the parameters specified in the model given by Equation~\ref{eq:joint_prior_model}. Specifically, we employ expectation maximization (EM) algorithm~\citep{Dempster:1977} to obtain mode of the posterior distribution for the parameters in the model specified in Equation~\ref{eq:joint_prior_model}. Note that the algorithm is computationally fast and can be useful to get some rough estimates of the parameter specially for large data-sets. Also, we may use this algorithm in specific way to select appropriate values of the hyperparameters (See Section~\ref{subsec:hyperparameter_singleML}) in the case of MCMC based posterior inference. Note that the rough estimates can also help find suitable initial values for the MCMC procedures, particularly for analyzing massive data. To describe the procedure, let us consider    
complete data log-likelihood (From Equation~\ref{eq:post_mix}) as follows 
\begin{eqnarray}
 & & \sum_{i=1}^N \sum_{c=1}^C Z_{ic} \log \MLDensity(X_i \,\mid \,\theta_c) + Z_{ic} \log \pi_{c} +\nonumber \\
 & & +  \sum_{c=1}^C \alpha_c \log \pi_{c}+  tr\left( \nu_c\,V_cD_cM_c^T\priorXzero\right)- \nu_c\log\left(\hypdc\right) ,
\label{eq:loglik_MDV}
\end{eqnarray}
where $Z_{ic}= \mathbb{I}(Z_i = c)$ and   
\begin{equation*}
\log \MLDensity(X_i \,\mid \,\theta_c)=  tr({(V_c D_cM_c^T)} X_i)) -  \log\left(\hypdc\right).
\end{equation*}
Let we start the iterative algorithm at an initial point $\left ( \btheta^{(0)}, \bpi^{(0)}\right)$. We construct a sequence of parameter values $\left\{ \left ( \btheta^{(t)}, \bpi^{(t)}\right)\right\}_{t\geq 1}$ where we move from $\left ( \btheta^{(t)}, \bpi^{(t)}\right)$ to $\left ( \btheta^{(t+1)}, \bpi^{(t+1)}\right)$ using the ``E-step" and ``M-step" described below.

%Let the value of the parameters $\btheta$ and $\bpi$ at the $t$-th iteration step is $\btheta^{(t)}$ and $\bpi^{(t)}$ respectively. 
%Then we move to the value at iteration $(t+1)$-th step  we follow the subsequent procedure. 

\paragraph{E-step:}
%Let us denote the expected value of $Z_{ic}$ when the expectation is taken with respect to the posterior distribution of   $\bZ $ given $\bX, \btheta^{(t)}$
%\begin{equation*}
%{\langle Z_{ic} \rangle}:=E(Z_{ic} \mid \bX, \btheta^{(t) }) = \frac{ \MLDensity(X_i\mid \theta_c^{(t)})\pi_c^{(t)}}{\sum_{k=1}^C \MLDensity(X_i\mid \theta_c^{(t)})\pi_k^{(t)}}
%\end{equation*}

%To compute the parameter values at the $(t+1)$-th step,

We construct the objective function 
\begin{eqnarray} 
& & {Q(\btheta, \, \bpi \mid \bX, \btheta^{(t)}, \bpi^{(t)})}\nonumber\\
 & &  :=   \sum_{i=1}^N \sum_{c=1}^C \langle Z_{ic} \rangle  \log \MLDensity(X_i\mid \theta_c) +  \langle Z_{ic} \rangle \log \pi_{c}  +\nonumber \\
 & & +  \sum_{c=1}^C \alpha_c \log \pi_{c}+  tr\left( \nu_c\,V_c D_c M_c^T\priorXzero\right)- \nu_c\log\left(\hypdc\right) ,
\end{eqnarray}
where 
\begin{equation*}
{\langle Z_{ic} \rangle}:=\mathbb{E}(Z_{ic} \mid \bX, \btheta^{(t) }) = \frac{ \pi_c^{(t)}\,\MLDensity(X_i\mid \theta_c^{(t)})}{\sum_{k=1}^C \pi_k^{(t)}\,\MLDensity(X_i\mid \theta_c^{(t)})}
\end{equation*}
\paragraph{M-step:}
In this step, we maximize $Q(\btheta, \, \bpi \mid \bX, \btheta^{(t)}, \bpi^{(t)})$ with respect to the  $\btheta^{}, \bpi^{}$.  It is easy to see that, $Q(\btheta, \, \bpi \mid \bX, \btheta^{(t)}, \bpi^{(t)})$ is maximized when we set  $\bpi=\hat{\bpi}$ where the $c$-th component of the vector ${\hat{\bpi}_c}$ , 
\begin{equation*}
\hat{\pi}_c = \frac{\alpha_c+\sum_{i=1}^N \langle z_{ic} \rangle}{N+\sum_{c=1}^{C}\alpha_c} \quad \mbox{ for } c=1, \ldots C.
\end{equation*}

Note that $\btheta:=\{\theta_c\}_{c=1}^{C}$ where $\theta_c=\{M_c,\bd_c,  V_c\}$. Hence, the function $Q(\btheta, \, \bpi \mid \bX, \btheta^{(t)}, \bpi^{(t)})$ can be maximized by maximizing  the function
\begin{eqnarray}
tr\left( V_c D_c {M_c}^T \left[ \EmVar{X}+\priorXzero \right] \right)- \nu_c\log\left(\hypdc\right), 
\label{eq:EmOptimization1}
\end{eqnarray}
with respect to the variables $M_c \in \SpaceM, V_c \in \SpaceV$ and $\bd_c \in \SpaceD$ for each $c=1,\ldots, C$ separately where  $\EmVar{X} =\sum_{i=1}^N \fracProbZ{X_i}$.

Let $\EmVar{M}, \EmVar{D}$ and $\EmVar{V}$ be the unique singular value decomposition \citep{Chikuse:2012} for the matrix $\left[ \EmVar{X}+\priorXzero \right] $. Let $\EmVar{\bd}$ be the diagonal elements of the matrix $ \EmVar{D}$ and 
$\widehat{\bd}_c$ be the solution of the set of equations  $h(\widehat{\bd}_c)=\EmVar{\bd}$ where $h(\variableX):={\left(\frac{\partial}{\partial\,\bd}\,{}_0F_1 \left(\frac{n}{2}, \frac{D^2}{4} \right)\right)}/{_0F_1 \left(\frac{n}{2}, \frac{D^2}{4} \right)}$. Standard Newton-Raphson (NR)~\citep{Wright1999} method can be used to solve for $\widehat{\bd}_c$ the from the equation $h(\widehat{\bd}_c)=\EmVar{\bd}$. In the case of $p=2$, we derive the explicit expression of the Hessian matrix  and show the steps by NR to solve for $\widehat{\bd}_c$ in Section~\ref{subsubsec:hessian_comp}.
 
From~\cite{Chikuse:2012} we get that the objective function in Equation~\ref{eq:EmOptimization1} is maximized at  $\widehat{M}_c=\EmVar{M}$, $\widehat{V}_c=\EmVar{V}$ and $\widehat{D}_c$ where  $\widehat{D}_c$ is the diagonal matrix with diagonal elements $\widehat{\bd}_c$.

Finally we move to the values $ \left ( \btheta^{(t+1)}, \bpi^{(t+1)}\right) $ by the setting,
$$  \btheta^{(t+1)}:= \left\{ \left(  \widehat{M}_c, \widehat{\bd}_c, \widehat{V}_c  \right)\right\}_{c=1}^{C} \mbox{ and }  \bpi^{(t+1)}:=\widehat{\bpi}.$$
  
We stop the iteration when we achieve convergence, i.e. the values of the parameters in the two consecutive iterations are very close.

%%%%%%%%%%%%%%%%%%%%%%%%%
%\subsubsection{Derivatives of ${}_0F_1(\cdot,\cdot)$ for $2 \times 2 $ diagonal matrix and Hessian computation}
\subsubsection{Hessian computation and NR method}
\label{subsubsec:hessian_comp}
For this subsection we omit the subscript $c$ for ease of notation. Now observe that, 
\begin{eqnarray}
{}_0F_1\left(p+2k,\frac{d_1^2+d_2^2}{4}\right)=\frac{\Gamma\left(p+2k\right)}{4^{-\frac{p+2k-1}{2}}}\left(\sqrt{d_1^2+d_2^2}\right)^{-(p+2k-1)}I_{p+2k-1}\left(\sqrt{d_1^2+d_2^2}\right)\nonumber.
\end{eqnarray}
where $I_\nu(\cdot)$ is the modified Bessel function of first kind with order $\nu$.
Taking partial derivative with respect to $d_1$ we have,
\begin{eqnarray}
\pdv{d_1} \, {}_0F_1\left(p+2k, \frac{d_1^2+d_2^2}{4}\right)
= \frac{ d_1}{2(p+2k)}\; {}_0F_1\left(p+2k+1, \frac{d_1^2+d_2^2}{2}\right)\nonumber.
\end{eqnarray}

Consider the expression for the hypergeometric function of the Matrix argument with $2\times 2$ matrix~\citep{Muirhead:1975}
\begin{eqnarray}
{}_0F_1\left(p, \frac{D^2}{4}\right)=\sum_{k=0}^{\infty}\frac{d_1^{2k}d_2^{2k}}{4^{2k}\left(p-\frac{1}{2}\right)_k\left(p\right)_{2k}k!}\; _0F_1\left(p+2k,\frac{d_1^2+d_2^2}{4}\right).
\end{eqnarray}
This representation is also useful as we can get a good idea on error bound by approximating the number of terms for this infinite series. 

Let us use the following notations
\begin{eqnarray*}
T_0 = \frac{\left(\frac{d_1^2}{4}\right)^k \left(\frac{d_2^2}{4}\right)^k}{\left(p-\frac{1}{2}\right)_k\left(p\right)_{2k}k!} \;\;&&\;\;T_1 = {}_0F_1\left(p+2k,\frac{d_1^2+d_2^2}{4}\right)\\
T_2 = {}_0F_1\left(p+2k+1, \frac{d_1^2+d_2^2}{4}\right) \;\;&&\;\;
T_3 = {}_0F_1\left(p+2k+2, \frac{d_1^2+d_2^2}{4}\right).
\end{eqnarray*}

We derive  
\begin{eqnarray}
\pdv{d_1} \,\left({}_0F_1\left(p, \frac{D^2}{4}\right) \right)
&=&\sum_{k=0}^{\infty}
T_0\; \Bigg\{\frac{2k}{d_1}\;T_1 + \frac{ d_1}{2(p+2k)}\; T_2\Bigg\},\nonumber\\
%\label{eq:d_d1_0F1}
\pdv{d_2} \,\left({}_0F_1\left(p, \frac{D^2}{4}\right)\right) 
&=&\sum_{k=0}^{\infty}
T_0\; \Bigg\{\frac{2k}{d_2}\;T_1 + \frac{ d_2}{2(p+2k)}\; T_2\Bigg\},\nonumber\\
%\label{eq:d_d2_0F1}
\pdv{d_1}\pdv{d_2} \,\left({}_0F_1\left(p, \frac{D^2}{4}\right)\right) 
&=& \sum_{k=0}^{\infty} T_0\; \Bigg\{ \frac{2k}{d_1}\,\frac{2k}{d_2}\;T_1 + \frac{k\,(\frac{d_2}{d_1}+\frac{d_1}{d_2})}{(p+2k)} T_2 \nonumber \\
&&\;\;\qquad + \; \frac{ d_1}{2(p+2k)}\; \frac{d_2}{2(p+2k+1)} T_3\Bigg\},\nonumber \\
%\label{eq:d_d1_d2_0F1}
\pdvtwo{d_1} \,\left({}_0F_1\left(p, \frac{D^2}{4}\right)\right) 
&=& \sum_{k=0}^{\infty} T_0\; \Bigg\{  \frac{2k}{d_1}\,\frac{2k-1}{d_1}\;T_1 + \frac{4k+1}{2(p+2k)} T_2 \nonumber \\
&&\;\qquad + \;\frac{ d_1}{2(p+2k)}\; \frac{d_1}{2(p+2k+1)} T_3\Bigg\},\nonumber\\
%\label{eq:d_d1_d1_0F1}
\pdvtwo{d_2} \,\left({}_0F_1\left(p, \frac{D^2}{4}\right)\right) 
&=& \sum_{k=0}^{\infty} T_0\; \Bigg\{  \frac{2k}{d_2}\,\frac{2k-1}{d_2}\;T_1 + \frac{4k+1}{2(p+2k)} T_2 \nonumber \\
&&\;\qquad +\; \frac{ d_2}{2(p+2k)}\; \frac{d_2}{2(p+2k+1)} T_3\Bigg\}.
%\label{eq:d_d2_d2_0F1}
\label{eq:all_derivatives_0F1}
\end{eqnarray}

Denoting $R(d_1,d_2) = \left(\hyp\right)$, the Hessian matrix is written with the help of set of Equations in~\ref{eq:all_derivatives_0F1} as 
%\begin{equation*}
$H = \begin{bmatrix}
   		H_{11} & H_{12} \\
    	H_{21} & H_{22}
   \end{bmatrix}$,
%\end{equation*}  
where 
%\begin{equation*}
%H = \begin{bmatrix}
%   \frac{\partial R(d_1,d_2)}{\partial\,d_1\,\partial d_1} & \frac{\partial R(d_1,d_2)}{\partial\,d_1\,\partial d_2}  \\
%    \frac{\partial R(d_1,d_2)}{\partial\,d_2\,\partial d_1} & \frac{\partial R(d_1,d_2)}{\partial\,d_2\,\partial d_2} 
%    \end{bmatrix}
%\end{equation*}  
\begin{eqnarray*}
H_{11} &=& \frac{\partial}{\partial d_1} \left( \frac{\frac{\partial R}{\partial d_1}}{R} \right) = \left( \frac{\frac{\partial^2 R}{\partial d_1^2}}{R} \right) - {\left( \frac{\frac{\partial R}{\partial d_1}}{R} \right)}^2, \\
H_{12} &=& \frac{\partial}{\partial d_1} \left( \frac{\frac{\partial R}{\partial d_2}}{R} \right) = \left( \frac{\frac{\partial R}{\partial d_1}\frac{\partial R}{\partial d_2}}{R} \right) - {\left( \frac{\frac{\partial R}{\partial d_1}}{R} \right)} {\left( \frac{\frac{\partial R}{\partial d_2}}{R} \right)},\\
H_{21} &=& H_{12},\\
H_{22} &=& \frac{\partial}{\partial d_2} \left( \frac{\frac{\partial R}{\partial d_2}}{R} \right) = \left( \frac{\frac{\partial^2 R}{\partial d_2^2}}{R} \right) - {\left( \frac{\frac{\partial R}{\partial d_2}}{R} \right)}^2.
\end{eqnarray*}
where $R$ is used in the places of $R(d_1,d_2)$ for brevity of symbol.

Now, the update equation for NR method is given below
\begin{equation*}
\begin{bmatrix}
  	\widehat{d}_{1}^{new}  \\
   	\widehat{d}_{2}^{new}
    \end{bmatrix} = \begin{bmatrix}
  	\widehat{d}_{1}^{old}  \\
   	\widehat{d}_{2}^{old}
    \end{bmatrix} - H^{-1}_{\widehat{d}_{1}^{old},\widehat{d}_{2}^{old}}\,\begin{bmatrix}
  	\frac{\partial R(\widetilde{d}_{1}^{old},\widetilde{d}_{2}^{old})}{\partial\,d_1} \\
   	\frac{\partial R(\widetilde{d}_{1}^{old},\widetilde{d}_{2}^{old})}{\partial\,d_2}
    \end{bmatrix}. 
\end{equation*}

%%%%%%%%%%%%%%%%%%%%%%%%%%%%%%%%%%%%
\section{Experiments with simulated data}
\label{sec:sim}
We carry out two sets of simulation to investigate the clustering framework with our proposed Bayesian mixture model. In order to evaluate the performance of our clustering method, we consider the following three criteria -- 
\begin{enumerate}[(a)]
\item identification of the correct number of clusters,
\item correct assignment for each data point to the appropriate cluster and thus evaluate a measure of goodness for clustering using some well established metrics, 
\item %once the correct number of clusters is identified, accurate 
accuracy in estimation of cluster specific parameters. 
\end{enumerate}
%Observe that these criteria have been successfully used to evaluate efficiency of several clustering algorithms in~\cite{REF}.

%through the following criteria (1) correctly identifying the true number of underlying cluster [DIC] (2) retrieving clustering proportion and the correct cluster assignment for each data point [cluster evaluation metrics](3) calculating the accuracy of cluster specific parameters [asymptotic for F-F\_hat].

In order to evaluate the criterion $(a)$, we start with two simulation scenarios where the true numbers of clusters are three and four, respectively. In each case, we have $50$ individual datasets where number of data points is $400$ and $500$, respectively. For rest of the section we refer the these two simulation scenarios as simulation $(i)$ and simulation $(ii)$.
In simulation $(i)$, three parameter matrices are set 

%In order to find a suitable starting point we employ a hierarchical clustering of the data and assign them to the corresponding cluster. Using maximum likelihood estimate (Chikuse's ref) for each cluster, we initialize the cluster specific parameters. For each parameters in a cluster we run MCMC iterations to generate samples from their respective posterior distributions. It is important to note that MCMC convergence is independent of this initial value selection procedure. This procedure is followed in order to decrease the running time only.
 
We select appropriate values of hyperparameters for prior distributions in~\ref{eq:indep_prior_model} empirically using the procedure developed in Section~\ref{subsec:hype_sel}. Note that, the value of $K^\dagger$ is set to $20$ to reflect the concentration similar to $5\%$ of the size of the respective cluster.

%For $D$ we set the hyperparameters of the Gamma distribution to be $\alpha=1.0$ and $\beta=0.15$. 
%For $\bd$ we set the hyperparameters of the $\IMDY$ class of distribution empirically 
%and set to be $\nu=1.0$ and $\BoEta = (0.15,0.15)$. %{\attn{do we need to talk about $\JMDY$ or skip ?}}
%For the Dirichlet distribution prior, used for mixing proportion, we use a symmetric Dirichlet distribution with each hyperparameter equal to $1$ to make it a non-informative one. 

In general, for MCMC procedure, choice of a good initial point expedite the convergence for practical purposes. Therefore, we use the procedure described in Section~\ref{subsec:EM_part} to set the initial value of the parameters $M$, $\bd$ and $V$.

Optimal number of cluster is chosen based on $DIC$ criteria described in Section~\ref{subsec:model_sel}. We performed numerous experiments with several score functions for $DIC$~\citep{Robert:2006} apart from standard definition of $DIC$. We run our model with number of clusters equal to $2, 3, 4$ and $5$ for simulation $(i)$ and $2, 3, 4, 5$ and $6$ for simulation $(ii)$. We present a summary of our result (see Table~\ref{tbl:DIC_DIC5_for_sim}) for $DIC$ and ${DIC}_5$ values, where we have shown that in almost all the cases ($94\%$ for original ${DIC}$, $95\%$ for original ${DIC}_5$) we are able to select the correct number of clusters. The computation time for ${DIC}_5$ is significantly less than that of original $DIC$.
%For a detailed DIC/DIC\_5 table for all the dataset please refer Table S\_T1. 
%{\attn{ ${DIC}_4, {DIC}_6, {DIC}_7, {DIC}_8$ is not making much sense from the values.}}

\begin{table}[htb!]
\resizebox{\columnwidth}{!}{
\begin{tabular}{cccc}
\hline\\
Method & True number of clusters & Total number of datasets & Number of datasets with correct \\
& & & number of estimated clusters \\
\hline\\
$DIC$ & 3 & 50 & 48 \\
$DIC$ & 4 & 50 & 46 \\
${DIC}_5$ & 3 & 50 & 47\\
${DIC}_5$ & 4 & 50 & 48 \\
\hline
\end{tabular}
}
\caption{Number of datasets where correct number of clusters is identified with $DIC$ and ${DIC}_5$.}
\label{tbl:DIC_DIC5_for_sim}
\end{table}
We notice that in simulation, whenever the model fails to identify the true number of clusters, it always overestimates the number of clusters. Realizing this, we appropriately design a penalized version of the standard $DIC$ criterion with which we significantly improve the estimation of correct model.

%{\footnote{In discussion section: we need to say the the we are always selecting more number of cluster as our true model when we have misclassification. In Remark section: At the end define our own version of penalized version of DIC (how to validate this?) and we also find that DIC\_5 performs similar to DIC and computation time for DIC\_5 is much lesser.}}

\paragraph{Common metrics for evaluating clustering methods}
It is important to measure the assignment of each data point to the appropriate cluster. Note that, even if the number of clusters is right, the performance of the clustering method could be low because of incorrect cluster assignments. In order to evaluate clustering efficiency one could calculate several external cluster evaluation metrics. Here in this study we compute purity, Normalized mutual information (NMI), rand index (RI), adjusted rand index (ARI), Jaccard Index (JI) and F-measure~\citep{rand1971objective, vinh2010information}. 

We build up some notations for introducing those metric briefly.
Let us assume we have N data points denoted ${\{X_j\}}_{j=1}^N$. Set of $C$ true classes is given by $\mathcal{A} = \{A_1, A_2, \cdots, A_C\}$ where $A_c = \{j : X_j  \mbox{ belongs to $c$-th cluster} \}$ for $c=1, 2, \cdots, C$ and clustering method returns $K$ number of clusters and the set of clusters is given by $\mathcal{B} = \{B_1,B_2,\cdots, B_K\}$ where $B_k = \{j : X_j  \mbox{ assigned to $k$-th cluster} \}$ for $k = 1, 2, \cdots, K$. Note that we use $\mid \cdot \mid$ to denote the number of elements in a set.

\begin{itemize}
\item Purity is defined (see~\cite{rand1971objective, vinh2010information}) as 
$$
\mbox{purity}(\mathcal{A},\mathcal{B}) = \frac{\sum_{k} \underset{c}{\max} \mid B_k \cap A_c\mid}{N}.
$$
It is the most simple evaluation measure. To compute purity each cluster is assigned to the class which is most prevalent in the cluster and the accuracy of the assignment is measured by counting the number of correctly assigned data to the cluster and dividing by total number of data in the dataset. Clearly, Purity lies between $0$ and $1$ where perfect clustering has a purity of $1$. 

\item NMI is an information-theoretic measure which is defined as
$$
\mbox{NMI}(\mathcal{A},\mathcal{B}) = \frac{I(\mathcal{A},\mathcal{B})}{[H(\mathcal{A}) + H(\mathcal{B})]/2},
$$
where, $I(\cdot,\cdot)$ and $H(\cdot)$ stand for mutual information and entropy, respectively with
$$
I(\mathcal{A},\mathcal{B}) = \sum_k \sum_c \frac{\mid B_k \cap A_c\mid}{N} \log \frac{N\,\mid B_k \cap A_c\mid}{\mid B_k \mid\,\mid A_c \mid},
$$
and 
$$
H(\mathcal{A}) = -\sum_k \frac{\mid B_k \mid}{N} \log \frac{\mid B_k \mid}{N} ;\;\;H(M) =-\sum_c \frac{\mid A_c \mid}{N} \log \frac{\mid A_c \mid}{N}.
$$
NMI reaches its maximum value $1$ only when the two sets $\mathcal{A}$ and $\mathcal{B}$ have a perfect one-to-one correspondence.

\item RI is written as
$$
RI = \frac{TP+TN}{TP+FP+TN+FN},
$$ 
where TP is the number of true positives, TN is the number of true negatives, FP is the number of false positives, and FN is the number of false negatives. This can be viewed as a measure of the percentage of correct decisions. Note that, here false positives and false negatives are equally weighted.
The Rand index also lies between $0$ and $1$. When clustering results agree with the class perfectly, the Rand index is $1$.

\item ARI is a chance-corrected version of RI. A problem with RI is that the expected value of the RI between two random clustering methods is not a constant. This problem is corrected in ARI which assumes the generalized hyper-geometric distribution as the model of randomness. The ARI has the maximum value $1$, and its expected value is 0 in the case of random clusters. A larger ARI means a higher agreement between two clustering methods.

\item JI is defined by the following formula --
$$JI = \frac{TP}{TP+FP+FN}.$$
It is also known as intersection over union used to quantify the similarity between two sets. It takes a value between $0$ and $1$. Index value $1$ or $0$ means two sets are identical or two sets have no common elements, respectively. 

\item F-measure is defined by 
$$ F_\beta = \frac{(\beta^2+1)\cdot Precision\cdot Recall}{\beta^2\,Precision + Recall},$$
where
%where $P$ is precision and $R$ is recall defines as
$$
Precision = \frac{TP}{TP+FP} \;\;\mbox{ and } \;\; Recall =\frac{TP}{TP+FN}.
$$
F-measure can be used to penalize false negatives more strongly than false positives by selecting  $\beta > 1$. On the other hand, when $\beta=0$, recall has no impact on F-measure.
\end{itemize}

We summarize all the evaluation metrics for the two simulation scenarios in the following Table~\ref{tbl:ext_metric_3} and~\ref{tbl:ext_metric_4}. We observe that most of the metrics are close to the maximum possible value $1$, which indicates an overall success of our clustering method.

\begin{table}[htb!]
\centering
\begin{tabular}{ccccccccccc}
	\hline\\
Metrics & PUR    & RI     & ARI   & JI     & NMI    & F05    & F1     & F2     & F5     \\
        \hline\\
%Min.    & 0.9642 & 0.9506 & 0.8888 & 0.8894 & 0.862  & 0.8412 & 0.9251 & 0.9259 & 0.9267 & 0.9271 \\
%1st Qu. & 0.9801 & 0.9738 & 0.941  & 0.9413 & 0.9243 & 0.9085 & 0.9604 & 0.9607 & 0.9609 & 0.9608 \\
%Median  & 0.9851 & 0.9801 & 0.9552 & 0.9554 & 0.9419 & 0.926  & 0.97   & 0.9701 & 0.9701 & 0.9701 \\
Mean    & 0.984 & 0.979 & 0.952 & 0.938 & 0.923  & 0.968 & 0.968 & 0.968 & 0.968 \\
%3rd Qu. & 0.99   & 0.9864 & 0.9694 & 0.9696 & 0.9601 & 0.9462 & 0.9794 & 0.9797 & 0.9798 & 0.9798 \\
%Max.    & 0.9975 & 0.9968 & 0.9928 & 0.9928 & 0.9904 & 0.9866 & 0.995  & 0.9952 & 0.9954 & 0.9955 \\
Std. dev.      & 0.008 & 0.010   & 0.023  & 0.028  & 0.031 & 0.015  & 0.015  & 0.015 & 0.015\\
\hline\\
\end{tabular}
\caption{Clustering evaluation metrics when true number of clusters equal to three}
\label{tbl:ext_metric_3}
\end{table}

\begin{table}[htb!]
\centering
\begin{tabular}{ccccccccccc}
\hline\\
 Metrics & PUR    & RI     & ARI     & JI     & NMI    & F05    & F1     & F2     & F5     \\
        \hline\\
%Min.    & 0.9526 & 0.9552 & 0.8804 & 0.8811 & 0.8353 & 0.847  & 0.9108 & 0.9103 & 0.9097 & 0.9094 \\
%1st Qu. & 0.9759 & 0.9758 & 0.9354 & 0.9358 & 0.9075 & 0.9102 & 0.9515 & 0.9515 & 0.9516 & 0.952  \\
%Median  & 0.9796 & 0.9795 & 0.9456 & 0.9459 & 0.9217 & 0.9224 & 0.9584 & 0.9592 & 0.9594 & 0.9594 \\
Mean    & 0.978 & 0.978 & 0.942  & 0.918 & 0.921 & 0.957 & 0.957 & 0.957 & 0.957 \\
%3rd Qu. & 0.9827 & 0.9827 & 0.9539 & 0.9542 & 0.9332 & 0.936  & 0.9652 & 0.9654 & 0.9659 & 0.9661 \\
%Max.    & 0.9901 & 0.9899 & 0.9731 & 0.9733 & 0.9605 & 0.9598 & 0.9803 & 0.9798 & 0.9797 & 0.9797 \\
Std. dev.      & 0.008 & 0.008 & 0.020  & 0.027 & 0.024 & 0.015 & 0.015 & 0.015 & 0.015\\
\hline\\
\end{tabular}
\caption{Clustering evaluation metrics when true number of clusters equal to four}
\label{tbl:ext_metric_4}
\end{table}

%Also note that, average misclassification error is  $1.58\%$  
%($7$  misclassifications out of $400$ on an average)  and  $2.16\%$  
%($11$ misclassifications  out of $500$ on an average) for the simulation scenarios $(i)$ and $(ii)$, respectively. 

\begin{figure}[htb!]
\centering
\includegraphics[scale=0.72]{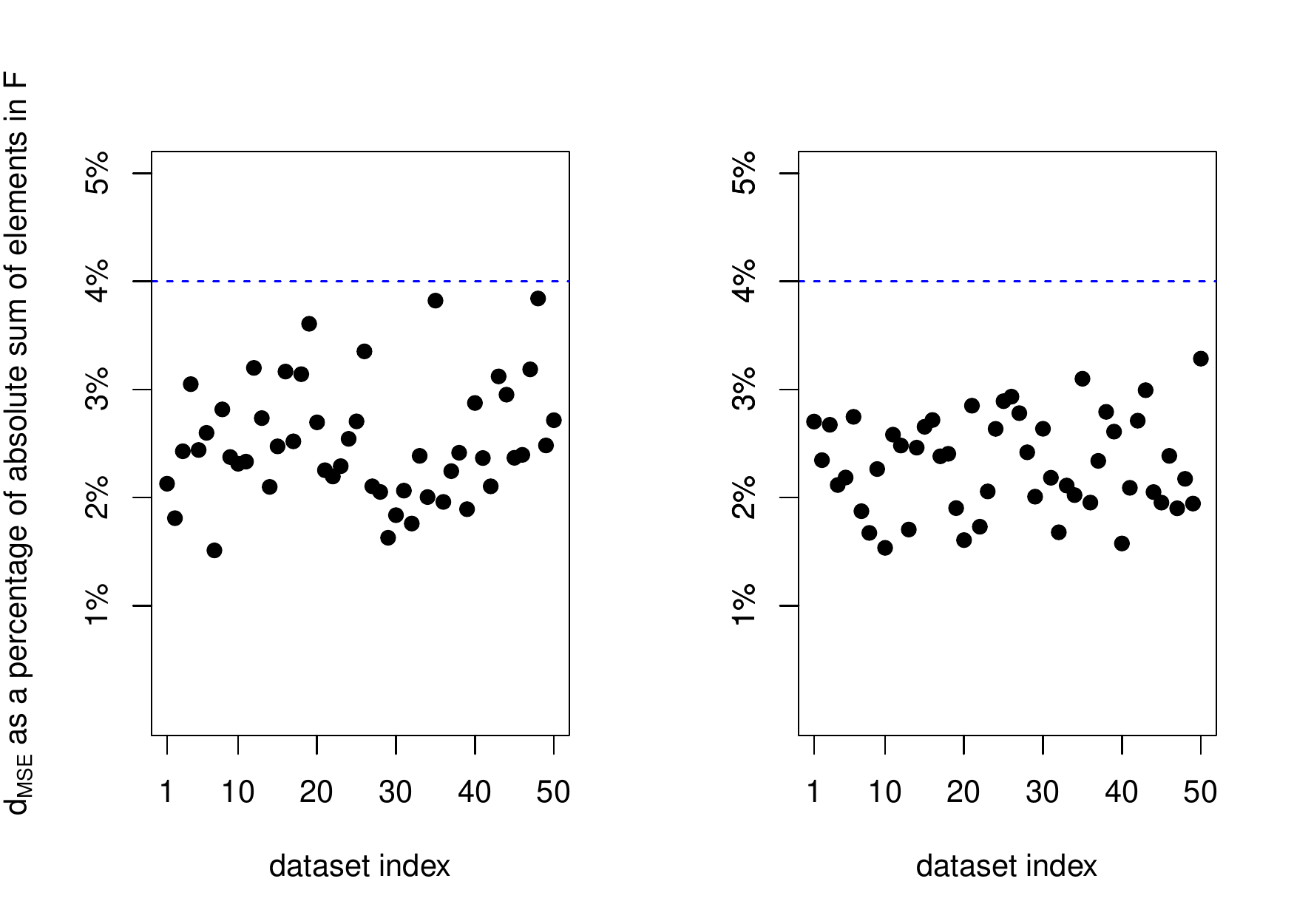}
\caption{Relative error (in percentage) for parameter $F$ in simulation (i) (left panel) and simulation (ii) (right panel).}
\label{fig:d_MSE_rel_change}
\end{figure}

\begin{figure}
\begin{center}
\begin{tabular}{ll}
\includegraphics[scale=0.3]{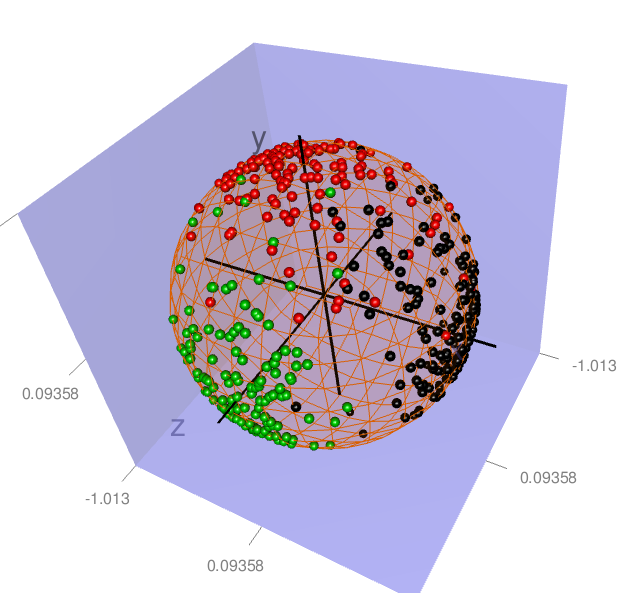} & 
\includegraphics[scale=0.3]{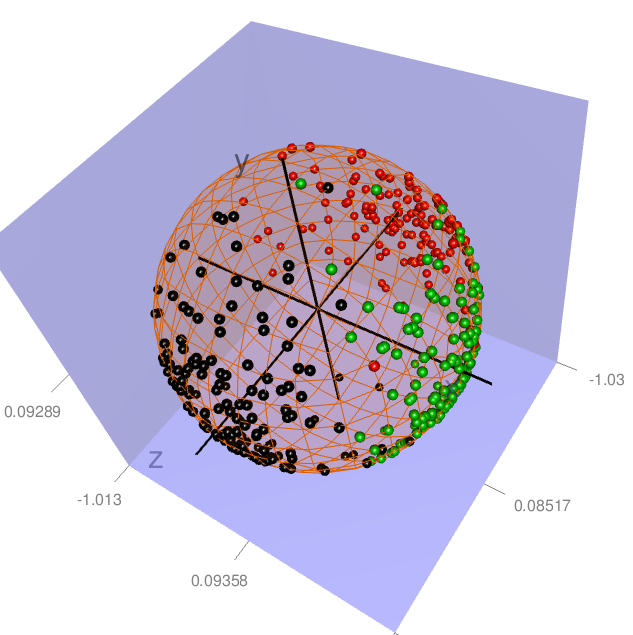} \\
\end{tabular}
\end{center}
\caption{First and second eigenvectors using one of the datasets in simulation $(i)$.}
\label{fig:sim_clust3_embed}
\end{figure}

\begin{figure}
\begin{center}
\begin{tabular}{ll}
\includegraphics[scale=0.34]{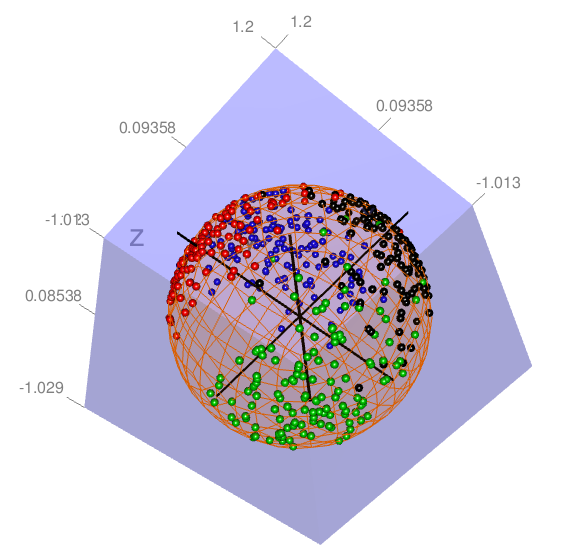} &
\includegraphics[scale=0.34]{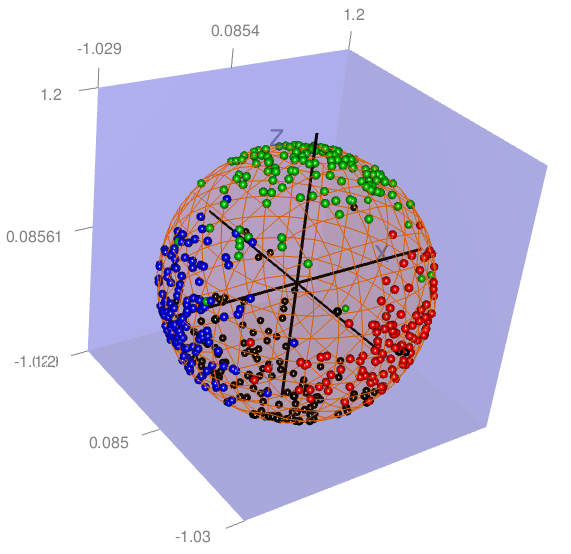} \\
\end{tabular}
\end{center}
\caption{First and second eigenvectors using one of the datasets in simulation $(ii)$.}
\label{fig:sim_clust4_embed}
\end{figure}

In order to evaluate parameter values for each clusters let us denote the true parameter set for $C$ classes by $\{F_1,F_2, \cdots, F_C\}$ where $F_c = M_{c} {D}_c V_{c}^T$. We find out $d_{MSE} = \sum_c {\lVert\hat{F}_c - F_c\rVert}_F$, where $\hat{F}_c$ is the estimate of the parameter matrix for the $c$-th cluster, where ${\lVert\cdot\rVert}_F$ denotes the matrix Frobenious norm. 
%In order to calculate the total mean squared error (MSE) between original and estimated parameters we choose to use the original parametrization. 
%In this way, we can treat the parameter matrix as a six dimensional vector and compute $L_2$ distance between estimated and original matrices. 
We plot below (in Figure~\ref{fig:d_MSE_rel_change}) the relative error (in percentage) in estimating the true parameter $F$. From the plot we  observe that procedure is efficient in estimating the parameter as the maximum relative error is below $4\%$. 
%For (3) accurately estimating the parameters of the clusters is one of the important criteria for evaluating any clustering techniques. We here perform an experiment and show that mean squared error (MSE) of our parameters (F\_hat) from the true parameters (F) decrease as we increase the size of the dataset.
We show the simulation results for one particular dataset from simulation (i) and (ii) for both the eigenvectors in Figures~\ref{fig:sim_clust3_embed} and~\ref{fig:sim_clust4_embed}, respectively.

\begin{figure}
\begin{center}
\begin{tabular}{cc}
\includegraphics[scale=0.32]{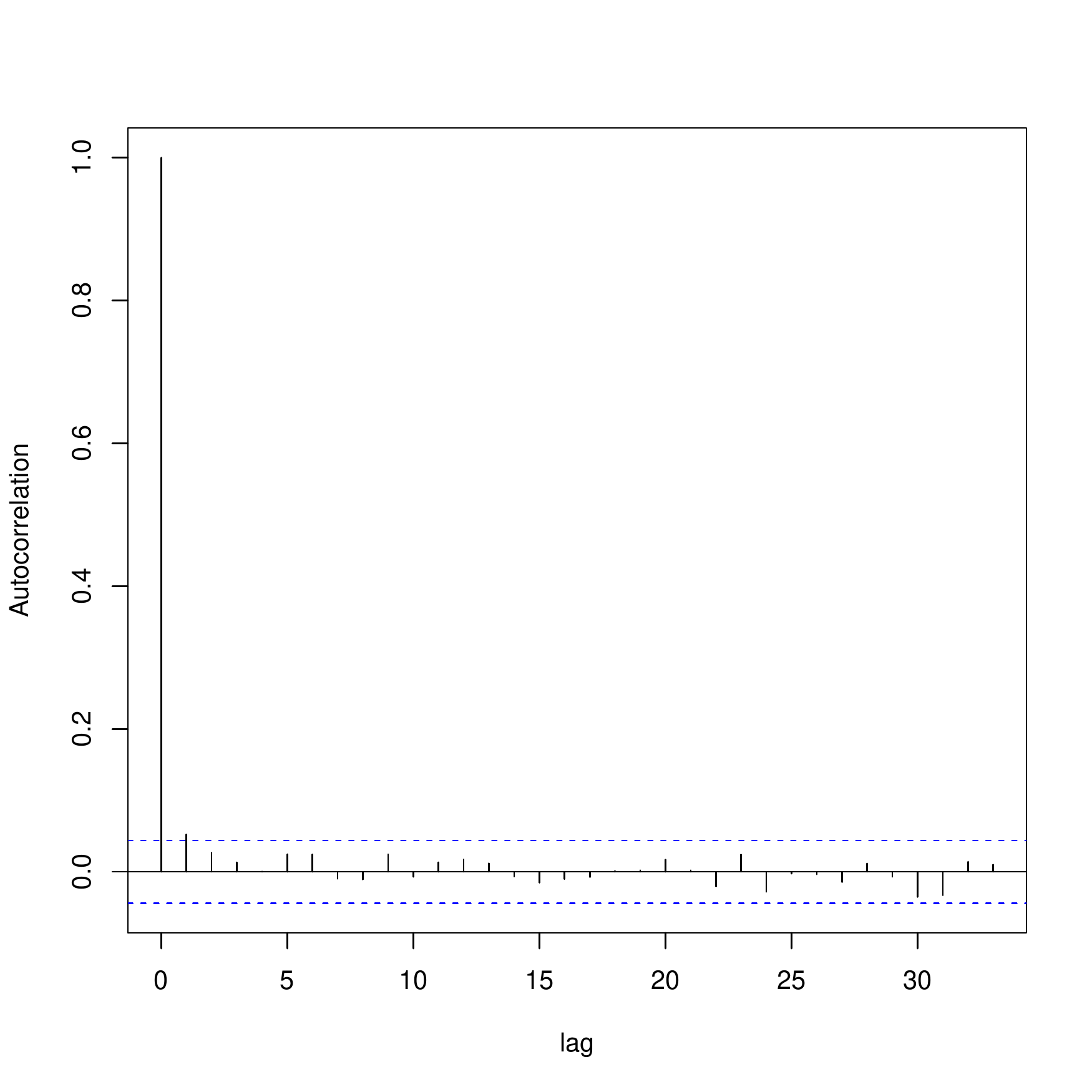} & 
\includegraphics[scale=0.32]{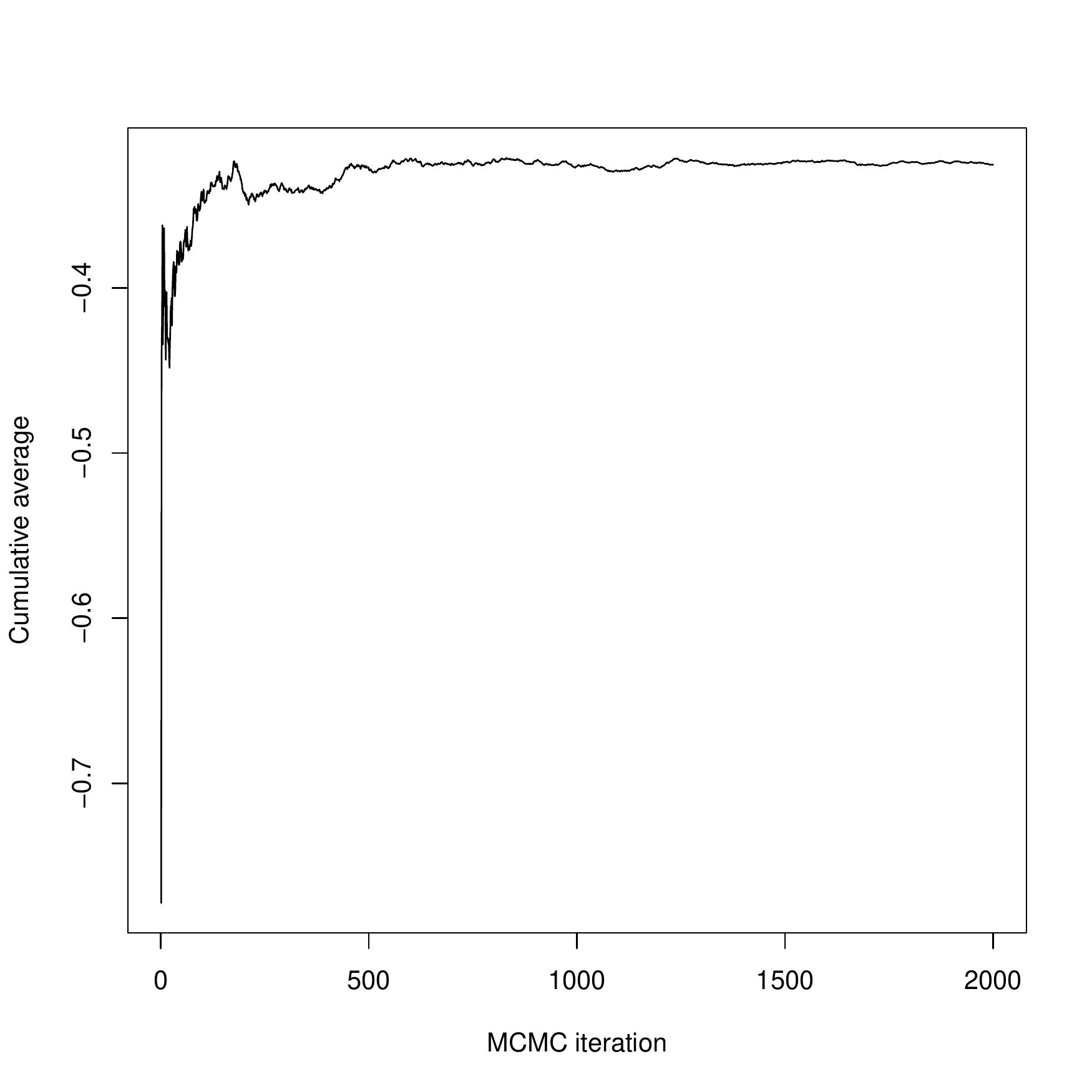} \\
(a) ACF of $F_{11}$ & (b) Cumulative average of $F_{11}$ \\
\end{tabular}
\end{center}
\caption{Diagnostic plots of MCMC for simulation $(i)$.}
\label{fig:MCMC_ACF_and_cumsum_avg}
\end{figure}

As our Bayesian inference technique involves MCMC sampling scheme, it is customary to check the standard MCMC convergence and efficiency diagnostics~\citep{Cowles:1996}. We investigate the convergence by carefully observing the MCMC cumulative average plot and auto-correlation function (ACF) plot for one of the elements of parameter $F$ for one of the clusters in the dataset. Here in Figure~\ref{fig:MCMC_ACF_and_cumsum_avg}, we show both the plots for the simulation scenario $(i)$. By looking at the cumulative average (Figure~\ref{fig:MCMC_ACF_and_cumsum_avg}(b)) we set the value for number of burn-in iteration to $800$. The small values in the ACF plot (Figure~\ref{fig:MCMC_ACF_and_cumsum_avg}(a)) indicates high efficiency for parameter estimation based on the MCMC samples. Note that, these plots have very similar characteristics for all the other scenarios. 
%We also summarize the result for estimated parameter values.

%TODO: finding trend of MSE with dataset size - experimental consistency with true \# of clusters
%%%%%%%%%%%%%%%%%%%%%%%%%%%%%%%%%%%%%%%%%%%

\section{Application}
\label{sec:real_data_app}
%Justification for MCMC based bayesian analysis: it would be great if we find multimodal density plot, then our analysis would be warranted.
In this section we show two real data based applications with our model. The first one is associated with medical image analysis while the second one is related to astronomical data. We present each application in different subsections below.
 
%%%%%%%%%%%%%%%%%
\subsection{Diffusion tensor imaging data}
\label{subsec:dti}
%Data source, generation, processing and results\\
%Plots embedded in real brain structure
The human brain consists of more than $100$ billion neurons, and it is arguably the most complex structure in our body~\citep{Basser:2002, Mori:2006}. Magnetic resonance imaging (MRI) is powerful noninvasive and three-dimensional imaging technique to characterize the entire brain anatomy. Diffusion tensor imaging (DTI) is a relatively new MRI technique which helps to reconstruct the underlying 3D structures of axonal bundles in the brain. Using a technique called tractography using the data collected by DTI the voxels that belong to the same white matter tract are grouped together. This is used to investigate brain connectivity, for example, cortex-white matter connectivity~\citep{Catani:2002, Lazar:2005} or corticothalamic connectivity~\citep{Guy:2004}.

DTI technique was introduced in the mid 1990s~\citep{Basser:1994}. The diffusion term represents translational motion of water molecules and this motions is used as a probe to estimate the axonal organization of the brain. The water molecules move relatively easily along the axonal bundles compared to the perpendicular to these bundles because there are fewer obstacles to prevent movement along the fibers which carry rich anatomical information about the white matter~\citep{Mori:2006}. Fiber orientations are estimated from three independent diffusion measurements along the $x, y$ and $z$ axes. However, these three measurements are not enough as the fiber orientation is not always along one of these three axes. But to accurately construct apparent diffusion coefficient where the intensity of each voxel is proportional to the extent of diffusion, we need to measure diffusion along many directions, which is difficult. In order to give a practical solution to this, the concept of DT was introduced~\citep{Basser:1994}.

\begin{figure}[htb!]
\centering
\includegraphics[scale=0.55]{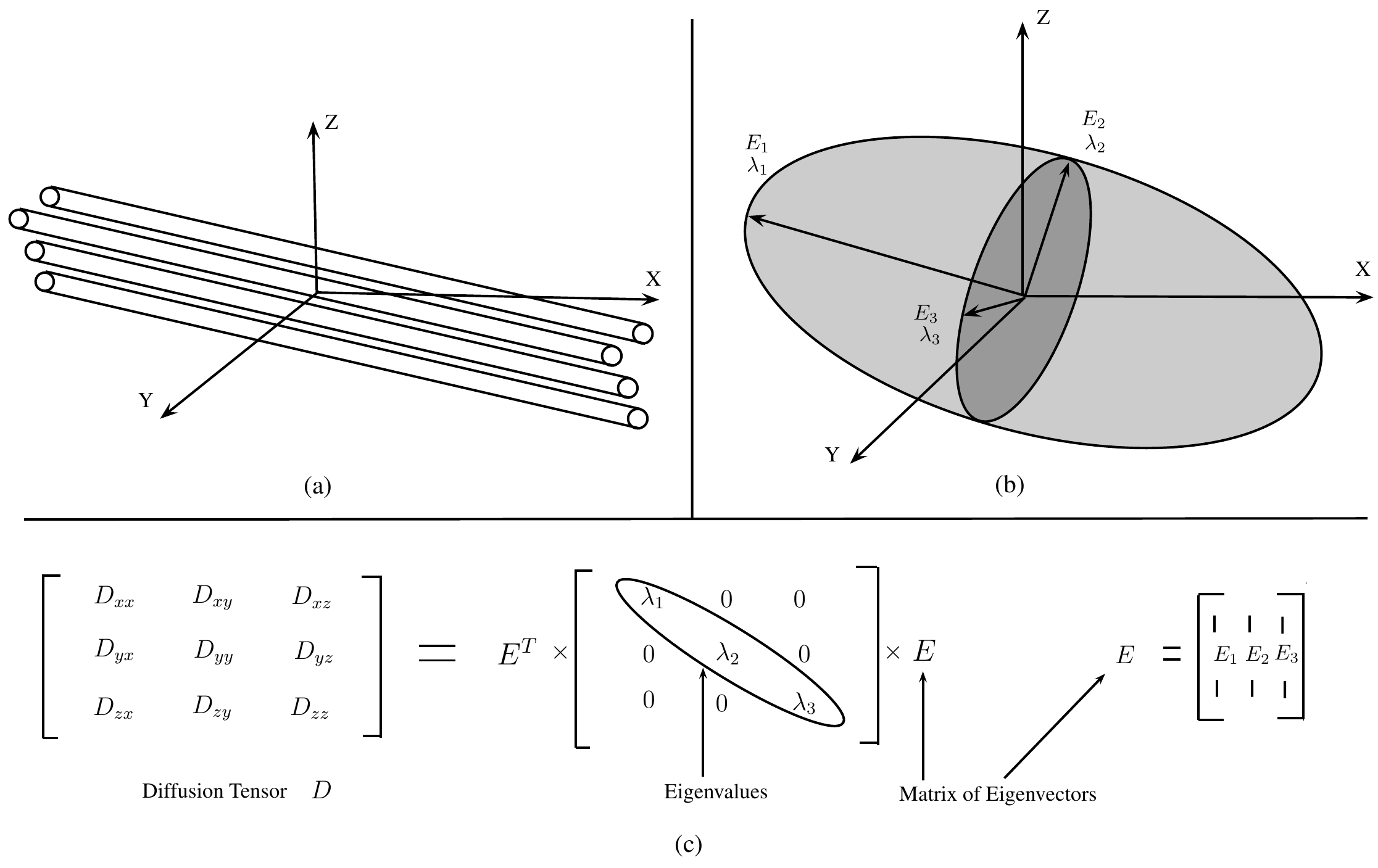}
\caption{A schematic representation of (a) fibers and (b) estimated ellipsoids with the corresponding (c) eigenvectors and eigenvalues.}
\label{fig:DTI_sch}
\end{figure}

In this model, measurements along different axes (see Figure~\ref{fig:DTI_sch}a) are fitted to a 3D ellipsoid shown in Figure~\ref{fig:DTI_sch}b, which represents average diffusion distance in each direction~\citep{Mori:2006}. Note that the properties of a 3D ellipsoid can be defined by six parameters -- three of its eigenvalues and corresponding eigenvectors (mutually perpendicular), which can compactly represented by a $3 \times 3$ symmetric, positive-definite matrix (SPD) and this is known as DT. In anisotropic fibrous tissues the major eigenvector also defines the fiber tract axis of the tissue.
The three positive eigenvalues of DT ($\lambda_1,\lambda_2$ and $\lambda_3$) give the diffusivity in the direction of each eigenvector, denoted by  $E_1, E_2$ and $E_3$ in Figure~\ref{fig:DTI_sch}c.
 
According to our knowledge, this is the very first work with DTI data which is modeled with a mixture of $\ML$ distributions. Also, we consider a final dataset after selecting the voxels in the white matter region of the brain containing information from almost $63,000$ voxels. Our implementation is very efficient in handling this large amount of data.
We model diffusion tensors by elements in $\StiefelS$. Note that, for the scope of this project we are only interested in the direction of the eigenvectors of DT. Also, we only need to model $E_1$ and $E_2$ as direction of $E_3$ will be totally governed by the rest of the two eigenvectors. Therefore, we have two orthonormal eigenvectors in three dimensions i.e. a $3 \times 2$ matrix which has two orthonormal vectors as columns - this is precisely the space of $3 \times 2$ orthonormal matrices i.e. $\StiefelS$. 

%%%
%%% MOVE THE BELOW PARA TO RESULTS SECTION. IT IS NOT APPROPRIATE HERE.
%%%
In practice, Wishart distribution is commonly used to analyze DT, a $3\times 3$ positive definite matrix. 
%This choice does not particularly model the directional components though. 
It could be argued that one can use a mixture of Wishart distributions directly on the space of SPD matrices. However, note that, in the case of Wishart distribution the sense of directionality is difficult to comprehend. The directional aspect of eigenvectors from DTI data can be therefore better suited to model by using a mixture of $\ML$ distributions. It is easier to find  interpretations of the parameters for $\ML$ distribution in terms of direction of the data. Therefore our Bayesian mixture model is relatively more flexible in terms of handling DTI data which have directional components. Also our inference mechanism can handle a very large number of DTI data from each voxels. 
To the extent of our knowledge, this is the first paper that develops the framework to analyze DTI data when they are modeled as objects on $\StiefelS$.

Before presenting the results, we would like to point out that our results could be improved by incorporating eigenvalues along with the eigenvectors. However, that requires more complicated statistical model which we currently reserve for our future work and it is outside of the scope of current paper as we mainly focusing on building the appropriate framework for analyzing DTI data. Nevertheless, we show in in Section~\ref{subsubsec:results_dti} that we have found evidences of meaningful clusters by only investigating the directional part of the data.

% data source
\subsubsection{Data source and pre-processing}
The Philadelphia Neurodevelopmental Cohort (PNC) is a large-scale initiative to understand how genetics impact trajectories of brain development and cognitive functioning in adolescence, and understand how abnormal trajectories of development are associated with psychiatric symptomatology~\citep{Satterthwaite:2014}. As part of the PNC, 1,445 children ages 8-21 received multi-modal neuroimaging in order to evaluate with a detailed cognitive and psychiatric assessment.
Data is pre-processed with the comprehensive DTI data processing software library FSL~\citep{Woolrich:2009}.

Some of the important features of this dataset is that all imaging data was acquired at a single site, on a single scanner, in a short period of time that did not span any software or hardware upgrades. Quality of the images of the DTI data was primarily assessed by visual inspection and rarely, two artifacts were noted in the DTI data~\citep{Satterthwaite:2014}. 

\begin{figure}
\begin{tabular}{ccc}
%\begin{center}
\includegraphics[scale=0.3]{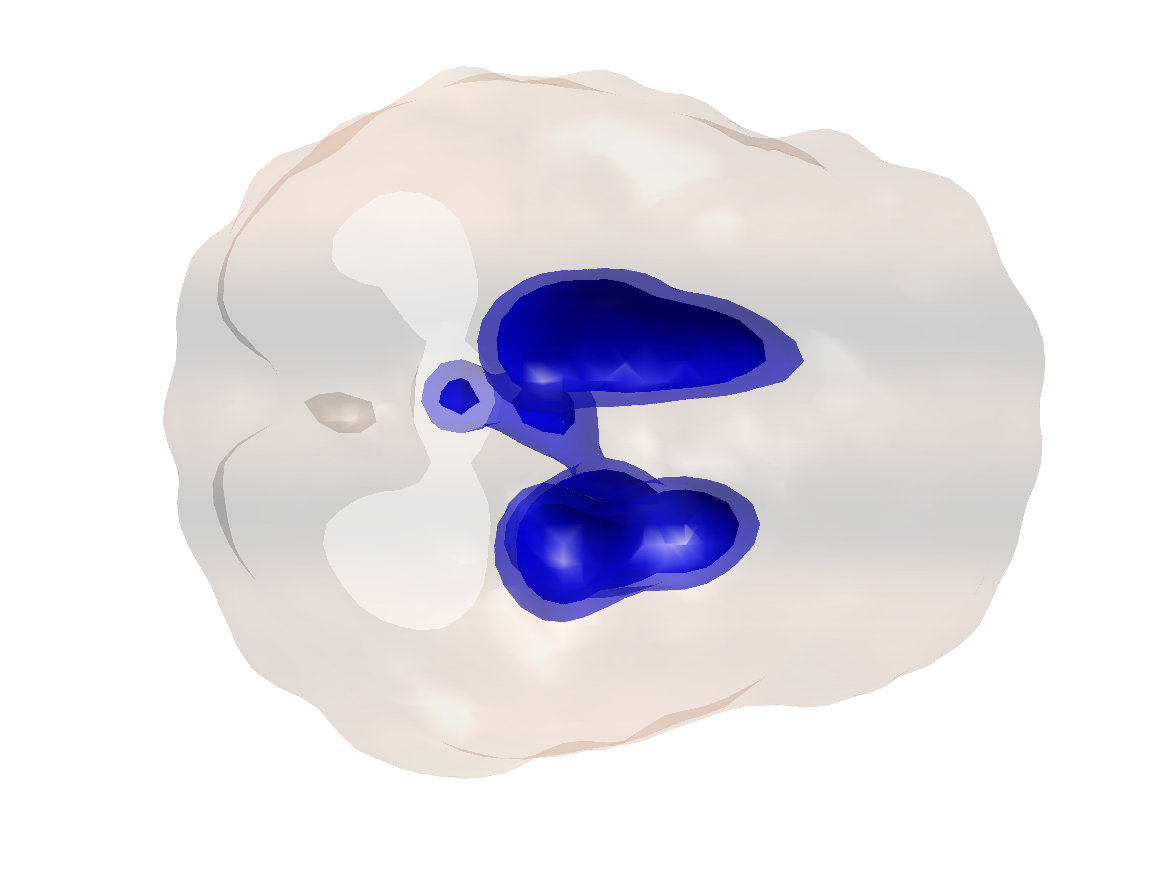} & \includegraphics[scale=0.3]{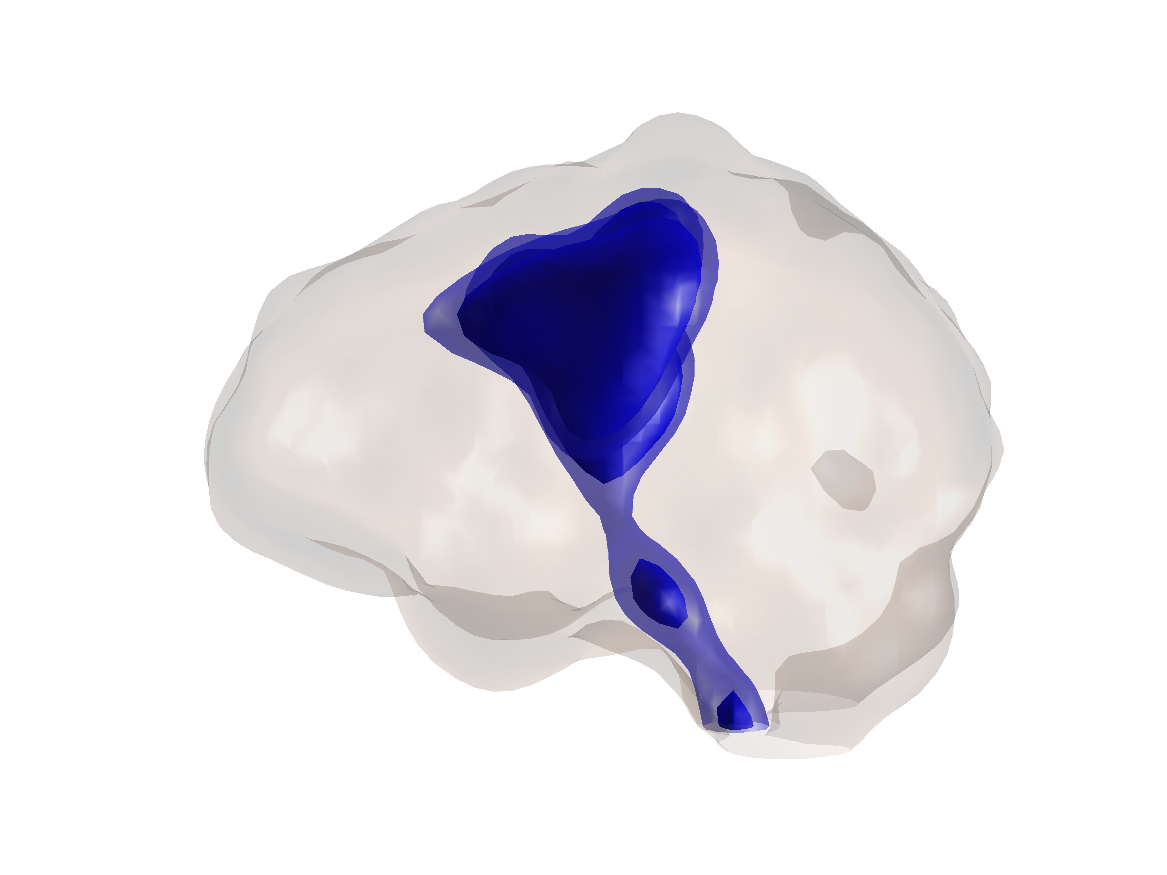} &
\includegraphics[scale=0.3]{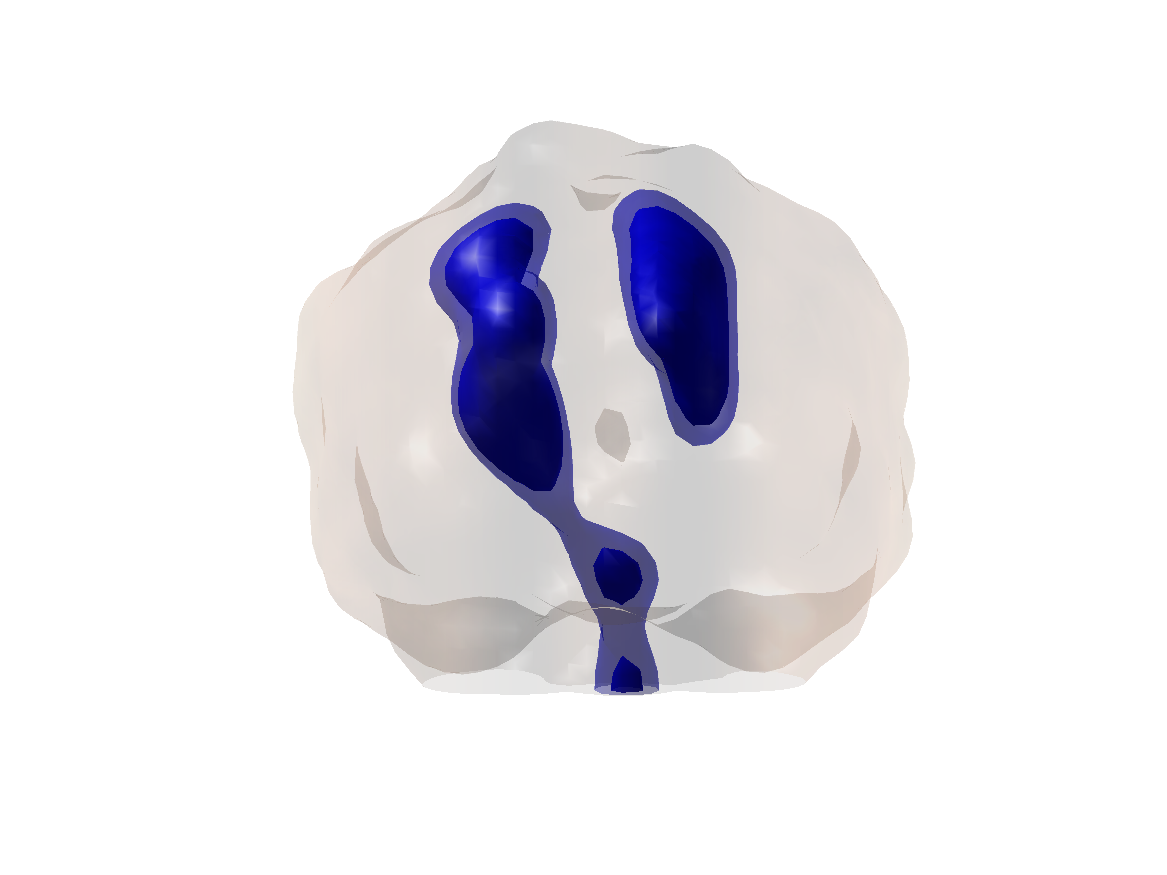} \\
\includegraphics[scale=0.3]{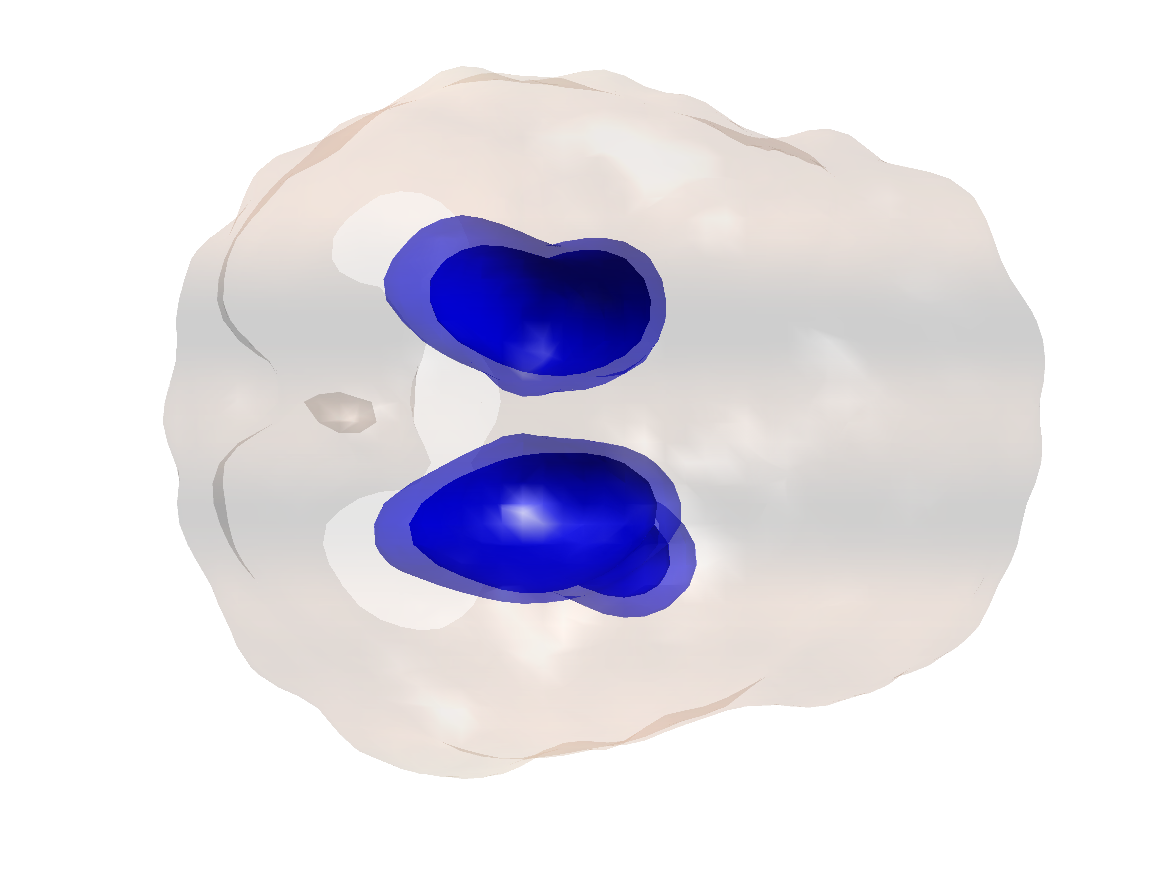} & \includegraphics[scale=0.3]{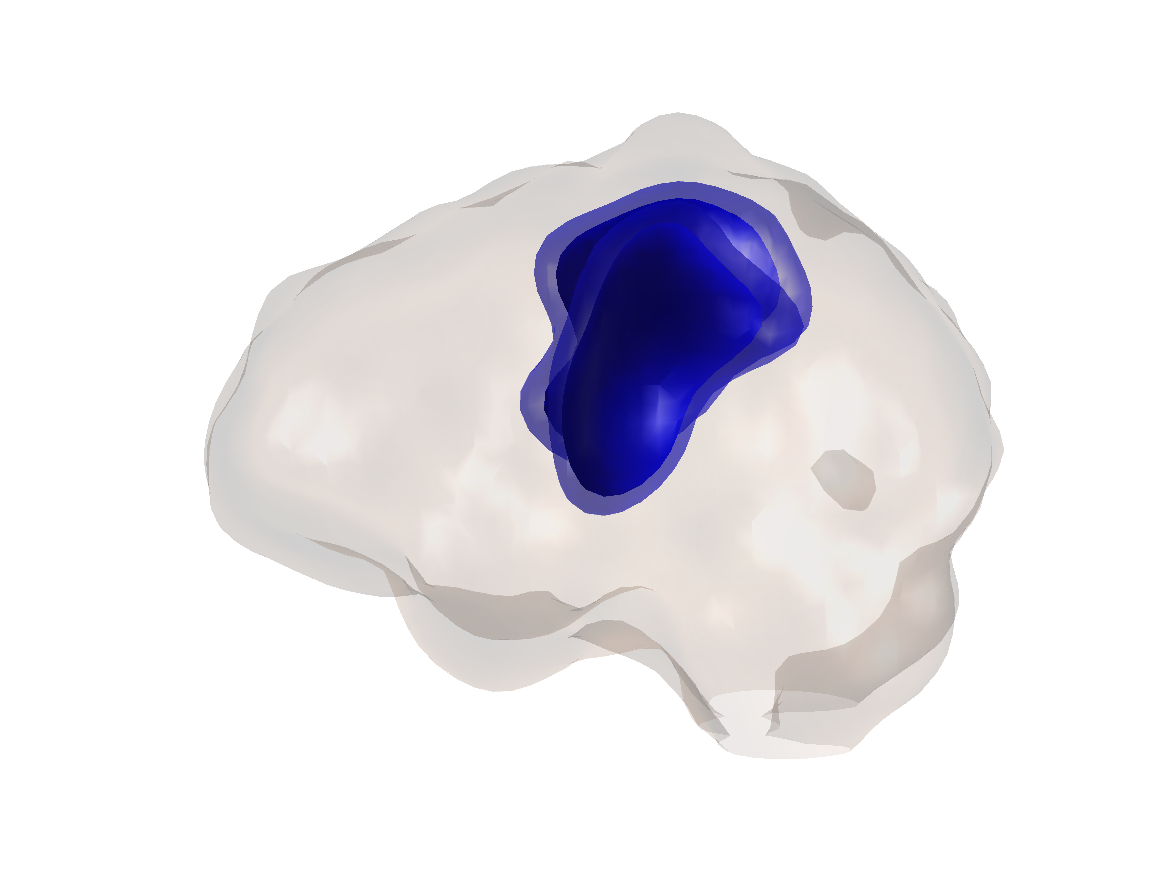} &
\includegraphics[scale=0.3]{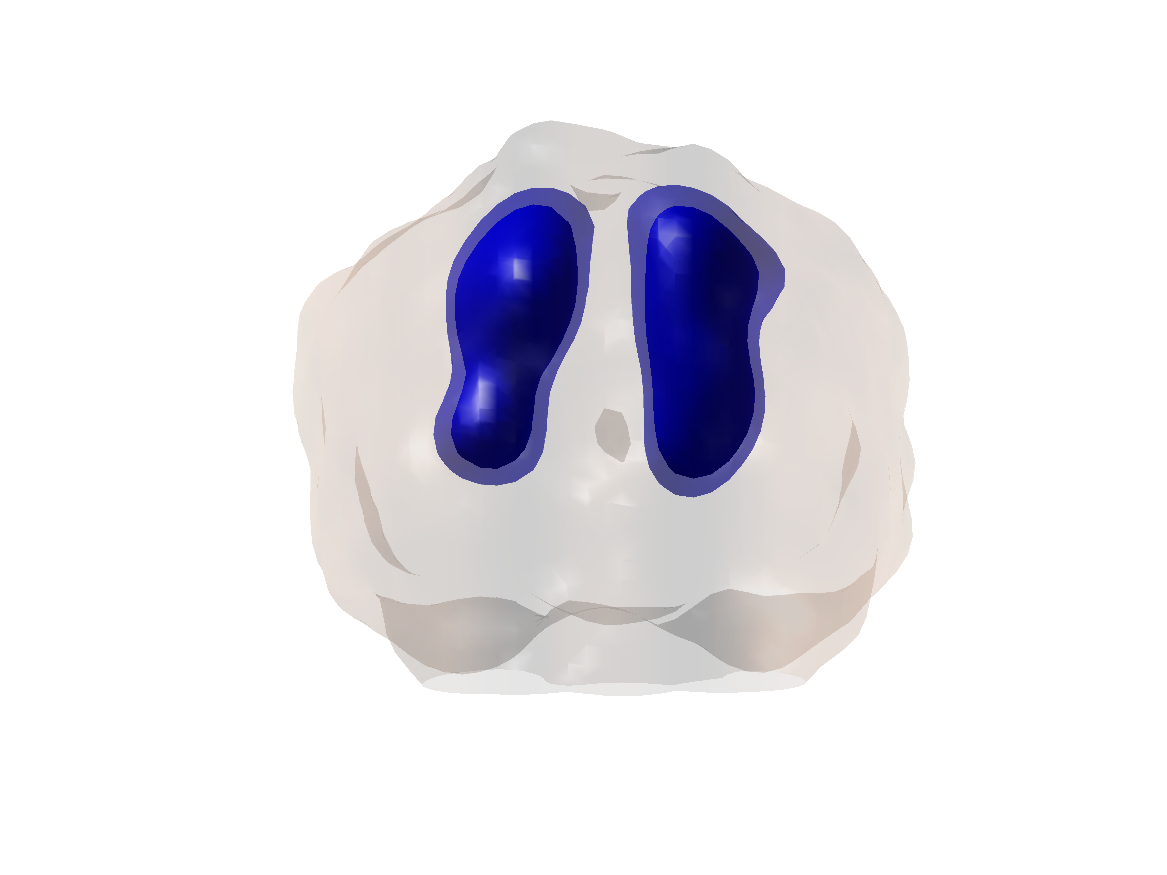} \\
\includegraphics[scale=0.3]{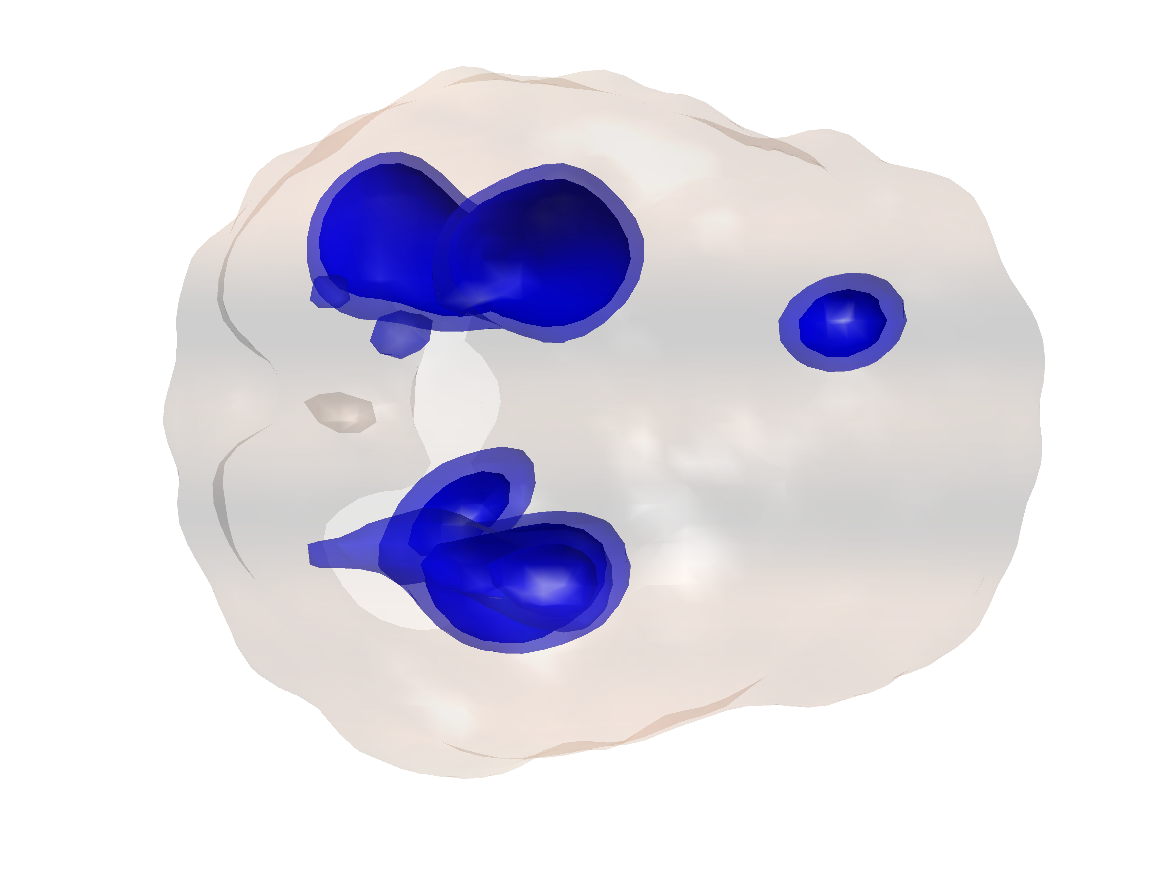} & \includegraphics[scale=0.3]{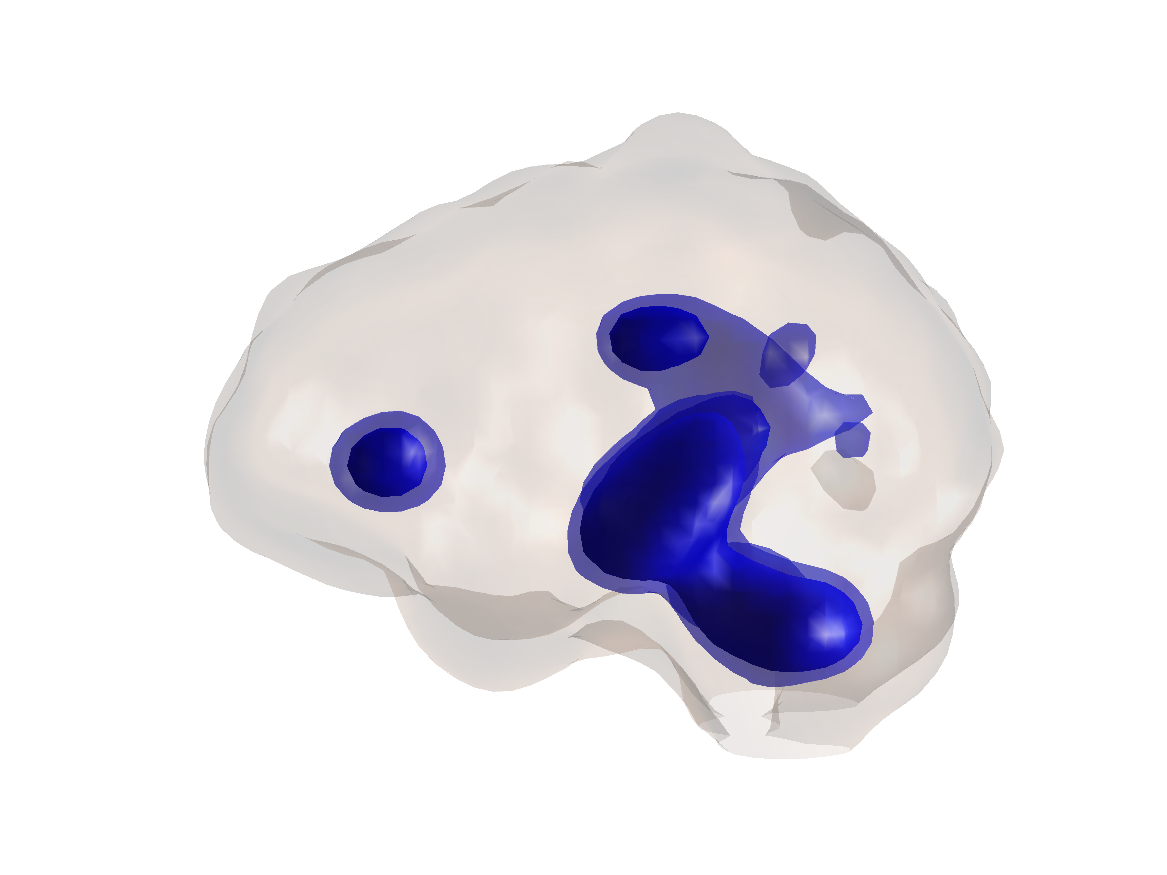} &
\includegraphics[scale=0.3]{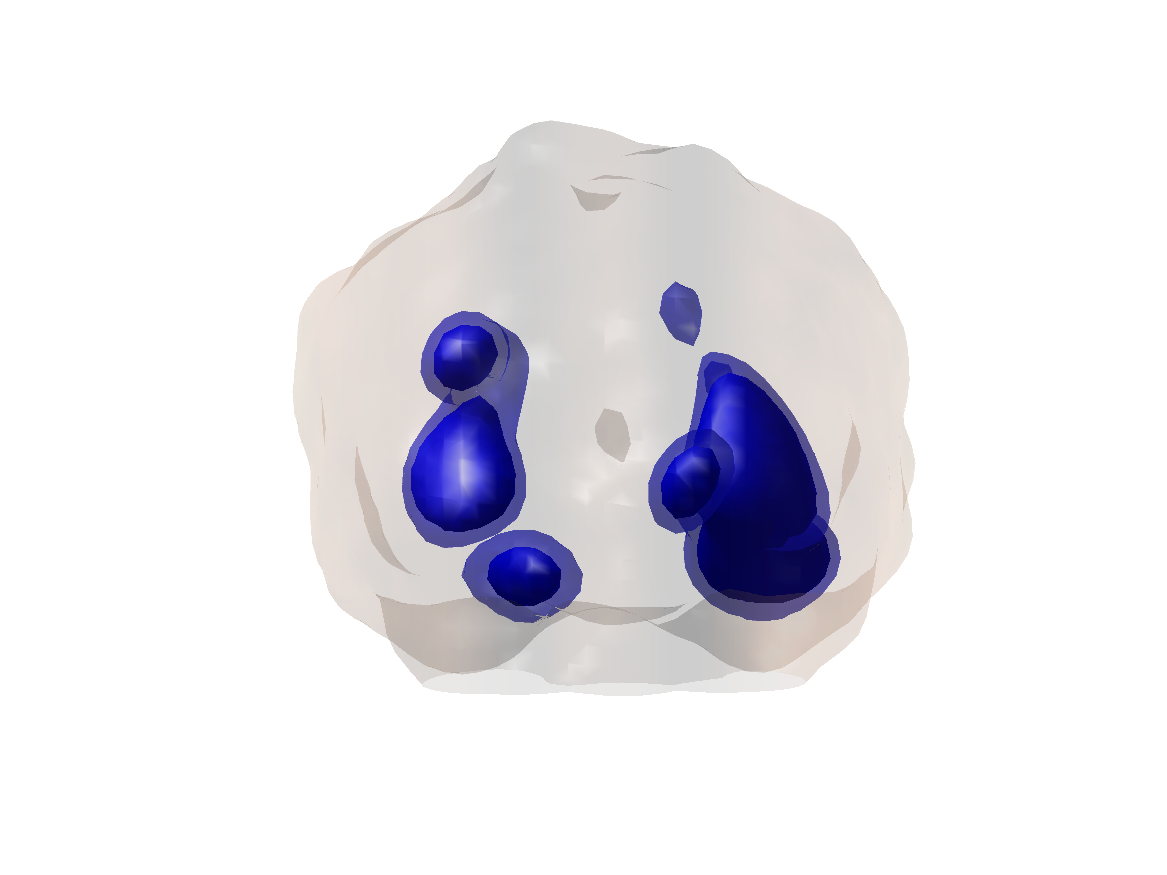} \\
(a) Top view &(b) Side view & (c) Front view \\

%\end{center}
\end{tabular}
\caption{DTI clustering results for three major clusters}
\label{fig:dti_results}
\end{figure}

%%%% what to summarize ?
\subsubsection{Results}
\label{subsubsec:results_dti}
We take one anonymous subject from this dataset consisting of $62,667$ measurements. 
%After pre-processing~\citep{} each measurement is converted to a $3 \times 2$ orthonormal matrix i.e. an element of $\StiefelS$ and 
We use a finite mixture of $\ML$ distributions to cluster this large dataset.

We use conditional conjugate prior distributions defined in Equation~\ref{eq:indep_prior_model}.
We select appropriate values of hyperparameters using the procedure developed for empirical prior in Section~\ref{subsec:hype_sel}. Note that, the value of $K^\dagger$ is set to $100$ as we expect relatively large number of data points in each cluster.
Here we use the procedure described in Section~\ref{subsec:EM_part} to set the initial value of the parameters $M$, $\bd$ and $V$ for the MCMC algorithm. First $1000$ MCMC samples are discarded as burn-in samples.

We use different number of clusters  
%-- $6, 7, 8, 10, 11$ and $12$ 
to fit the dataset with our Bayesian model and choose $12$ as the estimated number of clusters by DIC criterion described in Section~\ref{subsec:model_sel}. 
%We use the following values for the hyperparameters. 
%{\attn{the fact that 12 has the highest DIC might be difficult to argue}}

%We show the the table for DIC values for different values of $C$ below.

In Figure~\ref{fig:dti_results} we present the top three clusters with their voxel locations mapped inside the anatomical structure of brain (see \url{http://www.compgenome.org/stiefel} for 3D version of these figures). Note that in this figure, panel $(a)$, $(b)$ and $(c)$ represent top, side and front view, respectively. It is important to notice that we are successfully able to locate few important fiber structures in the dataset from the sample. 
%{\attn{describe the figures in detail if possible}}

%%% Matlab Figures @Pal

%{\attn{
%\begin{itemize}
%\item First work with $\ML$ for DTI in the big-data context. 
%\item represent diffusion pattern of water molecule inside a fiber track using $3 \times 3$ positive definite matrix.
%\item Why $\StiefelS$ ?
%\item Why not Wishart - interpretation, flexibility in handling DTI data.
%\item Performance could be improved by incorporating eigenvalues as well. However, that requires more involved Statistical model which is reserved for our future work. This paper is mainly to build framework.
%\item In fact, we have meaningful clusters with only directional part    
%\item Data source, generation, processing and results
%\end{itemize}
%}}

%%%%%%%%%%%%%%%%%%%%%%%%%
\subsection{Near Earth comet dataset}
\label{subsec:neo}
%%%http://www.uapress.arizona.edu/onlinebks/ResourcesNearEarthSpace/resources22.pdf
The Near Earth Object (NEO) population is defined as a group of small bodies with perihelion distance less than $1.3$ astronomical unit (AU) and aphelion distance greater than $0.983$ AU~\citep{Donnison:2006}. NEOs are NEAs (near-Earth asteroids) and NECs (near-Earth comets). NEAs are asteroids whose perihelion distance is less than $1.3$ AU. NECs are comets whose perihelion distance is less than $1.3$ AU and whose orbital period is less than $200$ years~(\url{https://cneos.jpl.nasa.gov/faq/}). A detailed categorization of NEO can also be found in~\url{https://cneos.jpl.nasa.gov/about/neo_groups.html}. They are also called short-period (SP) comets, which are generally confined to direct orbits with angle of inclination with respect to a reference plane, less than approximately $35^{\circ}$. The SP comets are in well determined orbits with modest eccentricities and inclinations. This make them a possible resource for space developments~\citep{Lewis:1993}. 
%Scientifically, observing a sample of these near-Earth objects (NEOs) can tell us more about the primordial materials of our Solar System. Another major goal also is to detect those objects that may collide with Earth. 

The NEC dataset was built by the Near Earth Object Program of the National Aeronautics and Space Administration(NASA). Each data point characterizes the orientation of a two-dimensional elliptical orbit in three-dimensional space, and thus lies on the Stiefel manifold $\mathcal{V}_{3,2}$. For our experiment we have downloaded NEC dataset containing 175 entries. Orientation of SP comet's orbit can be specified by the following quantities. We could find the definition of these three important quantities in \url{https://ssd.jpl.nasa.gov/?glossary}.
\begin{itemize}
\item Celestial longitude ($L$)
\item Latitude of the perihelion ($\theta$)
\item Longitude of the ascending node ($\Omega$)
\end{itemize} 

Celestial longitude of the comet ($L$)~\citep{Hughes:1985} and latitude of the perihelion ($\theta$)~\citep{Yabushita:1979} are computed by the following formula, respectively.
$$
L = \Omega + tan^{-1}\left(\frac{\sin\,\omega \, \cos\,i}{\cos\,\omega}\right) \qquad\qquad \sin\, \theta = \sin\,i \, \sin\,\omega
$$

\begin{figure}
\includegraphics[scale=1.0]{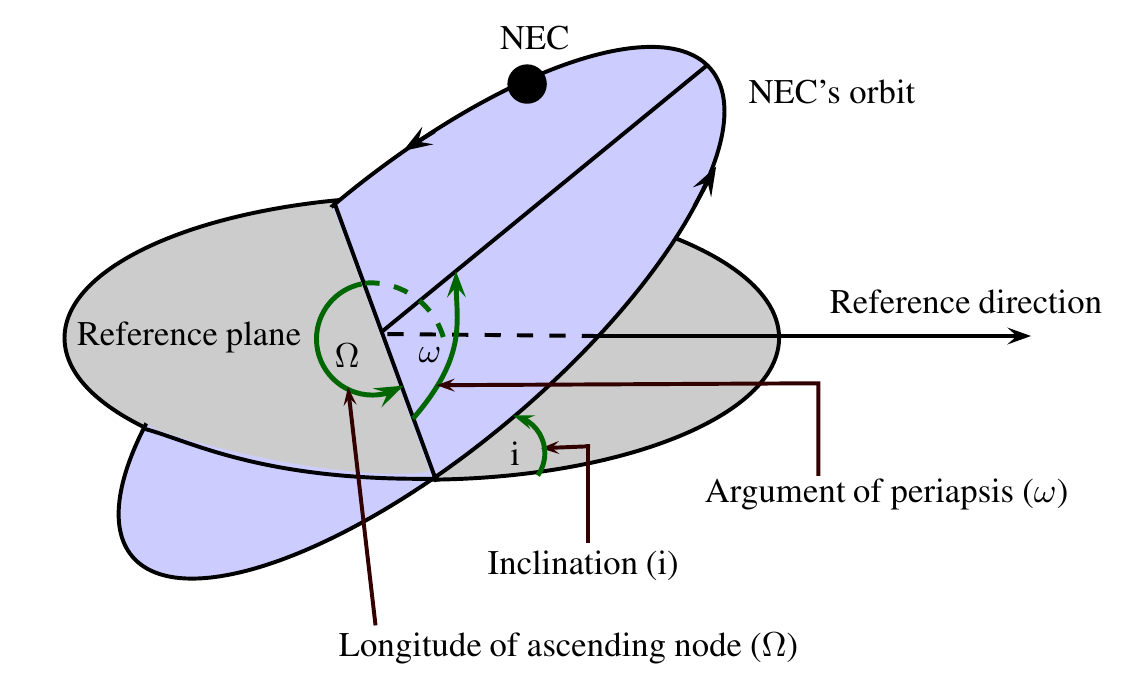}
\caption{Near Earth comet's orbit and orbital elements}
\label{fig:NEC_angles}
\end{figure}
From the dataset we could find the values of orbital inclination ($i$), longitude of the ascending node ($\Omega$), argument of periapsis (perihelion)($\omega$) as shown in Figure~\ref{fig:NEC_angles} Using the appropriate transformations given in~\cite{Jupp:1979,Yabushita:1979} we find $L$, $\theta$ and $\Omega$ for each comet. The direction of the perihelion is ${\bx}_1 = (cos\,\theta\, cos\,L, cos\,\theta\,sin\,L, sin\,\theta)$ and the directed unit normal to the orbit given by the right hand rule is
$$
{\bx}_2 = (sin\,\theta\, sin\,\Omega -sin\,\theta\,cos\,\Omega -cos\,\theta\,sin\,(\Omega-L))/r
$$
where $r^2 = sin^2\,\theta + cos^2\,\theta\,sin^2\,(\Omega-L)$. The orientation of the orbit therefore can be represented by the matrix ${X} \in \mathcal{V}_{3,2}$ given by $X = [{\bx}_1^T \; {\bx}_2^T]$. An appropriate model for the distribution of these matrices is the $\ML$ family~\citep{Jupp:1979}. 

Here we model NEC dataset as a finite mixture of $\ML$ distributions. We ran our model for number of clusters equals to $3, 4, 5, 6$.
In each situation we use $2000$ MCMC samples out of which we set initial $1000$ iterations as burn-ins. 

We select appropriate values of hyperparameters for prior distributions in Equation~\ref{eq:indep_prior_model} empirically using the procedure developed in Section~\ref{subsec:hype_sel}.

We choose number of burn-in iterations ($1000$ in this case) by observing the MCMC convergence diagnostic plot. Below we report the DIC for selecting the model. Our DIC (shown in Table~\ref{tbl:DIC_NEC}) is minimized at number of clusters equals to four. Note that, also from the reported results in~\cite{Lin:2017}, 
%(see left panel of Figure $2$), 
four seems to be the most likely number of clusters.
\begin{table}[htb!]
\centering
\begin{tabular}{|c|c|}
\hline
Number of Clusters & DIC Value \\
\hline
 3 & 3074.91 \\
 4 & 2607.04 \\ 
 5 & 2712.96 \\
 6 & 2685.94 \\
 \hline                        
\end{tabular}
\caption{DIC table for NEC dataset}
\label{tbl:DIC_NEC}
\end{table}

\begin{figure}[htb!]
\begin{center}
\includegraphics[scale=0.5]{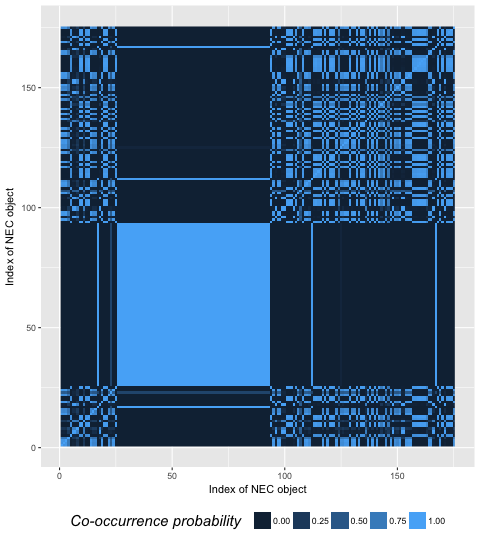}
\end{center}
\caption{Cluster co-occurrence probability matrix for NEC dataset.}
\label{fig:cocluster_NEC}
\end{figure}

We compute the probabilities for any two NEC data to belong to the same cluster for all the NEC data. This is also called as cluster co-occurrence probability matrix~\citep{Hofmann:1998}. We draw the corresponding heatmap in Figure~\ref{fig:cocluster_NEC} to show this.  

\begin{figure}[htb!]
\includegraphics[scale=0.52]{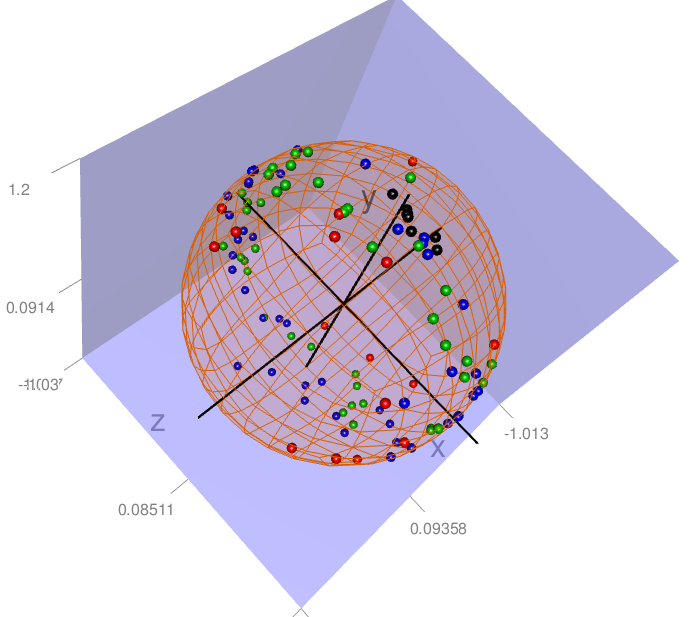}
\caption{First eigenvector of a data point is embedded in a sphere.}
\label{fig:sphere-embed_1}
\end{figure}

\begin{figure}[htb!]
\includegraphics[scale=0.52]{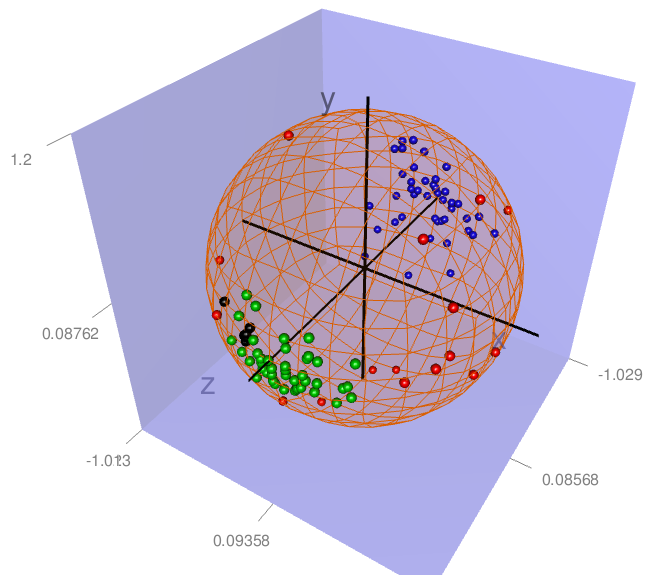}
\caption{Second eigenvector of a data point is embedded in a sphere.}
\label{fig:sphere-embed_2}
\end{figure}

Finally, we plot each eigenvector from a data point of $\mathcal{V}_{3,2}$ in a sphere (Figure~\ref{fig:sphere-embed_1} and~\ref{fig:sphere-embed_2}).
%(panel (a) and panel (b) in Figure~\ref{fig:sphere-embed}). 
We use different color (red, blue, green, black) to represent four different clusters. The NECs denoted by the points with same color indicates the group of comets with similar orbital characteristics.

%\begin{figure}[htb!]
%\begin{tabular}{cc}
%\includegraphics[scale=0.61]{./figures/eig1.png} &
%\includegraphics[scale=0.525]{./figures/eig2.png} \\
%(a) First eigenvector & (b) Second eigenvector \\
%\end{tabular}
%\caption{Each eigenvector of a data point is embedded in a sphere.}
%\label{fig:sphere-embed}
%\end{figure}

\section{Discussions and Future directions}
\label{sec:disc}
In this paper, we build a Bayesian framework for a mixture of $\ML$ distributions which could be applied to real world directional data. We construct two special  families of distributions to be used as prior distributions following the orginal conjugate prior construction in~\cite{Diaconis:Ylvisaker:1979}. We discuss few important properties for our prior class of distributions. For the mixture model we computed the posterior and also give insights on selection of hyperparameters,  which should be helpful for practitioners. Finally, we are able to handle a large amount of DTI data in the real data application and results look quite promising. 

For our future extension, instead of selecting the number of clusters by DIC criterion, we would like the number of clusters to be a random variable. A fully Bayesian model-based approach which assumes a parametric prior (e.g. Poisson) on the number of clusters, could be employed. The next natural step in this direction is to extend the existing model to a non-parametric framework. In fact, non-parametric version is more flexible in terms of modeling and experimenting with different types of underlying clustering structure. Note that, though~\cite{Lin:2017} opened the doors to such modeling, their  model space differs from ours in various respects. 
%({\attn{We need to carefully delineate the differences without much criticism}})\footnote{this is the tricky part we have to add in what what way.}

On a separate direction, we also plan to explore in depth the analytical properties of the hypergeometric function of matrix argument function (${}_0F_1(\cdot)$) for $p \ge 2$. Direct computation, as is done in our case studies, could create bottlenecks for data coming from higher dimension. Analytical bounds could help either in approximation or designing a good MCMC sampler. For example, one could borrow the importance sampling approach used for evaluating the normalizing constants in~\cite{Mitra:2013}. This would primarily rely on the ability to simulate efficiently from $\ML$ distributions, which is already ensured by~\cite{Hoff:2009}. Along this line, it would be nice to study the theoretical properties, particularly ergodicity of the MCMC schemes rigorously.  

%Finally, there is a scope of extending the newly proposed family of  priors distribution (Section~\ref{subsec:prior_construct}) to a larger class of  Bayesian models involving more general densities (e.g matrix-Bingham) on manifolds. %\footnote{hint at some  generalities here}. The properties of prior and posterior discovered in that section can also be seamlessly generalized.
%\footnote{wanted to also include spatial mods in brain application, but am hesitant because 1) we are anyway not emphasizing that application and 2) maybe we are giving them too many ideas -- {\attn{although we could emphasize again that our method runs on big data}}}

The coming together of state-of-the-art Bayesian methods incorporating topological properties of the space is a rich area that has been initiated only recently by~\cite{Bhatta:2012} and~\cite{Lin:2017}. We plan to continue along this direction and contribute to the Bayesian methodological development on general analytic manifolds, which would be appropriate to analyze large-scale data with complex structure. 

%\section{Appendix ({\attn{not for arXiv version}})}
%\label{sec:appendix}

%%%%%%%%%%%%%%%%%%%%%%

%\begin{eqnarray}
%&&\left\{\prod_{i=1}^N P(X_i, Z_i \mid \pi, {{\Theta}})\right\} P(\Bpi,\Theta)\nonumber\\
%&=&\left\{\prod_{i=1}^N \sum_{c=1}^C P(X_i | Z_i = c, \pi, {{\Theta}}) P(Z_i = c\mid \pi,\Theta)\right\} P(\Bpi,\Theta)\nonumber\\
%&=&\left\{\prod_{i=1}^N P(X_i | {{\Theta}}_{Z_i}) \pi_{Z_i}\right\} P(\Theta) P(\Bpi)\nonumber\\
%&=&\left\{\prod_{i=1}^N \prod_{c=1}^C \left( P(X_i \mid \theta_c) \pi_c \right)^{I(Z_i=c)} \right\} P({\Theta}) P(\Bpi)\nonumber\\
%&\propto& \prod_{c=1}^C \left({\pi_c}^{N_c+\alpha_c-1}\, \frac{\exp(trace({(V_c^tD_cM_c^t)} \tilde{S}_c))}{{}_0F_1(\frac{n}{2};D_c^2/4)^{N_c}} \times \frac{\exp(trace({G_0^t}M_c))}{{}_0F_1(\frac{n}{2};G_0^2/4)} \times \frac{\exp(trace({H_0^t}V_c))}{{}_0F_1(\frac{n}{2};H_0^2/4)}\right.\nonumber\\
%&\times& \left.\prod_{k=1}^p d_{ck}^{\alpha-1} e^{(-\beta_k d_{ck})} \mathbb{I}(d_{c1} \geq \cdots \geq d_{cp})\right)\nonumber\\
%&\propto&\prod_{c=1}^C \left({\pi_c}^{N_c+\alpha_c-1}\, \frac{\exp(trace({(V_c^tD_cM_c^t)} \tilde{S}_c + G_0^t M_c + H_0^tV_c)} {{}_0F_1(\frac{n}{2};D_c^2/4)^{N_c}} \times \prod_{k=1}^p d_{ck}^{\alpha-1} e^{(-\beta_k d_{ck})} \mathbb{I}(d_{c1} \geq \cdots \geq d_{cp}) \right)\nonumber\\
%\label{eq:post_mix_appendix}
%\end{eqnarray}

%%%%%%%%%%%%%%%%% BIBLIOGRAPHY %%%%%%%%%%%%%

\bibliographystyle{imsart-nameyear}
\bibliography{ref_param_dti}

\begin{thebibliography}{88}
% BibTex style file: imsart-nameyear.bst, 2013-01-28
% Default style options (sort=1,type=nameyear).
% Used options (sort=1,type=nameyear).

\bibitem[\protect\citeauthoryear{Absil, Mahony and
  Sepulchre}{2009}]{Absil:2009}
\begin{bbook}[author]
\bauthor{\bsnm{Absil},~\bfnm{P-A}\binits{P.-A.}},
  \bauthor{\bsnm{Mahony},~\bfnm{Robert}\binits{R.}} \AND
  \bauthor{\bsnm{Sepulchre},~\bfnm{Rodolphe}\binits{R.}}
(\byear{2009}).
\btitle{Optimization algorithms on matrix manifolds}.
\bpublisher{Princeton University Press}.
\end{bbook}
\endbibitem

\bibitem[\protect\citeauthoryear{Anand, Mittal and Meer}{2016}]{Anand:2016}
\begin{bincollection}[author]
\bauthor{\bsnm{Anand},~\bfnm{Saket}\binits{S.}},
  \bauthor{\bsnm{Mittal},~\bfnm{Sushil}\binits{S.}} \AND
  \bauthor{\bsnm{Meer},~\bfnm{Peter}\binits{P.}}
(\byear{2016}).
\btitle{Robust Estimation for Computer Vision Using Grassmann Manifolds}.
In \bbooktitle{Riemannian Computing in Computer Vision}
\bpages{125--144}.
\bpublisher{Springer}.
\end{bincollection}
\endbibitem

\bibitem[\protect\citeauthoryear{Bangert, Hennig and
  Oelfke}{2010}]{Bangert:2010}
\begin{binproceedings}[author]
\bauthor{\bsnm{Bangert},~\bfnm{Mark}\binits{M.}},
  \bauthor{\bsnm{Hennig},~\bfnm{Philipp}\binits{P.}} \AND
  \bauthor{\bsnm{Oelfke},~\bfnm{Uwe}\binits{U.}}
(\byear{2010}).
\btitle{Using an infinite von Mises-Fisher mixture model to cluster treatment
  beam directions in external radiation therapy}.
In \bbooktitle{Machine Learning and Applications (ICMLA), 2010 Ninth
  International Conference on}
\bpages{746--751}.
\bpublisher{IEEE}.
\end{binproceedings}
\endbibitem

\bibitem[\protect\citeauthoryear{Basser and Jones}{2002}]{Basser:2002}
\begin{barticle}[author]
\bauthor{\bsnm{Basser},~\bfnm{Peter~J}\binits{P.~J.}} \AND
  \bauthor{\bsnm{Jones},~\bfnm{Derek~K}\binits{D.~K.}}
(\byear{2002}).
\btitle{Diffusion-tensor MRI: theory, experimental design and data analysis--a
  technical review}.
\bjournal{NMR in Biomedicine}
\bvolume{15}
\bpages{456--467}.
\end{barticle}
\endbibitem

\bibitem[\protect\citeauthoryear{Basser, Mattiello and
  LeBihan}{1994}]{Basser:1994}
\begin{barticle}[author]
\bauthor{\bsnm{Basser},~\bfnm{Peter~J}\binits{P.~J.}},
  \bauthor{\bsnm{Mattiello},~\bfnm{James}\binits{J.}} \AND
  \bauthor{\bsnm{LeBihan},~\bfnm{Denis}\binits{D.}}
(\byear{1994}).
\btitle{MR diffusion tensor spectroscopy and imaging}.
\bjournal{Biophysical journal}
\bvolume{66}
\bpages{259--267}.
\end{barticle}
\endbibitem

\bibitem[\protect\citeauthoryear{Berg, Meyer and Yu}{2004}]{Berg:Meyer:Yu:2004}
\begin{barticle}[author]
\bauthor{\bsnm{Berg},~\bfnm{Andreas}\binits{A.}},
  \bauthor{\bsnm{Meyer},~\bfnm{Renate}\binits{R.}} \AND
  \bauthor{\bsnm{Yu},~\bfnm{Jun}\binits{J.}}
(\byear{2004}).
\btitle{Deviance information criterion for comparing stochastic volatility
  models}.
\bjournal{Journal of Business \& Economic Statistics}
\bvolume{22}
\bpages{107--120}.
\end{barticle}
\endbibitem

\bibitem[\protect\citeauthoryear{Bhatia}{2007}]{Bhatia:2007}
\begin{barticle}[author]
\bauthor{\bsnm{Bhatia},~\bfnm{R}\binits{R.}}
(\byear{2007}).
\btitle{Positive Definite Matrices Princeton University Press}.
\bjournal{Princeton and Oxford}.
\end{barticle}
\endbibitem

\bibitem[\protect\citeauthoryear{Bhattacharya and Dunson}{2012}]{Bhatta:2012}
\begin{barticle}[author]
\bauthor{\bsnm{Bhattacharya},~\bfnm{Abhishek}\binits{A.}} \AND
  \bauthor{\bsnm{Dunson},~\bfnm{David~B}\binits{D.~B.}}
(\byear{2012}).
\btitle{Strong consistency of nonparametric Bayes density estimation on compact
  metric spaces with applications to specific manifolds}.
\bjournal{Annals of the Institute of Statistical Mathematics}
\bvolume{64}
\bpages{687--714}.
\end{barticle}
\endbibitem

\bibitem[\protect\citeauthoryear{Bishop}{2006}]{Bishop:2006}
\begin{bbook}[author]
\bauthor{\bsnm{Bishop},~\bfnm{Christopher~M}\binits{C.~M.}}
(\byear{2006}).
\btitle{Pattern recognition and machine learning}.
\bpublisher{springer}.
\end{bbook}
\endbibitem

\bibitem[\protect\citeauthoryear{Butler and Wood}{2003}]{Butler:2003}
\begin{barticle}[author]
\bauthor{\bsnm{Butler},~\bfnm{Ronald~W}\binits{R.~W.}} \AND
  \bauthor{\bsnm{Wood},~\bfnm{Andrew~TA}\binits{A.~T.}}
(\byear{2003}).
\btitle{Laplace approximation for Bessel functions of matrix argument}.
\bjournal{Journal of Computational and Applied Mathematics}
\bvolume{155}
\bpages{359--382}.
\end{barticle}
\endbibitem

\bibitem[\protect\citeauthoryear{Casella}{1985}]{Casella:1985}
\begin{barticle}[author]
\bauthor{\bsnm{Casella},~\bfnm{George}\binits{G.}}
(\byear{1985}).
\btitle{An introduction to empirical Bayes data analysis}.
\bjournal{The American Statistician}
\bvolume{39}
\bpages{83--87}.
\end{barticle}
\endbibitem

\bibitem[\protect\citeauthoryear{Casella and Berger}{2002}]{Casella:2002}
\begin{bbook}[author]
\bauthor{\bsnm{Casella},~\bfnm{George}\binits{G.}} \AND
  \bauthor{\bsnm{Berger},~\bfnm{Roger~L}\binits{R.~L.}}
(\byear{2002}).
\btitle{Statistical inference}
\bvolume{2}.
\bpublisher{Duxbury Pacific Grove, CA}.
\end{bbook}
\endbibitem

\bibitem[\protect\citeauthoryear{Catani et~al.}{2002}]{Catani:2002}
\begin{barticle}[author]
\bauthor{\bsnm{Catani},~\bfnm{Marco}\binits{M.}},
  \bauthor{\bsnm{Howard},~\bfnm{Robert~J}\binits{R.~J.}},
  \bauthor{\bsnm{Pajevic},~\bfnm{Sinisa}\binits{S.}} \AND
  \bauthor{\bsnm{Jones},~\bfnm{Derek~K}\binits{D.~K.}}
(\byear{2002}).
\btitle{Virtual in vivo interactive dissection of white matter fasciculi in the
  human brain}.
\bjournal{Neuroimage}
\bvolume{17}
\bpages{77--94}.
\end{barticle}
\endbibitem

\bibitem[\protect\citeauthoryear{Chikuse}{1991a}]{Chikuse:1991}
\begin{barticle}[author]
\bauthor{\bsnm{Chikuse},~\bfnm{Yasuko}\binits{Y.}}
(\byear{1991}a).
\btitle{High dimensional limit theorems and matrix decompositions on the
  Stiefel manifold}.
\bjournal{Journal of multivariate analysis}
\bvolume{36}
\bpages{145--162}.
\end{barticle}
\endbibitem

\bibitem[\protect\citeauthoryear{Chikuse}{1991b}]{Chikuse:1991:as}
\begin{barticle}[author]
\bauthor{\bsnm{Chikuse},~\bfnm{Yasuko}\binits{Y.}}
(\byear{1991}b).
\btitle{Asymptotic expansions for distributions of the large sample matrix
  resultant and related statistics on the Stiefel manifold}.
\bjournal{Journal of multivariate analysis}
\bvolume{39}
\bpages{270--283}.
\end{barticle}
\endbibitem

\bibitem[\protect\citeauthoryear{Chikuse}{1998}]{Chikuse:1998}
\begin{barticle}[author]
\bauthor{\bsnm{Chikuse},~\bfnm{Yasuko}\binits{Y.}}
(\byear{1998}).
\btitle{Density estimation on the Stiefel manifold}.
\bjournal{Journal of multivariate analysis}
\bvolume{66}
\bpages{188--206}.
\end{barticle}
\endbibitem

\bibitem[\protect\citeauthoryear{Chikuse}{2012}]{Chikuse:2012}
\begin{bbook}[author]
\bauthor{\bsnm{Chikuse},~\bfnm{Yasuko}\binits{Y.}}
(\byear{2012}).
\btitle{Statistics on special manifolds}
\bvolume{174}.
\bpublisher{Springer Science \& Business Media}.
\end{bbook}
\endbibitem

\bibitem[\protect\citeauthoryear{Conway}{1990}]{Conway:1990}
\begin{barticle}[author]
\bauthor{\bsnm{Conway},~\bfnm{JB}\binits{J.}}
(\byear{1990}).
\btitle{A $\{$Course$\}$ in $\{$Functional$\}$$\{$Analysis$\}$}.
\end{barticle}
\endbibitem

\bibitem[\protect\citeauthoryear{Cowles and Carlin}{1996}]{Cowles:1996}
\begin{barticle}[author]
\bauthor{\bsnm{Cowles},~\bfnm{Mary~Kathryn}\binits{M.~K.}} \AND
  \bauthor{\bsnm{Carlin},~\bfnm{Bradley~P}\binits{B.~P.}}
(\byear{1996}).
\btitle{Markov chain Monte Carlo convergence diagnostics: a comparative
  review}.
\bjournal{Journal of the American Statistical Association}
\bvolume{91}
\bpages{883--904}.
\end{barticle}
\endbibitem

\bibitem[\protect\citeauthoryear{DeIorio and
  Robert}{2002}]{DeIorio:Robert:2002}
\begin{barticle}[author]
\bauthor{\bsnm{DeIorio},~\bfnm{M.}\binits{M.}} \AND
  \bauthor{\bsnm{Robert},~\bfnm{C.~P}\binits{C.~P.}}
(\byear{2002}).
\btitle{Discussion of Bayesian measures of model complexity and fit}.
\bjournal{Journal of the Royal Statistical Society: Series B (Statistical
  Methodology)}
\bvolume{64}
\bpages{629--630}.
\bdoi{10.1111/1467-9868.00353}
\end{barticle}
\endbibitem

\bibitem[\protect\citeauthoryear{Dempster, Laird and
  Rubin}{1977}]{Dempster:1977}
\begin{barticle}[author]
\bauthor{\bsnm{Dempster},~\bfnm{Arthur~P}\binits{A.~P.}},
  \bauthor{\bsnm{Laird},~\bfnm{Nan~M}\binits{N.~M.}} \AND
  \bauthor{\bsnm{Rubin},~\bfnm{Donald~B}\binits{D.~B.}}
(\byear{1977}).
\btitle{Maximum likelihood from incomplete data via the EM algorithm}.
\bjournal{Journal of the royal statistical society. Series B (methodological)}
\bpages{1--38}.
\end{barticle}
\endbibitem

\bibitem[\protect\citeauthoryear{Diaconis and
  Ylvisaker}{1979}]{Diaconis:Ylvisaker:1979}
\begin{barticle}[author]
\bauthor{\bsnm{Diaconis},~\bfnm{Persi}\binits{P.}} \AND
  \bauthor{\bsnm{Ylvisaker},~\bfnm{Donald}\binits{D.}}
(\byear{1979}).
\btitle{Conjugate priors for exponential families}.
\bjournal{The Annals of statistics}
\bvolume{7}
\bpages{269--281}.
\end{barticle}
\endbibitem

\bibitem[\protect\citeauthoryear{Diebolt and
  Robert}{1994}]{DIEBOLT:ROBERT:1994}
\begin{barticle}[author]
\bauthor{\bsnm{Diebolt},~\bfnm{Jean}\binits{J.}} \AND
  \bauthor{\bsnm{Robert},~\bfnm{Christian~P.}\binits{C.~P.}}
(\byear{1994}).
\btitle{Estimation of Finite Mixture Distributions through Bayesian Sampling}.
\bjournal{Journal of the Royal Statistical Society. Series B (Methodological)}
\bvolume{56}
\bpages{363-375}.
\end{barticle}
\endbibitem

\bibitem[\protect\citeauthoryear{Donnison}{2006}]{Donnison:2006}
\begin{barticle}[author]
\bauthor{\bsnm{Donnison},~\bfnm{JR}\binits{J.}}
(\byear{2006}).
\btitle{Some aspects of the statistics of Near-Earth Objects}.
\bjournal{Proceedings of the International Astronomical Union}
\bvolume{2}
\bpages{69--76}.
\end{barticle}
\endbibitem

\bibitem[\protect\citeauthoryear{Downs}{1972}]{Downs:1972}
\begin{barticle}[author]
\bauthor{\bsnm{Downs},~\bfnm{Thomas~D}\binits{T.~D.}}
(\byear{1972}).
\btitle{Orientation statistics}.
\bjournal{Biometrika}
\bpages{665--676}.
\end{barticle}
\endbibitem

\bibitem[\protect\citeauthoryear{Edelman, Arias and Smith}{1998}]{Edelman:1998}
\begin{barticle}[author]
\bauthor{\bsnm{Edelman},~\bfnm{Alan}\binits{A.}},
  \bauthor{\bsnm{Arias},~\bfnm{Tom{\'a}s~A}\binits{T.~A.}} \AND
  \bauthor{\bsnm{Smith},~\bfnm{Steven~T}\binits{S.~T.}}
(\byear{1998}).
\btitle{The geometry of algorithms with orthogonality constraints}.
\bjournal{SIAM journal on Matrix Analysis and Applications}
\bvolume{20}
\bpages{303--353}.
\end{barticle}
\endbibitem

\bibitem[\protect\citeauthoryear{Fran{\c{c}}ois and
  Laval}{2011}]{Franccois:Laval:2011}
\begin{barticle}[author]
\bauthor{\bsnm{Fran{\c{c}}ois},~\bfnm{Olivier}\binits{O.}} \AND
  \bauthor{\bsnm{Laval},~\bfnm{Guillaume}\binits{G.}}
(\byear{2011}).
\btitle{Deviance information criteria for model selection in approximate
  Bayesian computation}.
\bjournal{arXiv preprint arXiv:1105.0269}.
\end{barticle}
\endbibitem

\bibitem[\protect\citeauthoryear{Gelman et~al.}{2003}]{Book:Gelman:Rubin:2003}
\begin{bbook}[author]
\bauthor{\bsnm{Gelman},~\bfnm{Andrew}\binits{A.}},
  \bauthor{\bsnm{Carlin},~\bfnm{John~B.}\binits{J.~B.}},
  \bauthor{\bsnm{Stern},~\bfnm{Hal~S.}\binits{H.~S.}} \AND
  \bauthor{\bsnm{Rubin},~\bfnm{Donald~B.}\binits{D.~B.}}
(\byear{2003}).
\btitle{{Bayesian Data Analysis, Second Edition (Chapman \& Hall/CRC Texts in
  Statistical Science)}},
\bedition{2} ed.
\bpublisher{Chapman and Hall/CRC}.
\end{bbook}
\endbibitem

\bibitem[\protect\citeauthoryear{Gelman et~al.}{2014}]{Gelman:2014}
\begin{bbook}[author]
\bauthor{\bsnm{Gelman},~\bfnm{Andrew}\binits{A.}},
  \bauthor{\bsnm{Carlin},~\bfnm{John~B}\binits{J.~B.}},
  \bauthor{\bsnm{Stern},~\bfnm{Hal~S}\binits{H.~S.}},
  \bauthor{\bsnm{Dunson},~\bfnm{David~B}\binits{D.~B.}},
  \bauthor{\bsnm{Vehtari},~\bfnm{Aki}\binits{A.}} \AND
  \bauthor{\bsnm{Rubin},~\bfnm{Donald~B}\binits{D.~B.}}
(\byear{2014}).
\btitle{Bayesian data analysis}
\bvolume{2}.
\bpublisher{CRC press Boca Raton, FL}.
\end{bbook}
\endbibitem

\bibitem[\protect\citeauthoryear{Gross and Richards}{1987}]{Gross:1987}
\begin{barticle}[author]
\bauthor{\bsnm{Gross},~\bfnm{Kenneth~I}\binits{K.~I.}} \AND
  \bauthor{\bsnm{Richards},~\bfnm{Donald St~P}\binits{D.~S.~P.}}
(\byear{1987}).
\btitle{Special functions of matrix argument. I. Algebraic induction, zonal
  polynomials, and hypergeometric functions}.
\bjournal{Transactions of the American Mathematical Society}
\bvolume{301}
\bpages{781--811}.
\end{barticle}
\endbibitem

\bibitem[\protect\citeauthoryear{Gross and Richards}{1989}]{Gross:1989}
\begin{barticle}[author]
\bauthor{\bsnm{Gross},~\bfnm{Kenneth~I}\binits{K.~I.}} \AND
  \bauthor{\bsnm{Richards},~\bfnm{Donald St~P}\binits{D.~S.~P.}}
(\byear{1989}).
\btitle{Total positivity, spherical series, and hypergeometric functions of
  matrix argument}.
\bjournal{Journal of Approximation theory}
\bvolume{59}
\bpages{224--246}.
\end{barticle}
\endbibitem

\bibitem[\protect\citeauthoryear{Gupta and Richards}{1985}]{Gupta:1985}
\begin{barticle}[author]
\bauthor{\bsnm{Gupta},~\bfnm{Rameshwar~D}\binits{R.~D.}} \AND
  \bauthor{\bsnm{Richards},~\bfnm{Donald St~P}\binits{D.~S.~P.}}
(\byear{1985}).
\btitle{Hypergeometric functions of scalar matrix argument are expressible in
  terms of classical hypergeometric functions}.
\bjournal{SIAM journal on mathematical analysis}
\bvolume{16}
\bpages{852--858}.
\end{barticle}
\endbibitem

\bibitem[\protect\citeauthoryear{Guy M~Mckhann}{2004}]{Guy:2004}
\begin{barticle}[author]
\bauthor{\bsnm{Guy M~Mckhann},~\bfnm{II}\binits{I.}}
(\byear{2004}).
\btitle{Non-invasive mapping of connections between human thalamus and cortex
  using diffusion imaging}.
\bjournal{Neurosurgery}
\bvolume{54}.
\end{barticle}
\endbibitem

\bibitem[\protect\citeauthoryear{Hardy, Littlewood and
  P{\'o}lya}{1952}]{Hardy:1952}
\begin{bbook}[author]
\bauthor{\bsnm{Hardy},~\bfnm{Godfrey~Harold}\binits{G.~H.}},
  \bauthor{\bsnm{Littlewood},~\bfnm{John~Edensor}\binits{J.~E.}} \AND
  \bauthor{\bsnm{P{\'o}lya},~\bfnm{George}\binits{G.}}
(\byear{1952}).
\btitle{Inequalities}.
\bpublisher{Cambridge university press}.
\end{bbook}
\endbibitem

\bibitem[\protect\citeauthoryear{Herz}{1955}]{Herz:1955}
\begin{barticle}[author]
\bauthor{\bsnm{Herz},~\bfnm{Carl~S}\binits{C.~S.}}
(\byear{1955}).
\btitle{Bessel functions of matrix argument}.
\bjournal{Annals of Mathematics}
\bpages{474--523}.
\end{barticle}
\endbibitem

\bibitem[\protect\citeauthoryear{Hill and Waters}{1987}]{Hill:1987}
\begin{barticle}[author]
\bauthor{\bsnm{Hill},~\bfnm{Richard~D}\binits{R.~D.}} \AND
  \bauthor{\bsnm{Waters},~\bfnm{Steven~R}\binits{S.~R.}}
(\byear{1987}).
\btitle{On the cone of positive semidefinite matrices}.
\bjournal{Linear Algebra and its Applications}
\bvolume{90}
\bpages{81--88}.
\end{barticle}
\endbibitem

\bibitem[\protect\citeauthoryear{Hoff}{2009}]{Hoff:2009}
\begin{barticle}[author]
\bauthor{\bsnm{Hoff},~\bfnm{Peter~D}\binits{P.~D.}}
(\byear{2009}).
\btitle{Simulation of the matrix Bingham--von Mises--Fisher distribution, with
  applications to multivariate and relational data}.
\bjournal{Journal of Computational and Graphical Statistics}
\bvolume{18}
\bpages{438--456}.
\end{barticle}
\endbibitem

\bibitem[\protect\citeauthoryear{Hofmann and Puzicha}{1998}]{Hofmann:1998}
\begin{barticle}[author]
\bauthor{\bsnm{Hofmann},~\bfnm{Thomas}\binits{T.}} \AND
  \bauthor{\bsnm{Puzicha},~\bfnm{Jan}\binits{J.}}
(\byear{1998}).
\btitle{Statistical models for co-occurrence data}.
\end{barticle}
\endbibitem

\bibitem[\protect\citeauthoryear{Hornik and Gr{\"u}n}{2013}]{Hornik:2013}
\begin{barticle}[author]
\bauthor{\bsnm{Hornik},~\bfnm{K}\binits{K.}} \AND
  \bauthor{\bsnm{Gr{\"u}n},~\bfnm{B}\binits{B.}}
(\byear{2013}).
\btitle{On conjugate families and Jeffreys priors for von Mises-Fisher
  distributions}.
\bjournal{J Stat Plan Inference}
\bvolume{143}
\bpages{992-999}.
\bdoi{10.1016/j.jspi.2012.11.003}
\end{barticle}
\endbibitem

\bibitem[\protect\citeauthoryear{Hornik and Gr{\"u}n}{2014}]{Hornik:2014}
\begin{barticle}[author]
\bauthor{\bsnm{Hornik},~\bfnm{Kurt}\binits{K.}} \AND
  \bauthor{\bsnm{Gr{\"u}n},~\bfnm{Bettina}\binits{B.}}
(\byear{2014}).
\btitle{movMF: An R package for fitting mixtures of von Mises-Fisher
  distributions}.
\bjournal{Journal of Statistical Software}
\bvolume{58}
\bpages{1--31}.
\end{barticle}
\endbibitem

\bibitem[\protect\citeauthoryear{Hughes}{1985}]{Hughes:1985}
\begin{barticle}[author]
\bauthor{\bsnm{Hughes},~\bfnm{David~W}\binits{D.~W.}}
(\byear{1985}).
\btitle{The position of earth at previous apparitions of Halley's comet}.
\bjournal{Quarterly Journal of the Royal Astronomical Society}
\bvolume{26}
\bpages{513--520}.
\end{barticle}
\endbibitem

\bibitem[\protect\citeauthoryear{James}{1964}]{James:1964}
\begin{barticle}[author]
\bauthor{\bsnm{James},~\bfnm{Alan~T}\binits{A.~T.}}
(\byear{1964}).
\btitle{Distributions of matrix variates and latent roots derived from normal
  samples}.
\bjournal{The Annals of Mathematical Statistics}
\bpages{475--501}.
\end{barticle}
\endbibitem

\bibitem[\protect\citeauthoryear{James}{1976}]{James:1976}
\begin{bbook}[author]
\bauthor{\bsnm{James},~\bfnm{Ioan~Mackenzie}\binits{I.~M.}}
(\byear{1976}).
\btitle{The topology of Stiefel manifolds}
\bvolume{24}.
\bpublisher{Cambridge University Press}.
\end{bbook}
\endbibitem

\bibitem[\protect\citeauthoryear{Jupp and Mardia}{1979}]{Jupp:1979}
\begin{barticle}[author]
\bauthor{\bsnm{Jupp},~\bfnm{Peter~E}\binits{P.~E.}} \AND
  \bauthor{\bsnm{Mardia},~\bfnm{Kanti~V}\binits{K.~V.}}
(\byear{1979}).
\btitle{Maximum likelihood estimators for the matrix von Mises-Fisher and
  Bingham distributions}.
\bjournal{The Annals of Statistics}
\bpages{599--606}.
\end{barticle}
\endbibitem

\bibitem[\protect\citeauthoryear{Jupp and Mardia}{1980}]{Jupp:1980}
\begin{barticle}[author]
\bauthor{\bsnm{Jupp},~\bfnm{PE}\binits{P.}} \AND
  \bauthor{\bsnm{Mardia},~\bfnm{KV}\binits{K.}}
(\byear{1980}).
\btitle{A general correlation coefficient for directional data and related
  regression problems}.
\bjournal{Biometrika}
\bpages{163--173}.
\end{barticle}
\endbibitem

\bibitem[\protect\citeauthoryear{KaewTraKulPong and
  Bowden}{2002}]{Kaewtrakulpong:2002}
\begin{barticle}[author]
\bauthor{\bsnm{KaewTraKulPong},~\bfnm{Pakorn}\binits{P.}} \AND
  \bauthor{\bsnm{Bowden},~\bfnm{Richard}\binits{R.}}
(\byear{2002}).
\btitle{An improved adaptive background mixture model for real-time tracking
  with shadow detection}.
\bjournal{Video-based surveillance systems}
\bvolume{1}
\bpages{135--144}.
\end{barticle}
\endbibitem

\bibitem[\protect\citeauthoryear{Khare, Pal and Su}{2017}]{Khare:Pal:Su:2017}
\begin{barticle}[author]
\bauthor{\bsnm{Khare},~\bfnm{Kshitij}\binits{K.}},
  \bauthor{\bsnm{Pal},~\bfnm{Subhadip}\binits{S.}} \AND
  \bauthor{\bsnm{Su},~\bfnm{Zhihua}\binits{Z.}}
(\byear{2017}).
\btitle{A bayesian approach for envelope models}.
\bjournal{The Annals of Statistics}
\bvolume{45}
\bpages{196--222}.
\end{barticle}
\endbibitem

\bibitem[\protect\citeauthoryear{Khatri and Mardia}{1977}]{Khatri:1977}
\begin{barticle}[author]
\bauthor{\bsnm{Khatri},~\bfnm{CG}\binits{C.}} \AND
  \bauthor{\bsnm{Mardia},~\bfnm{KV}\binits{K.}}
(\byear{1977}).
\btitle{The von Mises-Fisher matrix distribution in orientation statistics}.
\bjournal{Journal of the Royal Statistical Society. Series B (Methodological)}
\bpages{95--106}.
\end{barticle}
\endbibitem

\bibitem[\protect\citeauthoryear{Koev and Edelman}{2006}]{Koev:2006}
\begin{barticle}[author]
\bauthor{\bsnm{Koev},~\bfnm{Plamen}\binits{P.}} \AND
  \bauthor{\bsnm{Edelman},~\bfnm{Alan}\binits{A.}}
(\byear{2006}).
\btitle{The efficient evaluation of the hypergeometric function of a matrix
  argument}.
\bjournal{Mathematics of Computation}
\bvolume{75}
\bpages{833--846}.
\end{barticle}
\endbibitem

\bibitem[\protect\citeauthoryear{Kristof}{1969}]{Kristof:1969}
\begin{barticle}[author]
\bauthor{\bsnm{Kristof},~\bfnm{Walter}\binits{W.}}
(\byear{1969}).
\btitle{A theorem on the trace of certain matrix products and some
  applications}.
\bjournal{ETS Research Report Series}
\bvolume{1969}.
\end{barticle}
\endbibitem

\bibitem[\protect\citeauthoryear{Lattin, Carroll and Green}{2003}]{Lattin:2003}
\begin{bbook}[author]
\bauthor{\bsnm{Lattin},~\bfnm{James~M}\binits{J.~M.}},
  \bauthor{\bsnm{Carroll},~\bfnm{J~Douglas}\binits{J.~D.}} \AND
  \bauthor{\bsnm{Green},~\bfnm{Paul~E}\binits{P.~E.}}
(\byear{2003}).
\btitle{Analyzing multivariate data}.
\bpublisher{Thomson Brooks/Cole Pacific Grove, CA}.
\end{bbook}
\endbibitem

\bibitem[\protect\citeauthoryear{Lazar and Alexander}{2005}]{Lazar:2005}
\begin{barticle}[author]
\bauthor{\bsnm{Lazar},~\bfnm{Mariana}\binits{M.}} \AND
  \bauthor{\bsnm{Alexander},~\bfnm{Andrew~L}\binits{A.~L.}}
(\byear{2005}).
\btitle{Bootstrap white matter tractography (BOOT-TRAC)}.
\bjournal{NeuroImage}
\bvolume{24}
\bpages{524--532}.
\end{barticle}
\endbibitem

\bibitem[\protect\citeauthoryear{Lewicki}{1998}]{Lewicki:1998}
\begin{barticle}[author]
\bauthor{\bsnm{Lewicki},~\bfnm{Michael~S}\binits{M.~S.}}
(\byear{1998}).
\btitle{A review of methods for spike sorting: the detection and classification
  of neural action potentials}.
\bjournal{Network: Computation in Neural Systems}
\bvolume{9}
\bpages{R53--R78}.
\end{barticle}
\endbibitem

\bibitem[\protect\citeauthoryear{Lewis, Matthews and
  Guerrieri}{1993}]{Lewis:1993}
\begin{barticle}[author]
\bauthor{\bsnm{Lewis},~\bfnm{John~S}\binits{J.~S.}},
  \bauthor{\bsnm{Matthews},~\bfnm{Mildred~Shapley}\binits{M.~S.}} \AND
  \bauthor{\bsnm{Guerrieri},~\bfnm{Mary~L}\binits{M.~L.}}
(\byear{1993}).
\btitle{Resources of near-Earth space}.
\bjournal{Resources of near-earth space}.
\end{barticle}
\endbibitem

\bibitem[\protect\citeauthoryear{Lin, Rao and Dunson}{2017}]{Lin:2017}
\begin{barticle}[author]
\bauthor{\bsnm{Lin},~\bfnm{Lizhen}\binits{L.}},
  \bauthor{\bsnm{Rao},~\bfnm{Vinayak}\binits{V.}} \AND
  \bauthor{\bsnm{Dunson},~\bfnm{David}\binits{D.}}
(\byear{2017}).
\btitle{BAYESIAN NONPARAMETRIC INFERENCE ON THE STIEFEL MANIFOLD}.
\bjournal{Statistica Sinica}
\bvolume{27}
\bpages{535--553}.
\end{barticle}
\endbibitem

\bibitem[\protect\citeauthoryear{Lui}{2012}]{Lui:2012}
\begin{barticle}[author]
\bauthor{\bsnm{Lui},~\bfnm{Yui~Man}\binits{Y.~M.}}
(\byear{2012}).
\btitle{Advances in matrix manifolds for computer vision}.
\bjournal{Image and Vision Computing}
\bvolume{30}
\bpages{380--388}.
\end{barticle}
\endbibitem

\bibitem[\protect\citeauthoryear{Lui and Beveridge}{2008}]{Lui:2008}
\begin{barticle}[author]
\bauthor{\bsnm{Lui},~\bfnm{Yui}\binits{Y.}} \AND
  \bauthor{\bsnm{Beveridge},~\bfnm{J}\binits{J.}}
(\byear{2008}).
\btitle{Grassmann registration manifolds for face recognition}.
\bjournal{Computer Vision--ECCV 2008}
\bpages{44--57}.
\end{barticle}
\endbibitem

\bibitem[\protect\citeauthoryear{Mardia and Jupp}{2009}]{Mardia:2009}
\begin{bbook}[author]
\bauthor{\bsnm{Mardia},~\bfnm{Kanti~V}\binits{K.~V.}} \AND
  \bauthor{\bsnm{Jupp},~\bfnm{Peter~E}\binits{P.~E.}}
(\byear{2009}).
\btitle{Directional statistics}
\bvolume{494}.
\bpublisher{John Wiley \& Sons}.
\end{bbook}
\endbibitem

\bibitem[\protect\citeauthoryear{Mardia and Khatri}{1977}]{Mardia:1977}
\begin{barticle}[author]
\bauthor{\bsnm{Mardia},~\bfnm{KV}\binits{K.}} \AND
  \bauthor{\bsnm{Khatri},~\bfnm{CG}\binits{C.}}
(\byear{1977}).
\btitle{Uniform distribution on a Stiefel manifold}.
\bjournal{Journal of Multivariate Analysis}
\bvolume{7}
\bpages{468--473}.
\end{barticle}
\endbibitem

\bibitem[\protect\citeauthoryear{Mardia, Taylor and
  Subramaniam}{2007}]{Mardia:2007}
\begin{barticle}[author]
\bauthor{\bsnm{Mardia},~\bfnm{Kanti~V}\binits{K.~V.}},
  \bauthor{\bsnm{Taylor},~\bfnm{Charles~C}\binits{C.~C.}} \AND
  \bauthor{\bsnm{Subramaniam},~\bfnm{Ganesh~K}\binits{G.~K.}}
(\byear{2007}).
\btitle{Protein bioinformatics and mixtures of bivariate von Mises
  distributions for angular data}.
\bjournal{Biometrics}
\bvolume{63}
\bpages{505--512}.
\end{barticle}
\endbibitem

\bibitem[\protect\citeauthoryear{McGraw et~al.}{2006}]{Mcgraw:2006}
\begin{binproceedings}[author]
\bauthor{\bsnm{McGraw},~\bfnm{Tim}\binits{T.}},
  \bauthor{\bsnm{Vemuri},~\bfnm{Baba}\binits{B.}},
  \bauthor{\bsnm{Yezierski},~\bfnm{Robert}\binits{R.}} \AND
  \bauthor{\bsnm{Mareci},~\bfnm{Thomas}\binits{T.}}
(\byear{2006}).
\btitle{Segmentation of high angular resolution diffusion MRI modeled as a
  field of von Mises-Fisher mixtures}.
In \bbooktitle{European Conference on Computer Vision}
\bpages{463--475}.
\bpublisher{Springer}.
\end{binproceedings}
\endbibitem

\bibitem[\protect\citeauthoryear{McKenna, Raja and Gong}{1999}]{Mckenna:1999}
\begin{barticle}[author]
\bauthor{\bsnm{McKenna},~\bfnm{Stephen~J}\binits{S.~J.}},
  \bauthor{\bsnm{Raja},~\bfnm{Yogesh}\binits{Y.}} \AND
  \bauthor{\bsnm{Gong},~\bfnm{Shaogang}\binits{S.}}
(\byear{1999}).
\btitle{Tracking colour objects using adaptive mixture models}.
\bjournal{Image and vision computing}
\bvolume{17}
\bpages{225--231}.
\end{barticle}
\endbibitem

\bibitem[\protect\citeauthoryear{McLachlan and Peel}{2004}]{Mclachlan:2004}
\begin{bbook}[author]
\bauthor{\bsnm{McLachlan},~\bfnm{Geoffrey}\binits{G.}} \AND
  \bauthor{\bsnm{Peel},~\bfnm{David}\binits{D.}}
(\byear{2004}).
\btitle{Finite mixture models}.
\bpublisher{John Wiley \& Sons}.
\end{bbook}
\endbibitem

\bibitem[\protect\citeauthoryear{Mitra et~al.}{2013}]{Mitra:2013}
\begin{barticle}[author]
\bauthor{\bsnm{Mitra},~\bfnm{Riten}\binits{R.}},
  \bauthor{\bsnm{M{\"u}ller},~\bfnm{Peter}\binits{P.}},
  \bauthor{\bsnm{Liang},~\bfnm{Shoudan}\binits{S.}},
  \bauthor{\bsnm{Yue},~\bfnm{Lu}\binits{L.}} \AND
  \bauthor{\bsnm{Ji},~\bfnm{Yuan}\binits{Y.}}
(\byear{2013}).
\btitle{A bayesian graphical model for chip-seq data on histone modifications}.
\bjournal{Journal of the American Statistical Association}
\bvolume{108}
\bpages{69--80}.
\end{barticle}
\endbibitem

\bibitem[\protect\citeauthoryear{Mori and Zhang}{2006}]{Mori:2006}
\begin{barticle}[author]
\bauthor{\bsnm{Mori},~\bfnm{Susumu}\binits{S.}} \AND
  \bauthor{\bsnm{Zhang},~\bfnm{Jiangyang}\binits{J.}}
(\byear{2006}).
\btitle{Principles of diffusion tensor imaging and its applications to basic
  neuroscience research}.
\bjournal{Neuron}
\bvolume{51}
\bpages{527--539}.
\end{barticle}
\endbibitem

\bibitem[\protect\citeauthoryear{Muirhead}{1975}]{Muirhead:1975}
\begin{barticle}[author]
\bauthor{\bsnm{Muirhead},~\bfnm{Robb~J}\binits{R.~J.}}
(\byear{1975}).
\btitle{Expressions for some hypergeometric functions of matrix argument with
  applications}.
\bjournal{Journal of multivariate analysis}
\bvolume{5}
\bpages{283--293}.
\end{barticle}
\endbibitem

\bibitem[\protect\citeauthoryear{Muirhead}{2009}]{Muirhead:2009}
\begin{bbook}[author]
\bauthor{\bsnm{Muirhead},~\bfnm{Robb~J}\binits{R.~J.}}
(\byear{2009}).
\btitle{Aspects of multivariate statistical theory}
\bvolume{197}.
\bpublisher{John Wiley \& Sons}.
\end{bbook}
\endbibitem

\bibitem[\protect\citeauthoryear{Picard}{2007}]{Picard:2007}
\begin{barticle}[author]
\bauthor{\bsnm{Picard},~\bfnm{Franck}\binits{F.}}
(\byear{2007}).
\btitle{An introduction to mixture models}.
\bjournal{Statistics for Systems Biology, Research Report}
\bvolume{7}.
\end{barticle}
\endbibitem

\bibitem[\protect\citeauthoryear{Rand}{1971}]{rand1971objective}
\begin{barticle}[author]
\bauthor{\bsnm{Rand},~\bfnm{William~M}\binits{W.~M.}}
(\byear{1971}).
\btitle{Objective criteria for the evaluation of clustering methods}.
\bjournal{Journal of the American Statistical association}
\bvolume{66}
\bpages{846--850}.
\end{barticle}
\endbibitem

\bibitem[\protect\citeauthoryear{Reisinger et~al.}{2010}]{Reisinger:2010}
\begin{binproceedings}[author]
\bauthor{\bsnm{Reisinger},~\bfnm{Joseph}\binits{J.}},
  \bauthor{\bsnm{Waters},~\bfnm{Austin}\binits{A.}},
  \bauthor{\bsnm{Silverthorn},~\bfnm{Bryan}\binits{B.}} \AND
  \bauthor{\bsnm{Mooney},~\bfnm{Raymond~J}\binits{R.~J.}}
(\byear{2010}).
\btitle{Spherical topic models}.
In \bbooktitle{Proceedings of the 27th international conference on machine
  learning (ICML-10)}
\bpages{903--910}.
\end{binproceedings}
\endbibitem

\bibitem[\protect\citeauthoryear{Robbins}{1985}]{Robbins:1985}
\begin{bincollection}[author]
\bauthor{\bsnm{Robbins},~\bfnm{Herbert}\binits{H.}}
(\byear{1985}).
\btitle{An empirical Bayes approach to statistics}.
In \bbooktitle{Herbert Robbins Selected Papers}
\bpages{41--47}.
\bpublisher{Springer}.
\end{bincollection}
\endbibitem

\bibitem[\protect\citeauthoryear{Rokach and Maimon}{2005}]{Rokach:2005}
\begin{bmisc}[author]
\bauthor{\bsnm{Rokach},~\bfnm{Lior}\binits{L.}} \AND
  \bauthor{\bsnm{Maimon},~\bfnm{Oded}\binits{O.}}
(\byear{2005}).
\btitle{The Data Mining and Knowledge Discovery Handbook: A Complete Guide for
  Researchers and Practitioners}.
\end{bmisc}
\endbibitem

\bibitem[\protect\citeauthoryear{Rudin et~al.}{1964}]{Rudin:1964}
\begin{bbook}[author]
\bauthor{\bsnm{Rudin},~\bfnm{Walter}\binits{W.}} \betal{et~al.}
(\byear{1964}).
\btitle{Principles of mathematical analysis}
\bvolume{3}.
\bpublisher{McGraw-Hill New York}.
\end{bbook}
\endbibitem

\bibitem[\protect\citeauthoryear{Satterthwaite
  et~al.}{2014}]{Satterthwaite:2014}
\begin{barticle}[author]
\bauthor{\bsnm{Satterthwaite},~\bfnm{Theodore~D}\binits{T.~D.}},
  \bauthor{\bsnm{Elliott},~\bfnm{Mark~A}\binits{M.~A.}},
  \bauthor{\bsnm{Ruparel},~\bfnm{Kosha}\binits{K.}},
  \bauthor{\bsnm{Loughead},~\bfnm{James}\binits{J.}},
  \bauthor{\bsnm{Prabhakaran},~\bfnm{Karthik}\binits{K.}},
  \bauthor{\bsnm{Calkins},~\bfnm{Monica~E}\binits{M.~E.}},
  \bauthor{\bsnm{Hopson},~\bfnm{Ryan}\binits{R.}},
  \bauthor{\bsnm{Jackson},~\bfnm{Chad}\binits{C.}},
  \bauthor{\bsnm{Keefe},~\bfnm{Jack}\binits{J.}},
  \bauthor{\bsnm{Riley},~\bfnm{Marisa}\binits{M.}} \betal{et~al.}
(\byear{2014}).
\btitle{Neuroimaging of the Philadelphia neurodevelopmental cohort}.
\bjournal{Neuroimage}
\bvolume{86}
\bpages{544--553}.
\end{barticle}
\endbibitem

\bibitem[\protect\citeauthoryear{Schwartz}{1965}]{Schwartz:1965}
\begin{barticle}[author]
\bauthor{\bsnm{Schwartz},~\bfnm{Lorraine}\binits{L.}}
(\byear{1965}).
\btitle{On bayes procedures}.
\bjournal{Probability Theory and Related Fields}
\bvolume{4}
\bpages{10--26}.
\end{barticle}
\endbibitem

\bibitem[\protect\citeauthoryear{Schwartzman}{2006}]{Schwartzman:2006}
\begin{bphdthesis}[author]
\bauthor{\bsnm{Schwartzman},~\bfnm{Armin}\binits{A.}}
(\byear{2006}).
\btitle{Random ellipsoids and false discovery rates: Statistics for diffusion
  tensor imaging data}
\btype{PhD thesis},
\bpublisher{Stanford University}.
\end{bphdthesis}
\endbibitem

\bibitem[\protect\citeauthoryear{Spiegelhalter
  et~al.}{2002}]{Spiegelhalter:2002}
\begin{barticle}[author]
\bauthor{\bsnm{Spiegelhalter},~\bfnm{David~J.}\binits{D.~J.}},
  \bauthor{\bsnm{Best},~\bfnm{Nicola~G.}\binits{N.~G.}},
  \bauthor{\bsnm{Carlin},~\bfnm{Bradley~P.}\binits{B.~P.}} \AND
  \bauthor{\bsnm{Van Der~Linde},~\bfnm{Angelika}\binits{A.}}
(\byear{2002}).
\btitle{Bayesian measures of model complexity and fit}.
\bjournal{Journal of the Royal Statistical Society: Series B (Statistical
  Methodology)}
\bvolume{64}
\bpages{583--639}.
\bdoi{10.1111/1467-9868.00353}
\end{barticle}
\endbibitem

\bibitem[\protect\citeauthoryear{Stauffer and Grimson}{1999}]{Stauffer:1999}
\begin{binproceedings}[author]
\bauthor{\bsnm{Stauffer},~\bfnm{Chris}\binits{C.}} \AND
  \bauthor{\bsnm{Grimson},~\bfnm{W~Eric~L}\binits{W.~E.~L.}}
(\byear{1999}).
\btitle{Adaptive background mixture models for real-time tracking}.
In \bbooktitle{Computer Vision and Pattern Recognition, 1999. IEEE Computer
  Society Conference on.}
\bvolume{2}
\bpages{246--252}.
\bpublisher{IEEE}.
\end{binproceedings}
\endbibitem

\bibitem[\protect\citeauthoryear{Tang, Chu and Huang}{2009}]{Tang:2009}
\begin{binproceedings}[author]
\bauthor{\bsnm{Tang},~\bfnm{Hao}\binits{H.}},
  \bauthor{\bsnm{Chu},~\bfnm{Stephen~M}\binits{S.~M.}} \AND
  \bauthor{\bsnm{Huang},~\bfnm{Thomas~S}\binits{T.~S.}}
(\byear{2009}).
\btitle{Generative model-based speaker clustering via mixture of von
  mises-fisher distributions}.
In \bbooktitle{Acoustics, Speech and Signal Processing, 2009. ICASSP 2009. IEEE
  International Conference on}
\bpages{4101--4104}.
\bpublisher{IEEE}.
\end{binproceedings}
\endbibitem

\bibitem[\protect\citeauthoryear{Titterington et~al.}{2006}]{Robert:2006}
\begin{barticle}[author]
\bauthor{\bsnm{Titterington},~\bfnm{D.~M.}\binits{D.~M.}},
  \bauthor{\bsnm{Robert},~\bfnm{C.~P.}\binits{C.~P.}},
  \bauthor{\bsnm{Forbes},~\bfnm{F.}\binits{F.}} \AND
  \bauthor{\bsnm{Celeux},~\bfnm{G.}\binits{G.}}
(\byear{2006}).
\btitle{Deviance information criteria for missing data models}.
\bjournal{Bayesian Analysis}
\bvolume{1}
\bpages{651--673}.
\end{barticle}
\endbibitem

\bibitem[\protect\citeauthoryear{Turaga, Veeraraghavan and
  Chellappa}{2008}]{Turaga:2008}
\begin{binproceedings}[author]
\bauthor{\bsnm{Turaga},~\bfnm{Pavan}\binits{P.}},
  \bauthor{\bsnm{Veeraraghavan},~\bfnm{Ashok}\binits{A.}} \AND
  \bauthor{\bsnm{Chellappa},~\bfnm{Rama}\binits{R.}}
(\byear{2008}).
\btitle{Statistical analysis on Stiefel and Grassmann manifolds with
  applications in computer vision}.
In \bbooktitle{Computer Vision and Pattern Recognition, 2008. CVPR 2008. IEEE
  Conference on}
\bpages{1--8}.
\bpublisher{IEEE}.
\end{binproceedings}
\endbibitem

\bibitem[\protect\citeauthoryear{Turaga et~al.}{2011}]{Turaga:2011}
\begin{barticle}[author]
\bauthor{\bsnm{Turaga},~\bfnm{Pavan}\binits{P.}},
  \bauthor{\bsnm{Veeraraghavan},~\bfnm{Ashok}\binits{A.}},
  \bauthor{\bsnm{Srivastava},~\bfnm{Anuj}\binits{A.}} \AND
  \bauthor{\bsnm{Chellappa},~\bfnm{Rama}\binits{R.}}
(\byear{2011}).
\btitle{Statistical computations on Grassmann and Stiefel manifolds for image
  and video-based recognition}.
\bjournal{IEEE Transactions on Pattern Analysis and Machine Intelligence}
\bvolume{33}
\bpages{2273--2286}.
\end{barticle}
\endbibitem

\bibitem[\protect\citeauthoryear{Vinh, Epps and
  Bailey}{2010}]{vinh2010information}
\begin{barticle}[author]
\bauthor{\bsnm{Vinh},~\bfnm{Nguyen~Xuan}\binits{N.~X.}},
  \bauthor{\bsnm{Epps},~\bfnm{Julien}\binits{J.}} \AND
  \bauthor{\bsnm{Bailey},~\bfnm{James}\binits{J.}}
(\byear{2010}).
\btitle{Information theoretic measures for clusterings comparison: Variants,
  properties, normalization and correction for chance}.
\bjournal{Journal of Machine Learning Research}
\bvolume{11}
\bpages{2837--2854}.
\end{barticle}
\endbibitem

\bibitem[\protect\citeauthoryear{Wood et~al.}{2004}]{Wood:2004}
\begin{binproceedings}[author]
\bauthor{\bsnm{Wood},~\bfnm{E}\binits{E.}},
  \bauthor{\bsnm{Fellows},~\bfnm{M}\binits{M.}},
  \bauthor{\bsnm{Donoghue},~\bfnm{JR}\binits{J.}} \AND
  \bauthor{\bsnm{Black},~\bfnm{MJ}\binits{M.}}
(\byear{2004}).
\btitle{Automatic spike sorting for neural decoding}.
In \bbooktitle{Engineering in Medicine and Biology Society, 2004. IEMBS'04.
  26th Annual International Conference of the IEEE}
\bvolume{2}
\bpages{4009--4012}.
\bpublisher{IEEE}.
\end{binproceedings}
\endbibitem

\bibitem[\protect\citeauthoryear{Woolrich et~al.}{2009}]{Woolrich:2009}
\begin{barticle}[author]
\bauthor{\bsnm{Woolrich},~\bfnm{Mark~W}\binits{M.~W.}},
  \bauthor{\bsnm{Jbabdi},~\bfnm{Saad}\binits{S.}},
  \bauthor{\bsnm{Patenaude},~\bfnm{Brian}\binits{B.}},
  \bauthor{\bsnm{Chappell},~\bfnm{Michael}\binits{M.}},
  \bauthor{\bsnm{Makni},~\bfnm{Salima}\binits{S.}},
  \bauthor{\bsnm{Behrens},~\bfnm{Timothy}\binits{T.}},
  \bauthor{\bsnm{Beckmann},~\bfnm{Christian}\binits{C.}},
  \bauthor{\bsnm{Jenkinson},~\bfnm{Mark}\binits{M.}} \AND
  \bauthor{\bsnm{Smith},~\bfnm{Stephen~M}\binits{S.~M.}}
(\byear{2009}).
\btitle{Bayesian analysis of neuroimaging data in FSL}.
\bjournal{Neuroimage}
\bvolume{45}
\bpages{S173--S186}.
\end{barticle}
\endbibitem

\bibitem[\protect\citeauthoryear{Wright and Nocedal}{1999}]{Wright1999}
\begin{barticle}[author]
\bauthor{\bsnm{Wright},~\bfnm{Stephen~J}\binits{S.~J.}} \AND
  \bauthor{\bsnm{Nocedal},~\bfnm{Jorge}\binits{J.}}
(\byear{1999}).
\btitle{Numerical optimization}.
\bjournal{Springer Science}
\bvolume{35}
\bpages{7}.
\end{barticle}
\endbibitem

\bibitem[\protect\citeauthoryear{Yabushita, Hasegawa and
  Kobayashi}{1979}]{Yabushita:1979}
\begin{barticle}[author]
\bauthor{\bsnm{Yabushita},~\bfnm{Shin}\binits{S.}},
  \bauthor{\bsnm{Hasegawa},~\bfnm{Ichiro}\binits{I.}} \AND
  \bauthor{\bsnm{Kobayashi},~\bfnm{Kazushi}\binits{K.}}
(\byear{1979}).
\btitle{The Distributions of Inclination and Perihelion Latitude of Long-Period
  Comets and Their Dynamical Implications}.
\bjournal{Publications of the Astronomical Society of Japan}
\bvolume{31}
\bpages{801}.
\end{barticle}
\endbibitem

\bibitem[\protect\citeauthoryear{Zeng et~al.}{2015}]{Zeng:2015}
\begin{barticle}[author]
\bauthor{\bsnm{Zeng},~\bfnm{Xianhua}\binits{X.}},
  \bauthor{\bsnm{Bian},~\bfnm{Wei}\binits{W.}},
  \bauthor{\bsnm{Liu},~\bfnm{Wei}\binits{W.}},
  \bauthor{\bsnm{Shen},~\bfnm{Jialie}\binits{J.}} \AND
  \bauthor{\bsnm{Tao},~\bfnm{Dacheng}\binits{D.}}
(\byear{2015}).
\btitle{Dictionary pair learning on Grassmann manifolds for image denoising}.
\bjournal{IEEE Transactions on Image Processing}
\bvolume{24}
\bpages{4556--4569}.
\end{barticle}
\endbibitem

\end{thebibliography}

%%%%%%%%%%%%%

\end{document}